\documentclass[longauth]{aa}  
\usepackage{graphicx}
\usepackage[varg]{txfonts}
\usepackage{hyperref}
\usepackage{mathtools}
\usepackage{longtable}
\usepackage{rotating}
\usepackage[utf8]{inputenc}
\usepackage{multicol}
\usepackage{color}
\usepackage{natbib}

\hypersetup{draft}

\begin{document} 

\title{LOFAR MSSS: Flattening low-frequency radio continuum spectra of nearby galaxies}
\author{K.\,T. Chy\.zy\inst{1}
\and W. Jurusik\inst{1}
\and J. Piotrowska\inst{1}
\and B. Nikiel-Wroczy\'nski\inst{1}
\and V. Heesen\inst{2,3} 
\and V. Vacca\inst{4}
\and N. Nowak\inst{1}
\and R. Paladino\inst{5}
\and P. Surma\inst{1}
\and S.\,S.~Sridhar\inst{6,7} 
\and G. Heald\inst{8}  
\and R. Beck\inst{9}
\and J. Conway\inst{10}
\and K. Sendlinger\inst{11}
\and M. Cury\l{}o\inst{1}
\and D. Mulcahy\inst{9,16}
\and J. W. Broderick\inst{7}
\and M. J. Hardcastle\inst{12}
\and J. R. Callingham\inst{7}
\and G. G\"urkan\inst{8}
\and M. Iacobelli\inst{7}
\and H.\,J.\,A. R\"ottgering\inst{13} 
\and B. Adebahr\inst{11}
\and A. Shulevski\inst{14}
\and R. -J. Dettmar\inst{11}
\and R. P. Breton\inst{15}
\and A. O. Clarke\inst{16}
\and J. S. Farnes\inst{17}
\and E. Orr\'u\inst{7}
\and V. N. Pandey\inst{7}
\and M. Pandey-Pommier\inst{18}
\and R. Pizzo\inst{7}
\and C. J. Riseley\inst{8}
\and A. Rowlinson\inst{7}
\and A. M. M. Scaife\inst{16}
\and A. J. Stewart\inst{19}
\and A. J. van der Horst\inst{20}
\and R. J. van Weeren\inst{13}
}
\offprints{Krzysztof T. Chy\.zy, \email{krzysztof.chyzy@uj.edu.pl}}
\institute{Astronomical Observatory of the Jagiellonian University, ul. Orla 171, 30-244 Krak\'ow, Poland\label{inst1}
\and Hamburger Sternwarte, Universit{\"a}t Hamburg, Gojenbergsweg 112, D-21029 Hamburg, Germany\label{inst2}
\and School of Physics and Astronomy, University of Southampton, Southampton SO17 1BJ, UK\label{inst3}
\and INAF–Osservatorio Astronomico di Cagliari, Via della Scienza 5, I-09047 Selargius (CA), Italy\label{inst4}
\and INAF-Istituto  di  Radioastronomia,  via  P.  Gobetti,  101,  40129
Bologna, Italy\label{inst5}
\and Kapteyn Astronomical Institute, University of Groningen, Postbus 800, 9700AV Groningen, The Netherlands\label{inst6}
\and ASTRON, the Netherlands Institute for Radio Astronomy, Postbus 2, 7990 AA, Dwingeloo, The Netherlands\label{inst7}
\and CSIRO Astronomy and Space Science, PO Box 1130, Bentley WA 6102, Australia\label{8}
\and Max-Planck-Institut für Radioastronomie, Auf dem Hügel 69, 53121 Bonn, Germany\label{inst9}
\and Department of Space, Earth and Environment, Chalmers University of Technology, Onsala Space Observatory, S-43992, Sweden \label{inst10}
\and Astronomisches Institut der Ruhr-Universit{\"a}t Bochum, Universit{\"a}tsstr.
150, 44801 Bochum, Germany\label{inst11}
\and Centre for Astrophysics Research, School of Physics, Astronomy and Mathematics, University of Hertfordshire, College Lane, Hatfield AL10 9AB, UK\label{12}
\and Leiden Observatory, Leiden University, PO Box 9513, 2300 RA Leiden, The Netherlands\label{13}
\and Anton Pannekoek Institute for Astronomy, University of Amsterdam, Postbus 94249, 1090 GE Amsterdam, The Netherlands\label{inst14}
\and Jodrell Bank Centre for Astrophysics, School of Physics and Astronomy, The University of Manchester, Manchester M13 9PL, UK\label{inst15}
\and University of Manchester, Jodrell Bank Centre for Astrophysics, Manchester M13 9PL, UK\label{inst16}
\and Oxford e-Research Centre, University of Oxford, Keble Road, Oxford OX1 3QG, England\label{inst17}
\and Univ Lyon, Univ Lyon1, Ens de Lyon, CNRS, Centre de Recherche Astrophysique de Lyon, UMR5574, F-69230, Saint-Genis-Laval, France\label{18}
\and Sydney Institute for Astronomy, School of Physics, The University of Sydney, NSW 2006, Australia\label{19}
\and Department of Physics, The George Washington University, 725 21st Street NW, Washington, DC 20052, USA\label{20}
}

\date{Received; accepted}

\abstract
{} 
{The shape of low-frequency radio continuum spectra of normal galaxies is not well understood, the key question being the role of physical processes such as thermal absorption in shaping them. In this work we take advantage of the LOFAR Multifrequency Snapshot Sky Survey (MSSS) to investigate such spectra for a large sample of nearby star-forming galaxies.}
{Using the measured 150\,MHz flux densities from the LOFAR MSSS survey and literature flux densities at various frequencies we have obtained integrated radio spectra for 106 galaxies characterised by different morphology and star formation rate. The spectra are explained through the use of a three-dimensional model of galaxy radio emission, and radiation transfer dependent on the galaxy viewing angle and absorption processes.}
{Our galaxies' spectra are generally flatter at lower compared to higher frequencies: the median spectral index $\alpha_{low}$ measured between $\approx$50\,MHz and 1.5\,GHz is $-0.57\pm0.01$ while the high-frequency one $\alpha_{high}$, calculated between 1.3\,GHz and 5\,GHz, is $-0.77\pm0.03$. As there is no tendency for the highly inclined galaxies to have more flattened low-frequency spectra, we argue that the observed flattening is not due to thermal absorption, contradicting the suggestion of \citet{israel90}. 
According to our modelled radio maps for M\,51-like galaxies, the free-free absorption effects can be seen only below 30\,MHz and in the global spectra just below 20\,MHz, while in the spectra of starburst galaxies, like M\,82, the flattening due to absorption is instead visible up to higher frequencies of about 150\,MHz. Starbursts are however scarce in the local Universe, in accordance with the weak spectral curvature seen in the galaxies of our sample. Locally, within galactic disks, the absorption effects are distinctly visible in M\,51-like galaxies as spectral flattening around 100-200\,MHz in the face-on objects, and as turnovers in the edge-on ones, while in M\,82-like galaxies there are strong turnovers at frequencies above 700\,MHz, regardless of viewing angle.
}
{Our modelling of galaxy spectra suggests that the weak spectral flattening observed in the nearby galaxies studied here results principally from synchrotron spectral curvature due to cosmic ray energy losses and propagation effects. We predict much stronger effects of thermal absorption in more distant galaxies with high star formation rates. Some influence exerted by the Milky Way's foreground on the spectra of all external galaxies is also expected at very low frequencies.}
\maketitle

\section{Introduction}
\label{s:intro}
The radio emission from normal star-forming galaxies traces the underlying distributions of thermal and relativistic plasmas, cosmic ray (CR) electrons, and magnetic fields, thus providing vital information about the physical processes at work in galaxies. Studying the radio emission at different frequencies via the radio continuum spectra of galaxies allows us to understand radio emission processes and the  structure and local properties of the galaxy's interstellar medium (ISM). 

The shape of radio continuum spectra can be characterised to first
order by their power-law  spectral index $\alpha$ ($S_\nu \approx
\nu^{\alpha}$), the value of which can be related to the various
radiation processes responsible for the emission. An optically thin plasma yields $\alpha =-0.1$ for thermal bremsstrahlung radiation, while synchrotron radiation gives $\alpha\approx -0.5$ for freshly accelerated CR electrons injected from supernova remnants into the star-forming disk. These CR electrons can sustain considerable synchrotron and inverse Compton radiation losses, giving rise to steeper spectra, with $\alpha$ also dependent on the structure and strength of the magnetic field and the confinement of CRs \citep{beck13,han17}. 

The transport of CR electrons away from supernova remnants can take the form of either diffusion, which depends on the magnetic field structure, or advection in a galactic wind \citep{{pohl91},heesen16,{heesen18}}. Therefore, galaxy spectra, particularly the integrated (global) ones, depend on  a complex interplay between thermal and nonthermal components, CR electron energy losses, and propagation effects \citep{lisenfeld00}.  

At low radio frequencies, the spectra of galaxies are expected to be modified by additional mechanisms. For instance, \ion{H}{II} regions become optically thick at low frequencies, which affects not only the propagation of free-free emitted photons but also the transmission of photons generated from synchrotron emission. At low frequencies, relativistic bremsstrahlung and especially ionisation losses can also be much more important than at higher frequencies, in particular in starburst galaxies \citep{murphy09}. Finally synchrotron self-absorption and Razin effects can further suppress the radio emission from regions of dense ISM and produce breaks in the galaxy spectra below 10\,MHz \citep{lacki13}.

Our current understanding of low-frequency spectra of galaxies and the
processes that shape them is limited, mostly due to the lack of high-quality radio observations at these frequencies. In the well-known work of \citet{israel90}, flux densities of 68 galaxies were observed with the Clarke Lake Telescope at 57\,MHz and compared with flux densities extrapolated from high-frequency measurements. The authors  found that flux densities at 57\,MHz were systematically lower than expected, with the difference largest for highly inclined objects, which they interpreted as evidence that the low-frequency flattening was due to free-free absorption from thermal gas concentrated in the plane of galaxy disks.  A new ISM phase was proposed, containing an ionised but relatively cool ($T<1000$\,K) gas, to account for this absorption. However, other authors have subsequently come to different conclusions: for example \citet{hummel91} after reanalysing the same data, confirmed the reduction of radio emission but questioned whether this flattening is correlated with galaxy inclination. The authors instead proposed that the observed spectral breaks were due to the steepening of high-frequency spectra caused by energy losses of the CR electrons propagating within the galaxy. Due to the difficulties of obtaining high-quality low-frequency measurements, the discussion on the shape of galaxy spectra and the role of thermal absorption has not yet been resolved \citep{{marvil15},{basu15}, {mulcahy18}}.

It is clear that thermal absorption is required to account for the observed spectra of local regions in the centres of galaxies, including the Milky Way \cite[e.g.][]{roy06}. Such an effect can also clearly be seen in the centre of M\,82 at 408\,MHz \citep{wills97} and at 150\,MHz \citep{varenius15} and likely also affects the integrated spectrum of M\,82, being visible as a weak flattening of the spectrum at frequencies below 300\,MHz  \citep{{condon92},{yoast13},{lacki13},{adebahr13}}. The Milky Way also shows a spectral turnover of the global spectrum, but at a much lower frequencies of about 3\,MHz \citep{brown73}. The frequency of such spectral breaks is related to the amount of ionised gas, specifically of the warm ionised medium, and therefore also to the age of star-forming regions, with recent star formation on a 10-Myr timescale being most important. Complex free-free absorption features can serve as an indicator of an early evolutionary state of a starburst, as shown by \citet{clemens10}. 

Free-free absorption can also strongly affect the relation between radio and far-infrared (FIR) emission \citep{schleicher13}. For the same reason it can modify the radio emission of high-redshift galaxies, thus influencing the source counts. Therefore, a better understanding of the role played by free-free absorption in galaxies is essential for studies of cosmological galaxy evolution.

A new observational facility, the Low Frequency Array \citep[LOFAR;][]{haarlem13}, opens up the possibility for
systematic studies of nearby galaxies at low frequencies and allows us
to reinvestigate the problems related to their low-frequency spectra.
The Multifrequency Snapshot Sky Survey \citep[MSSS;][]{heald15} covers the entire northern sky, 
enabling the detection of many catalogued nearby galaxies, which span a large range of star formation rates (SFRs), sizes, and morphological types.

In this paper, we determine flux densities at 150\,MHz from both the MSSS source catalogue and image mosaics for a large number of galaxies (Sect. \ref{s:data}). We then obtain global spectra of galaxies, combining our new MSSS flux densities with flux densities from the literature (Sect. \ref{s:results}). A three-dimensional (3D) model of galaxy radio emission is introduced (Sect. \ref{s:discussion}) for the purpose of analysing and interpreting these spectra. As cases in point, we analyse the spectra of M\,51 and M\,82 in detail, using spatially resolved information on the observed radio emission at different frequencies to constrain our model parameters. Synthetic radio intensity maps are then produced for various galaxy inclinations at different frequencies, which enable us to predict global spectra of galaxies and assess the role that free-free absorption plays in shaping them. In Sect. \ref{s:conclusions} we provide a summary and conclusions.

\section{Data and galaxy selection}
\label{s:data}

\subsection{Selection of the galaxy sample}
\label{s:sample}

\begin{figure*}
\centering
\includegraphics[clip,angle=0,width=6cm]{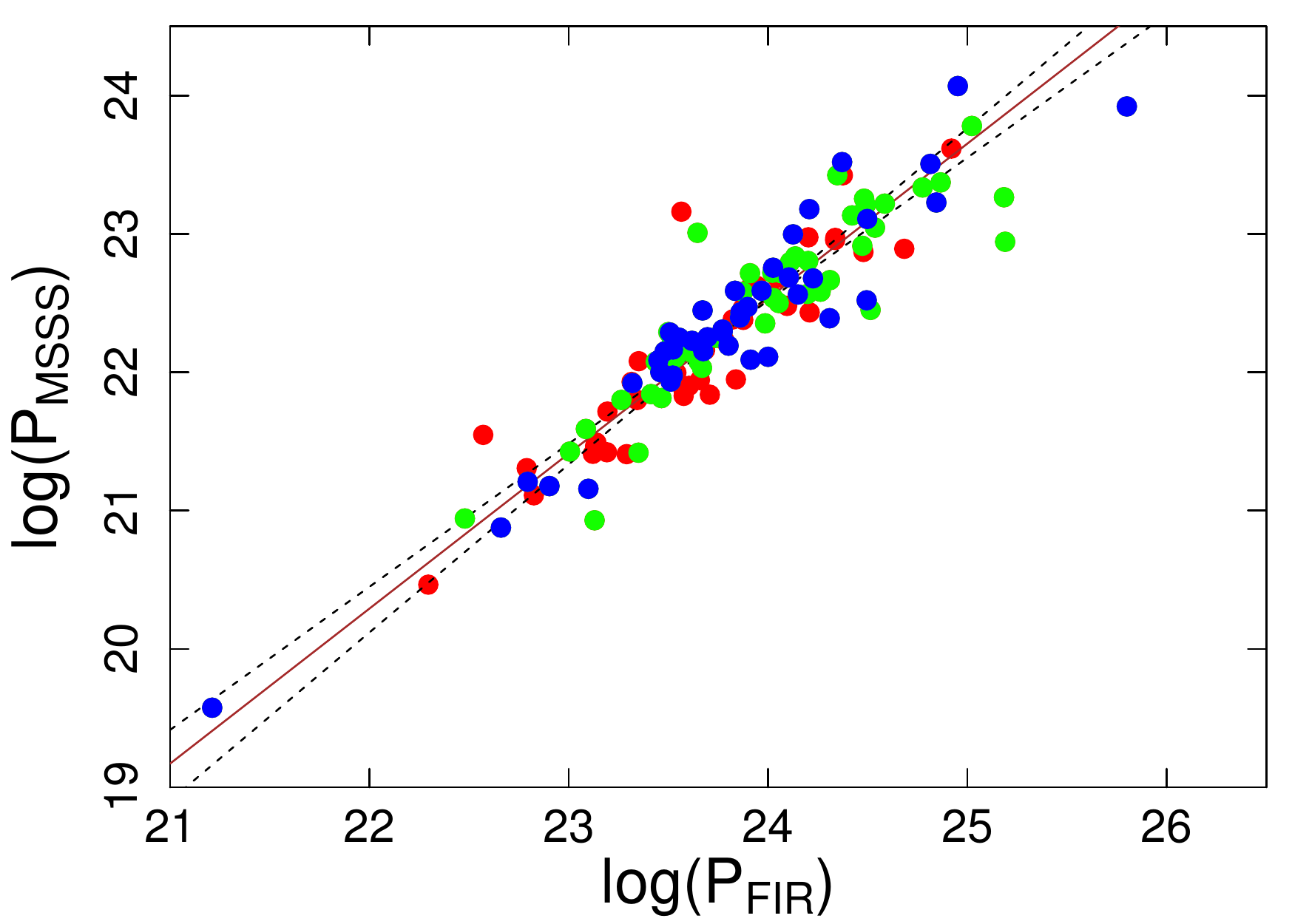}
\includegraphics[clip,angle=0,width=6cm]{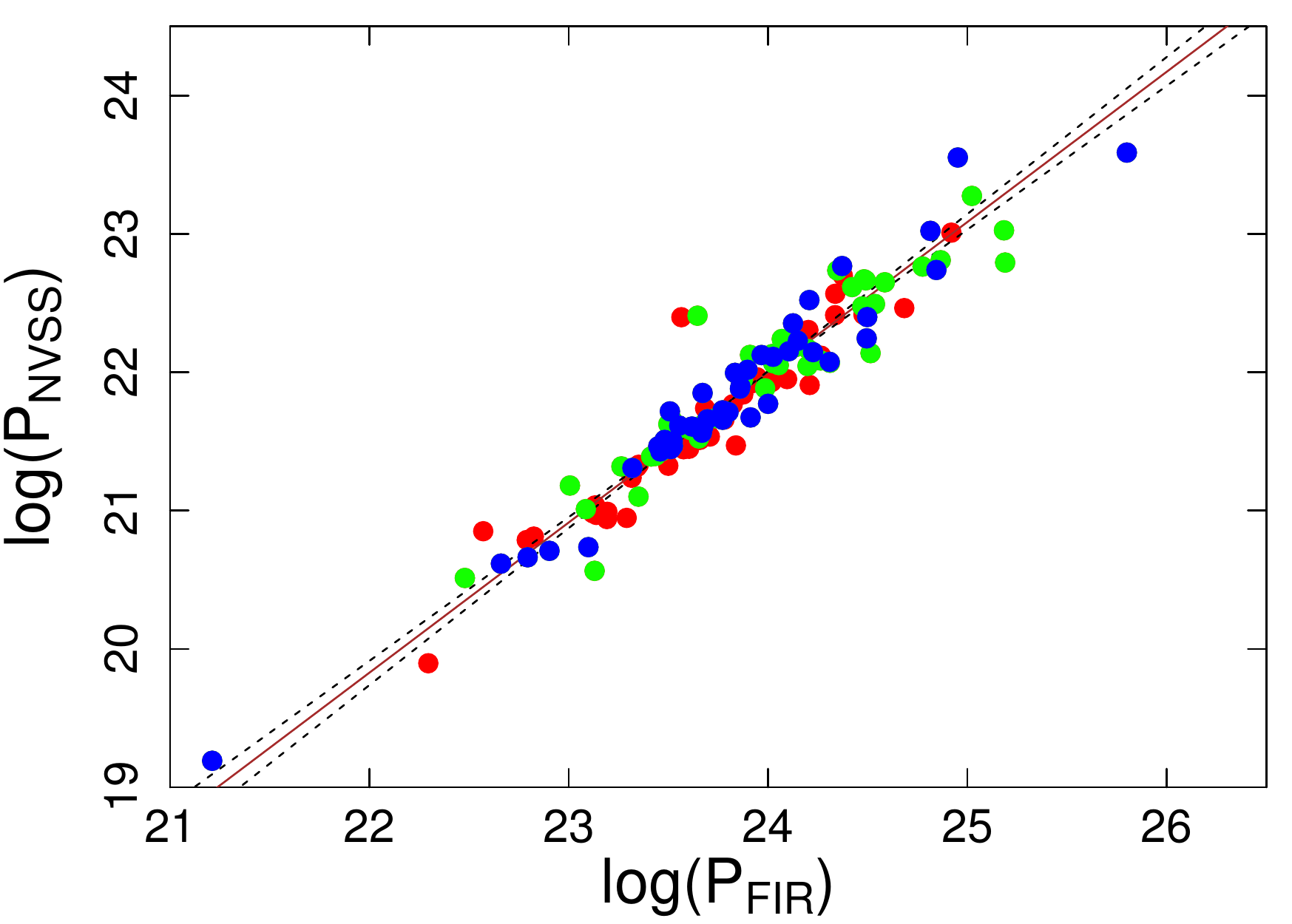}
\includegraphics[clip,angle=0,width=6cm]{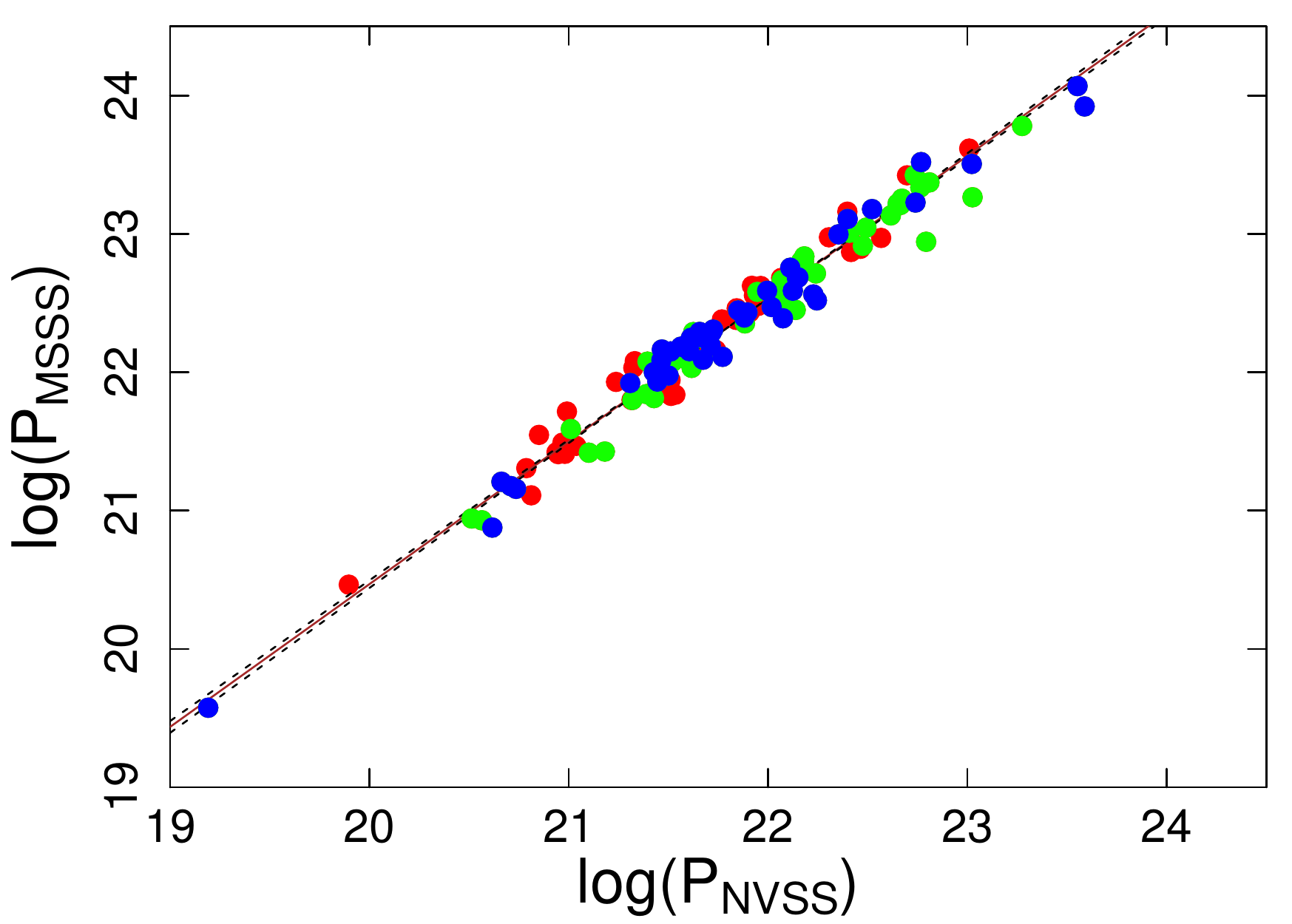}
\caption{Radio-FIR relation for the low-frequency MSSS (left) and high-frequency NVSS data (middle). The red, blue and green symbols
represent faint, medium bright, and bright NVSS galaxies, respectively. The MSSS-NVSS luminosity diagram is shown on the right. The solid line is a bisector fit, while the dashed lines represent simple X vs. Y and Y vs. X linear regressions in the logarithmic space of the  parameters.
}
\label{f:radio-FIR}
\end{figure*}

As the parent sample for our study we selected the compilation of \cite{yun01}, which contains radio counterparts to the IRAS
Redshift Survey galaxies detected in the NRAO VLA Sky Survey \citep[NVSS;][]{condon_etal_1998}.
The catalogue lists over 1800 IRAS flux-density-limited ($S_{60 \, \mu \rm{m}} 
\le 2$\,Jy) objects with known radio properties at 1.4\,GHz, and
constitutes the largest sample of this type within the local Universe.
It has been used to investigate the radio-luminosity function of galaxies, the radio-FIR correlation, and the extinction-free star-formation density for the local volume \citep{{yun01},{condon02}}.

The sample is not complete at low Galactic latitudes of $|b_{Gal}|<10\degr$. Therefore, we included a number of well-known galaxies from similar studies by \cite{condon87} and \cite{condon90}, namely: IC\,10, NGC\,628, UGC\,12914, NGC\,3646, NGC\,4217, NGC\,4449, and NGC\,5457.

In analysing galaxy spectra to study the role of thermal absorption, we were primarily interested in galaxies with Hubble types corresponding to spiral and irregular objects, galaxies that are known to have recent star formation. In order to have a uniform sample of such galaxies, we chose the following selection criteria:
\begin{itemize}
\item included in \cite{yun01}, 
supplemented with the above seven additional galaxies;
\item radio flux density $>50$\,mJy at 1.4\,GHz;
\item located in the northern hemisphere (Dec$>$0 degrees);
\item morphological type $T\ge0$ (according to the HyperLeda database) to avoid elliptical galaxies;
\item not dominated by an active galactic nucleus (AGN), for example, excluding NGC\,1275 and NGC\,4258, but taking into consideration M\,51 and M\,81 (with a low-luminosity LINER/Seyfert nucleus.)
\end{itemize}

Our initial sample fulfilling these criteria consisted of 200 galaxies.

\subsection{MSSS survey}
\label{s:MSSS}

We used the High Band Antenna (HBA) part of MSSS to measure low-frequency flux densities of our sample galaxies. Images from MSSS are available as sets of mosaics representing sky images of $10\degr\times10\degr$ in size. Each set consists of eight narrowband images generated with 2\,MHz bandwidth each, and with central frequencies ranging from 120 to 157\,MHz, as well as one broadband image obtained by averaging images from all bands.

We used the preliminary MSSS mosaics from the first internal version
of the MSSS source catalogue \citep[e.g. see details in][]{kokotanekov17}. The typical resolution of the mosaics is $3\arcmin$ and the root mean square (r.m.s.) noise level varies between about 15\,and 25\,mJy beam$^{-1}$. Differences in noise levels may 
result from fluctuations of the Milky Way foreground emission, 
distributions of strong sources in mosaics, imperfect calibration 
and ionospheric weather conditions during observations. The 
catalogue of automatically detected sources \citep[the MSSS source 
catalogue;][]{heald15} provides us with flux 
densities from the individual bands. The MSSS images, and thus the catalogued flux densities, were corrected as part of the MSSS analysis to mitigate the well-known LOFAR flux calibration transfer issues using the ``bootstrap'' method described by \cite{hardcastle16}; the residual flux calibration error should not exceed 10\%. A polynomial function was used with a Levenberg-Marquardt minimisation algorithm in order to determine the best-fit parameters of the multi-band spectrum in the logarithmic flux density-frequency space. The catalogue provided us with the parameters of the fit as well as the interpolated flux density of each galaxy at 150\,MHz used by us in this work.

\subsection{Flux density measurements}
\label{s:measurements}

Apart from the MSSS source catalogue, we also used our own flux density measurements based on the MSSS image mosaics; for these measurements we used the Common Astronomy Software Applications package \citep[CASA;][]{mcmullin07}. We applied a polygon mask to outline galactic emission in the averaged mosaic. This was then used to integrate flux densities in the eight individual mosaics of the various bands. The uncertainties of the flux densities were calculated as the local r.m.s. sensitivities $\sigma_{r.m.s}$  measured in the maps multiplied by the square root of the flux integration area $\Omega_{s}$ in units of the beam area $\Omega_{b}$: 
\begin{equation}
\sigma_{band}=\sigma_{r.m.s}\times \sqrt{\frac{\Omega_{s}}{\Omega_{b}}}.
\end{equation}

Finally, we employed a Levenberg-Marquardt minimisation method to fit
the multi-band power-law spectra to our measurements and to obtain the
interpolated flux densities at 150\,MHz $S_{150}$ together with their
uncertainties $\sigma_{L-M}$. We added an independent uncertainty of 5\% of the total flux density, stemming from the uncertainty of the absolute flux density scale and from calibration errors:
\begin{equation}
\sigma_{150}=\sqrt{\sigma_{L-M}^2+(0.05\,S_{150})^2}.
\end{equation}
Uncertainties of the flux densities from the MSSS source catalogue were calculated in a similar way.

The flux densities from CASA were generally similar to those from the MSSS catalogue, although we noticed that in some cases the CASA flux densities were slightly smaller, but still in agreement within the uncertainties. We attributed this to different ways of masking the area during the flux density integration: for the MSSS catalogue, each source comprises one or more Gaussian components identified within the boundaries of an automatically determined PyBDSF ``island mask'', whereas in CASA we manually specified polygon regions. The two distinct approaches caused slightly different pixel areas to be considered, with generally minimal discrepancy. If the flux densities from both methods were consistent, we used the catalogue flux density for further analysis. If, however, the flux densities differed significantly, for example due to a complicated source structure or due to a confusing background source close in projection to the target galaxy, we used the CASA measurements instead 
and applied a mask to omit background sources when integrating the flux. 
Due to the resolution of MSSS ($3\arcmin$) not being significantly finer than the typical angular sizes of the galaxies in our sample (mean value of $4\farcm4$), we were not able to deblend and remove background sources that are spatially coincident with a given galaxy.  
However, from our experience with observations of individual galaxies at various frequencies, the typical contribution from background sources is small, below 10\% of the integrated flux \citep[see e.g.][]{chyzy07}.

Table \ref{t:sample} presents the galaxies from our original sample that were detected in the MSSS survey and have reliable flux densities at 150\,MHz. All galaxies have redshifts less than 0.04 and constitute our sample of 129 sources. Some of these objects have been identified as galaxy pairs, unresolved in the MSSS mosaics. In these cases, the measured flux density refers to the system as a whole. Table \ref{t:sample} also provides the galaxy morphology (Hubble T-type), inclination angle of the galaxy disk ($90\degr$ corresponds to an edge-on object), the distance, and supplemental information concerning the flux density measurements and the constructed global spectra characterized by four different kinds of flags:
\begin{itemize}
\item interaction flag (according to the NED database): 1 - single source; 2 - luminous infrared galaxy (LIRG); 3 - strongly interacting galaxy but not LIRG;
\item flux density flag:  1 - flux density for a single source with an undisturbed disk; 2 - flux density for a single source but with unclear or amorphous disk plane, morphologically irregular dwarf, or starburst galaxy; 3 - flux density for a double or a triple source;
\item spectral flag (see Sect. \ref{s:spectra}): 1 - MSSS flux density fits well (within $\sigma_{150}$) the spectrum constructed from the literature data; 2 - MSSS flux density is within $1-2\sigma_{150}$ of the general spectral trend, or the spectral trend is estimated from only 2-3 flux density measurements; 3 - there is a lack of reliable/sufficient data to construct the global spectrum; 
\item flux density measurement method flag: this indicates whether the presented 150\,MHz flux densities come from: 1 - the MSSS source catalogue; or 2 - our CASA measurements from the MSSS image mosaics. 
\end{itemize}

\section{Results}
\label{s:results}

\subsection{Low-frequency radio-FIR correlation}
\label{s:radio-FIR}

In order to test the sample quality and identify possible objects with a substantial AGN contribution, we studied the radio-FIR relation for all 129 sources using galaxy luminosities at 150\,MHz (see Table \ref{t:sample} for the flux densities and distances we used) and at $60\,\mu$m \citep{yun01}. Furthermore the sample was divided into three subsamples, depending on the NVSS flux density at 1.4\,GHz \citep{yun01}: bright sources with $S_{NVSS}$>$143$\,mJy; medium bright sources with 87\,mJy<$S_{NVSS}$<$143$\,mJy; and faint sources with $S_{NVSS}$<$87$\,mJy. Each subsample consisted of 43 objects.

The low-frequency radio-FIR relation that we obtain is similar to the
high-frequency one (Fig. \ref{f:radio-FIR}), which confirms the high
quality of the MSSS data and corroborates the existence of a
well-defined relation between the FIR emission of heated dust grains
and the synchrotron emission of CR electrons, which dominates the radio
emission of galaxies at low frequencies. There are no distinct
outliers in this relation since we removed them from the sample
as described in Section 2.1. The spread of data
points in the diagram of MSSS and NVSS flux densities is even smaller
(Fig. \ref{f:radio-FIR}, right panel), which suggests that a
significant part of the dispersion in the radio-FIR relation comes
either from the uncertainties of the IRAS data or from the intrinsic
spread in the relation. Either way, our radio data do not appear to
be the limiting factor. 

Using the bisector method we estimated a slope in the radio-FIR power-law relation for the MSSS flux densities at 150\,MHz as $1.12 \pm 0.04$. For the same set of galaxies we obtained the slope for the NVSS data at 1.4\,GHz of $1.09\pm 0.03$. Our results are consistent with those found (once AGN
had been removed) by \cite{hardcastle16}, \cite{calistro-rivera17}, \cite{gurkan18}, and \cite{magnelli15}, who investigated the low-frequency radio-FIR relationships in
individual deep fields. The wide sky coverage of MSSS enables us
to explore this relationship in more detail in the local Universe (at $z\approx 0$), which will be presented in a forthcoming paper.

One would expect that the effects of absorption by dense thermal gas could be particularly evident in galaxies with high SFRs, which are also more luminous in the infrared band \citep{condon91}. Powerful infrared galaxies would thus become weaker in the low radio frequency range. Our galaxies do not show such behaviour, as can be seen in Fig. \ref{f:radio-FIR}, which may indicate that the effect of thermal absorption in our sample is relatively weak.

\subsection{Deriving galaxy spectra}
\label{s:spectra}
Our analysis of the spectral characteristics of radio emission of galaxies is based on their global spectra. We carried out an extensive search of the literature for galaxy flux densities at different frequencies, using mainly information from the NED and SIMBAD databases. We avoided old measurements, for example  early interferometric arrays before 1980. For 31 galaxies we found data in the GaLactic and Extragalactic All-sky Murchison Widefield Array \citep[GLEAM;][]{hurley17} survey, which provides a catalogue of integrated fluxes from 20 narrow-band images as well as a fitted 200\,MHz integrated flux density. For 19 galaxies from our sample, we did not find  literature data of sufficient quality to derive spectra. Such objects, marked in Table \ref{t:sample} by spectral flag 3, were excluded from further analysis. We also dropped double and multiple galaxy systems marked by the flux density flag 3 (Sect. \ref{s:measurements}). The resulting sample therefore consists of 106 galaxies.

\begin{figure*}
\centering
\includegraphics[clip,angle=0,width=1.0\textwidth,height=0.95\textheight]{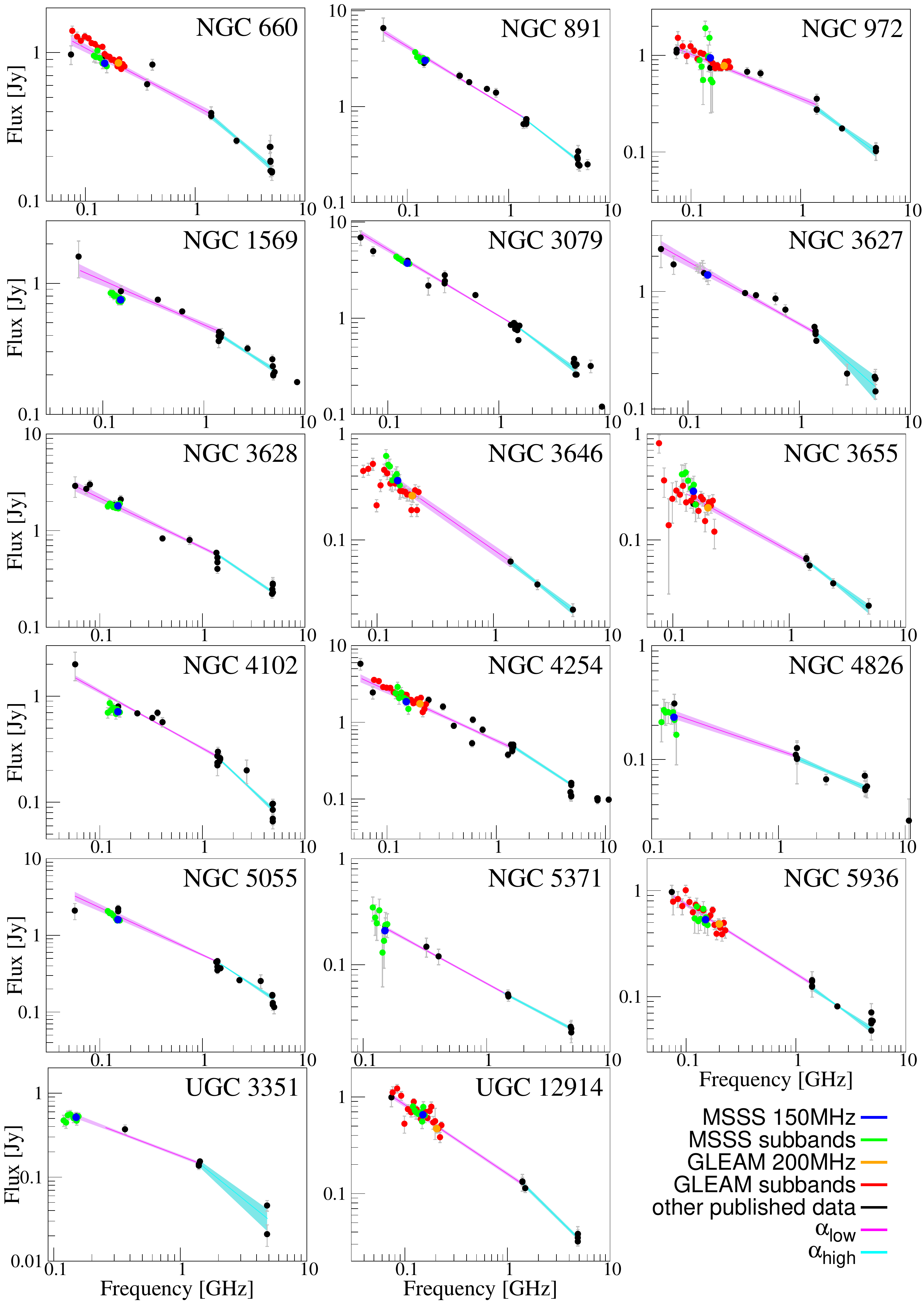}
\caption{Examples of radio spectra for a subset of MSSS galaxies. The flux densities from eight individual 
MSSS spectral bands are in green, the interpolated MSSS flux densities at 150\,MHz are in blue, the GLEAM flux densities from 20 individual spectral bands (if available) are in red, the GLEAM catalogue fitted flux densities at 200\,MHz are in orange, and other published measurements are in black. The pink and black lines represent the low- and high-frequency power-law fits, respectively. Highlighted regions show the 68\% confidence bands of the fits.
}
\label{f:spectra}
\end{figure*}

Some examples of spectral trends, integrated flux density measurements from the MSSS survey, and published data are presented in Fig. \ref{f:spectra}. The MSSS and GLEAM measurements are consistent with each other and follow the general spectral trends found in the published data. Our main interpretation of these diagrams is that the spectra tend to slightly flatten at low frequencies (e.g. NGC\,1569 and NGC\,4102). However, there are also spectra that appear to be quite straight (e.g. NGC\,3646 and NGC\,5936). 

\subsection{Spectral flattening}
\label{s:flattening}

The spectral curvatures seen in Fig. \ref{f:spectra} are small and usually not limited to just the lowest frequencies (e.g. NGC\,972 and NGC\,4102). Therefore, in order to evaluate spectral slopes, we constructed both low- and high-frequency spectral indices.
The low-frequency index ($\alpha_\mathrm{low}$) was obtained by fitting a power-law to the interpolated MSSS flux density at 150\,MHz and to
literature data from 50\,MHz up to 1.5\,GHz. When the GLEAM data were available we used the fitted flux density at 200\,MHz, and not the correlated data from individual sub-bands, which is similar to the approach with the MSSS data. The high-frequency index
($\alpha_\mathrm{high}$) between 1.3\,GHz and 5\,GHz was calculated in
a similar way. We did not consider frequencies higher than this so as to avoid the effects of spectral flattening due to the increasing thermal component: for example for a sample of galaxies, the average thermal fraction given by \cite{niklas97b} is $8\pm 1$\% at 1\,GHz, and by \cite{tabatabaei17} is ~10\% at 1\,GHz and ~23\% at 5\,GHz. Both the indices derived for all our galaxies are listed in Table \ref{t:sample} and some examples of the power-law spectral fits together with their 95\% confidence bands are presented in Fig. \ref{f:spectra}.

\begin{figure}
\centering
\includegraphics[clip,angle=0,width=9.0cm]{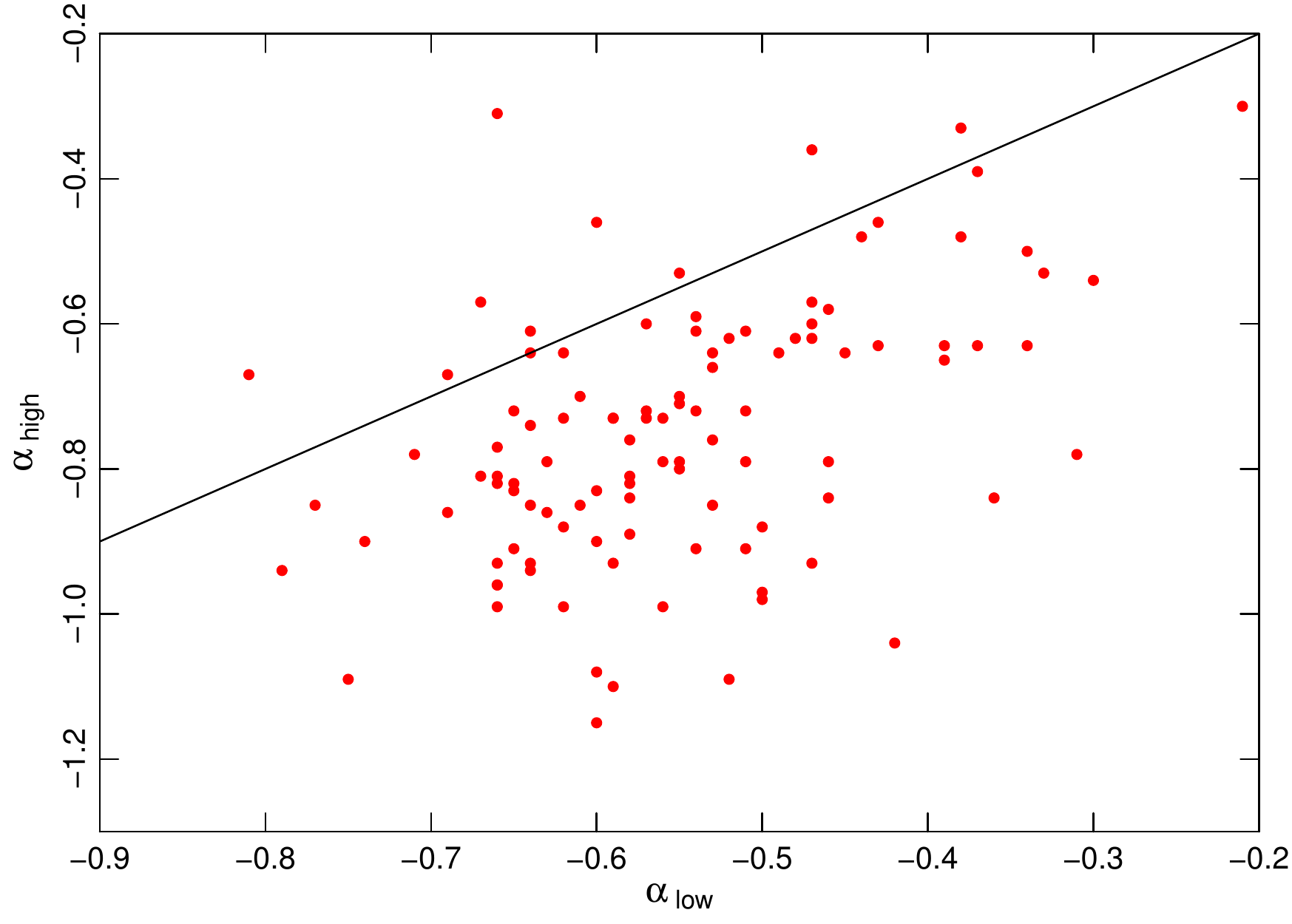}
\caption{Radio low-frequency $\alpha_\mathrm{low}$ and high-frequency $\alpha_\mathrm{high}$ spectral indices for our sample of 106 galaxies. The straight line corresponds to identical spectral index values, i.e. to simple power-law spectra.
}
\label{f:alpha_low_high}
\end{figure}

\begin{figure}
\centering
\includegraphics[angle=0,width=9cm]{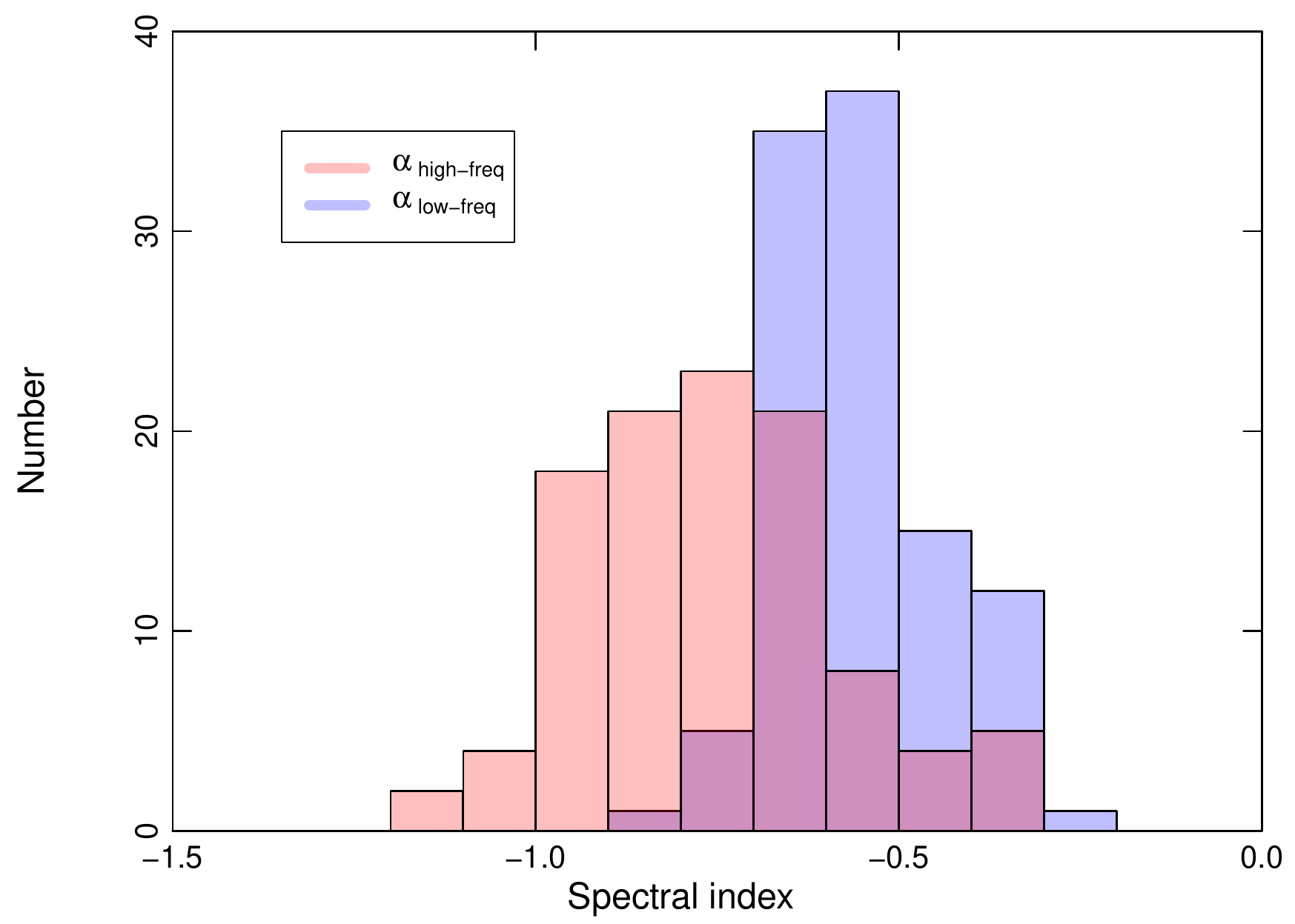}
\caption{Distributions of the radio low-frequency $\alpha_\mathrm{low}$ and high-frequency $\alpha_\mathrm{high}$ spectral indices for the MSSS galaxies (in blue and pink respectively). Overlapping parts of the histograms are in violet.
}
\label{f:alfa_hist}
\end{figure}

As can be seen in Fig. \ref{f:alpha_low_high}, there is a systematic difference between $\alpha_{\mathrm{low}}$ and $\alpha_\mathrm{high}$, which reflects the prevalence of galaxies that have flatter spectra at low frequencies than they have at high frequencies. The different spectral indices may result either from flattening of the low-frequency spectra or from steepening of the high-frequency ones; both have different underlying physical processes. The distributions of low- and high-frequency spectral indices are presented in Fig.~\ref{f:alfa_hist}. The distribution of the low-frequency index is wider and is shifted to smaller absolute values (indicating flatter spectra of galaxies). Accordingly, the high-frequency spectra can be observed as either steep or flat, while the low-frequency spectra are relatively flat. The two-sample Anderson-Darling, Kolmogoroff-Smirnov tests, and the Kruskal-Wallis rank test indicate that the hypothesis that the values of
$\alpha_\mathrm{high}$ and $\alpha_\mathrm{low}$ are derived from the
same population is highly unlikely since the  p-values{\footnote{The  p-value is the probability of obtaining by chance a result at least as extreme as that observed. We reject the statistical hypothesis if the p-value is equal or less than the standard significance level of 5\%.}} in all tests are
much smaller than 1\%.

The median value for $\alpha_\mathrm{low}$, with the uncertainty obtained using the bootstrap method as the standard deviation, is $-0.57\pm 0.01$. Analogously, for the high-frequency part of the spectrum, the spectral index $\alpha_\mathrm{high}$ is $-0.77\pm 0.03$. This high-frequency estimate is very close to the spectral index of $-0.74\pm 0.12$, as estimated by \cite{gioia82} between 408\,MHz and 10.7\,GHz for a sample of 57 galaxies, which confirms the statistical consistency of both samples. It is also close to the 1.4-10\,GHz spectral index of $-0.79\pm 0.15$ estimated for a sample of 36 galaxies by \cite{tabatabaei17}.
We can simply express the observed spectral curvature for each galaxy as the difference between the low- and high-frequency spectral indices $\Delta\alpha=\alpha_{\mathrm{low}}-\alpha_\mathrm{high}$. The estimated median value of this parameter for the whole sample of galaxies turns out to be relatively small but positive: $0.18\pm0.02$, which underpins the low-frequency spectral flattening. This value is statistically significant because a non-parametric sign test allows us to reject a null hypothesis that the median of $\Delta\alpha$ is equal to zero (i.e. the p-value appears much smaller than 1\%). In a similar way, we estimated a mean value of $\Delta\alpha$ as $0.20\pm0.02$, which according to a parametric Student's t-test is also statistically significant (the p-value is again much lower than 1\%).

\begin{table*}
\caption{Results of fitting two models (power-law and curved) to spectra of galaxies from Fig. \ref{f:spectra}. The best-fitted parameters for the power-law model ($A_1$) and the curved model ($A_2$) as well as the quality of the fits ($\chi^2_\nu$) and the p-value of the F-test are given. The last column shows the selected most reasonable model: P - denotes the power-law model, C - the curved model (see Sect. \ref{s:flattening}).
}
\begin{center}
\begin{tabular}{lcccccc}
\hline
\hline
Name       & Power-law: $A_1$ & $\chi^2_\nu$ & Curved: $A_2$ & $\chi^2_\nu$ & p-value (in \%) & Best\\
\hline
NGC660 & $-0.47 \pm 0.02$ & 3.39 & $-0.16 \pm 0.04$ & 1.73 & 0.22 & C \\
NGC891 & $-0.69 \pm 0.02$ & 2.44 & $-0.16 \pm 0.03$ & 0.94 & 0.003 & C \\
NGC972 & $-0.56 \pm 0.03$ & 2.66 & $-0.22 \pm 0.06$ & 1.13 & 0.26 & C \\
NGC1569 & $-0.40 \pm 0.03$ & 4.62 & $-0.18 \pm 0.04$ & 1.93 & 0.06 & C \\
NGC3079 & $-0.72 \pm 0.02$ & 7.88 & $-0.12 \pm 0.04$ & 5.30 & 0.16 & C \\
NGC3627 & $-0.57 \pm 0.04$ & 6.18 & $-0.19 \pm 0.09$ & 4.66 & 2.2 & C \\
NGC3628 & $-0.60 \pm 0.03$ & 6.20 & $-0.09 \pm 0.04$ & 4.93 & 3.5 & C \\
NGC3646 & $-0.79 \pm 0.02$ & 0.23 & $-0.03 \pm 0.07$ & 0.31 & 70 & P \\
NGC3655 & $-0.66 \pm 0.03$ & 0.84 & $-0.14 \pm 0.09$ & 0.66 & 16 & P \\
NGC4102 & $-0.63 \pm 0.03$ & 6.67 & $-0.30 \pm 0.04$ & 1.51 & $<0.001$ & C \\
NGC4254 & $-0.74 \pm 0.03$ & 6.24 & $-0.11 \pm 0.06$ & 5.62 & 8.3 & P \\
NGC4826 & $-0.43 \pm 0.03$ & 1.11 & $-0.04 \pm 0.06$ & 1.17 & 51 & P \\
NGC5055 & $-0.69 \pm 0.04$ & 9.04 & $-0.17 \pm 0.05$ & 4.46 & 0.12 & C \\
NGC5371 & $-0.63 \pm 0.01$ & 0.10 & $0.00 \pm 0.03$ & 0.12 & 90 & P \\
NGC5936 & $-0.70 \pm 0.02$ & 0.73 & $0.02 \pm 0.06$ & 0.80 & 77 & P \\
UGC3351 & $-0.69 \pm 0.07$ & 7.99 & $-0.42 \pm 0.11$ & 1.97 & 0.70 & C \\
UGC12914 & $-0.82 \pm 0.03$ & 1.59 & $-0.19 \pm 0.05$ & 0.64 & 0.85 & C \\
\hline
\end{tabular}
\end{center}
\label{t:fits}
\end{table*}

We also analysed galaxy spectra by fitting two models to their flux densities $S_\nu$ between 50\,MHz and 5\,GHz: a power-law spectrum of the form
\begin{equation}
\log S_\nu=\log A_0 + A_1 \log \nu
,\end{equation}
and a curved spectrum,
\begin{equation}
\log S_\nu=\log A_0 + A_1\,\log \nu + A_2\, {\log}^2 \nu
,\end{equation}
which are the simplest versions of polynomial models, having the lowest number of free parameters. In the model fitting we used the Marquardt-Levenberg algorithm to find the lowest reduced chi-square $\chi^2_\nu$ parameter. 

In Table \ref{t:fits} we show the best-fit parameters $A_1$ and $A_2$ from the power-law and curved models, respectively, as well as the obtained $\chi^2_\nu$ goodness of fit for all galaxies from Fig. \ref{f:spectra}. We also compared the models by the F-test, using the ratio of the residual sum of squares from the power-law and curved models, scaled by corresponding degrees of freedom. If the p-value obtained was small ($\leq 5\%$) we concluded that the curved model was statistically significantly better than the simple power-law model. Otherwise, we inferred that there is no convincing evidence to support the curved model and considered the power-law model as a satisfactory one (see the last column of Table \ref{t:fits}). 
We found that only about 35\% of 
galaxies had simple spectra described by the power-law model, while the 
other objects had curved spectra. For galaxies with curved spectra, 
the fitted values of $A_2$ were always negative. This corresponds to 
flatter spectra of galaxies at lower frequencies which is consistent 
with the statistical differences we found between distributions of
$\alpha_\mathrm{low}$ and $\alpha_\mathrm{high}$ parameters (Fig. 
\ref{f:alpha_low_high}), and their median values. Moreover, the model with a curved spectrum was always better fitted than the power-law model for galaxies with higher values of $\Delta\alpha$.

Therefore, our analysis shows that the spectral flattening towards low frequencies is small in nearby galaxies, yet is statistically significant. A similar conclusion was derived by \cite{israel90}, who noticed that flux densities at 57.5\,MHz are systematically lower than expected from an extrapolation from measurements at higher frequencies (Sect. \ref{s:intro}), and, for example, more
recently by \cite{calistro-rivera17} using LOFAR observations of the Bo\"otes deep field.

\begin{figure}
\centering
\includegraphics[clip,angle=0,width=9.2cm]{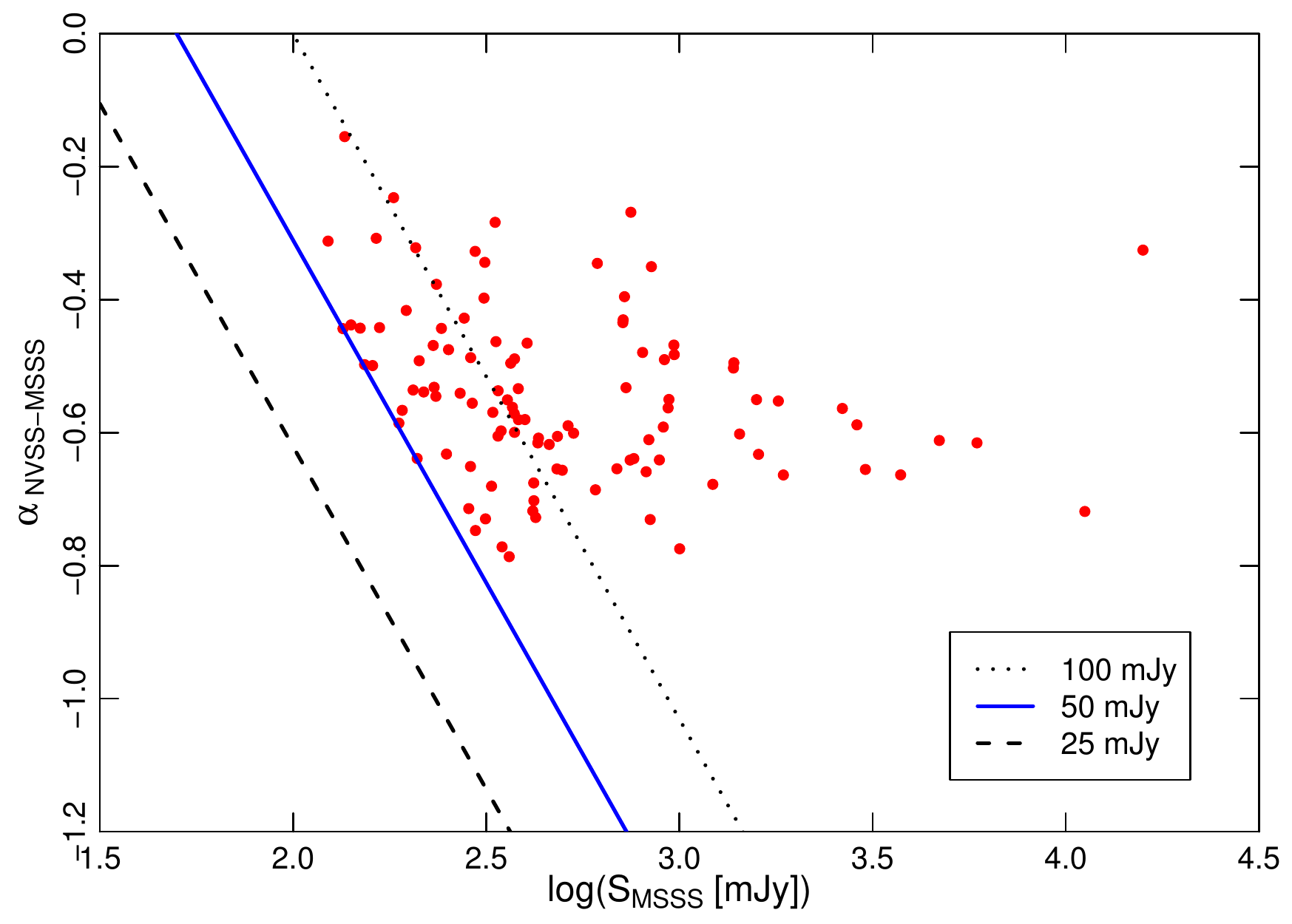}
\caption{Two-point spectral index between 150\,MHz and $1.4$\,GHz, as estimated from MSSS and NVSS data, for the sample of 106 galaxies. The dashed, solid, and dotted lines correspond to galaxy flux density limits of  25, 50, and 100\,mJy at 1.4\,GHz, respectively.
}
\label{f:alpha_fluxlimit}
\end{figure}

In Figure \ref{f:alpha_fluxlimit} we present the two-point spectral
index between 150\,MHz and $1.4$\,GHz derived from the MSSS and NVSS
data against the MSSS flux densities, and plot theoretical lines
indicating different flux density limits at 1.4\,GHz. The distribution
of data points shows a distinct cut corresponding to 50\,mJy at
1.4\,GHz. This is understandable, since this was actually the limit
applied as one of the selection criteria for our sample 
(Sect.~\ref{s:sample}). Although the scarcity of steep spectrum galaxies among the weakest sources ($\log(S_\mathrm{MSSS}/{\rm mJy})<2.5$) results from a selection bias, it cannot account for the curvature observed in the global galaxy spectra.

\subsection{Spectra versus galaxy inclination}
\label{s:inclination}

\begin{figure}[t]
\centering
\includegraphics[angle=0,width=9.2cm]{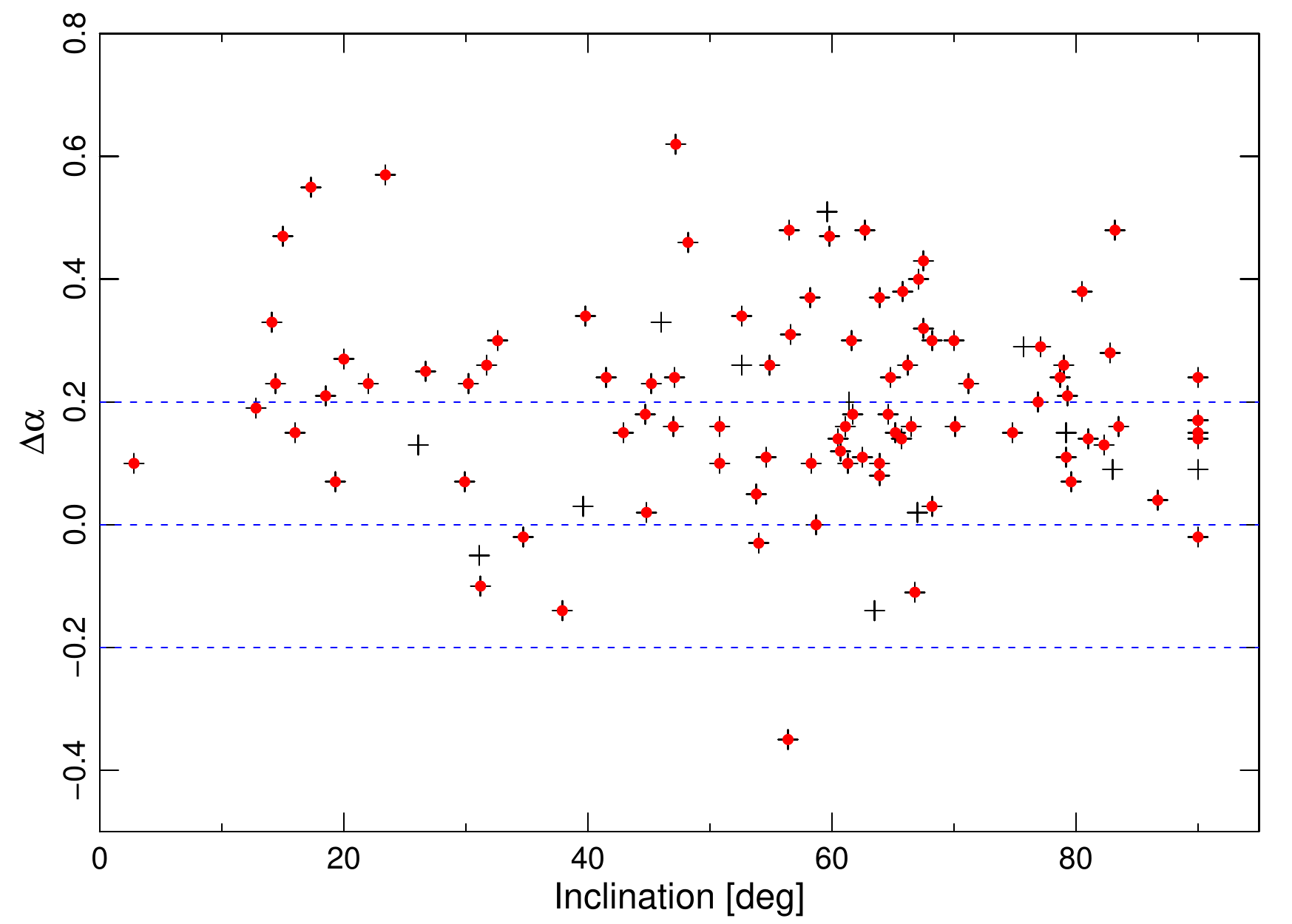}
\caption{Difference between low- and high-frequency spectral indices, $\Delta\alpha$, vs. inclination angle $i$ for the MSSS sample of 106 galaxies ($i=0\degr$ corresponds to a face-on galaxy). Galaxies with well-defined disks (93 objects) are indicated by red circles. 
}
\label{f:alfadiff_incl}
\end{figure}

\begin{figure}[t]
\centering
\includegraphics[clip,angle=0,width=9.2cm]{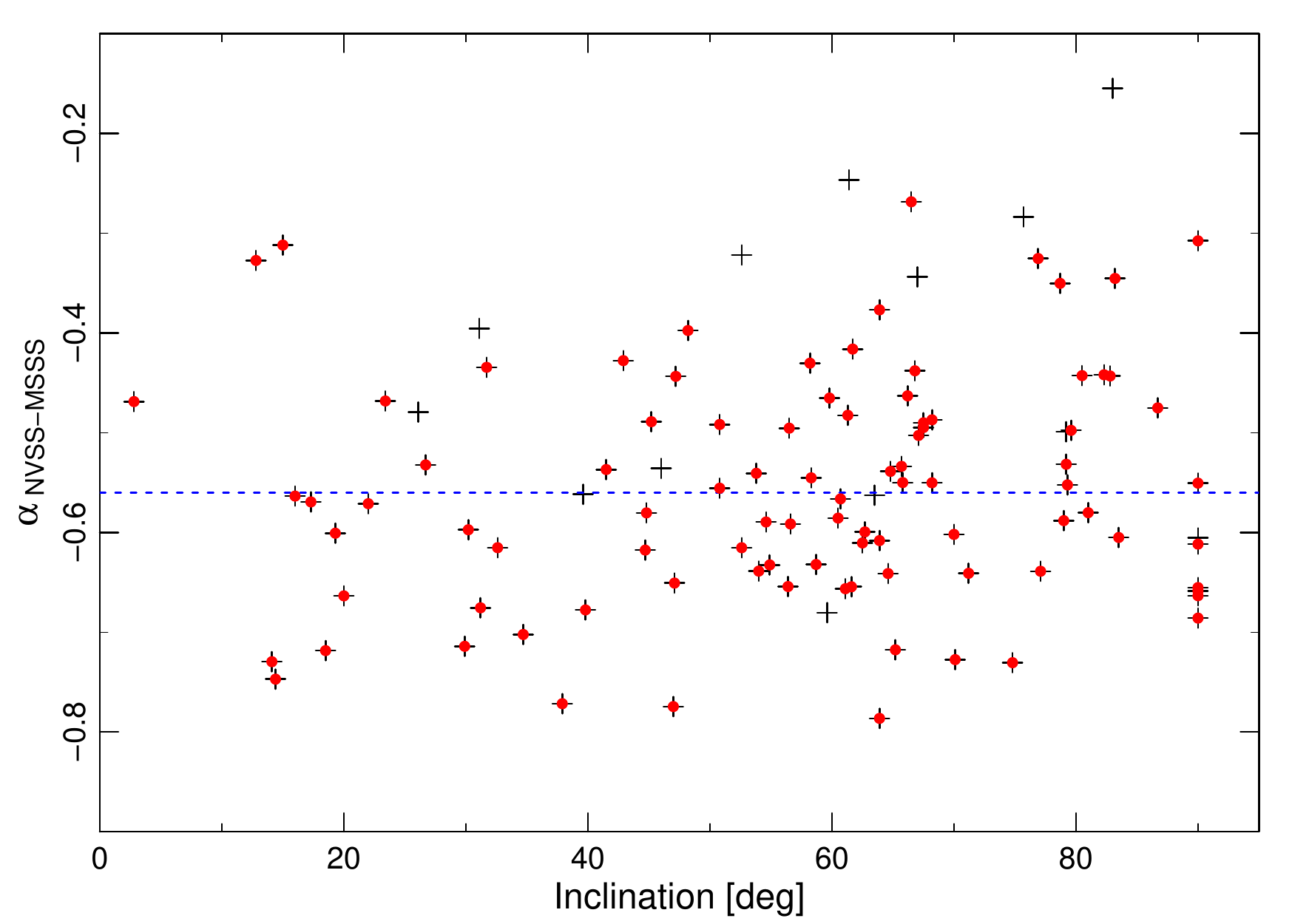}
\caption{Two-point spectral index between 150~MHz and $1.4$~GHz, determined from MSSS and NVSS data, vs. the inclination angle $i$ for the sample of 106 galaxies. Galaxies with well-defined disks (93 objects) are additionally marked by red circles. The horizontal line represent the median value of the spectral index of $-0.56$.
}
\label{f:incl_alfa}
\end{figure}

We noticed in Sect. \ref{s:spectra} that spectra of 
our galaxies are slightly curved, which could indicate some kind of 
underlying physical process. One possibility could be internal thermal 
absorption of nonthermal emission (Sect. \ref{s:intro}). We expect
strong absorption effects  to be at work for a well-mixed thermal and nonthermal plasma when galactic lines of sight
are long, for example in edge-on galaxies  (see Sect. \ref{s:model}). Also in the 
case when many localised \ion{H}{II} regions are situated along the line of sight, the synchrotron emission from all regions along the line of sight have to pass through them, and their number, and hence their absorption, increases with higher inclination.
To examine these predictions, we
constructed a diagram to show the difference between low- and
high-frequency spectral indices against the inclination angle $i$
(Fig. \ref{f:alfadiff_incl}). For strong absorption, we would find
galaxies in the top-right corner of this diagram. However, there are
no objects in this area and no distinct spectral dependence on galaxy
tilt can be observed. The Kendall correlation coefficient between
$\Delta \alpha$ and $i$ is just $-0.05$, which confirms this finding. A similar conclusion could be drawn from the two-point spectral index versus the inclination angle (Fig. \ref{f:incl_alfa}).

\begin{figure}[t]
\centering
\includegraphics[angle=0,width=9.2cm]{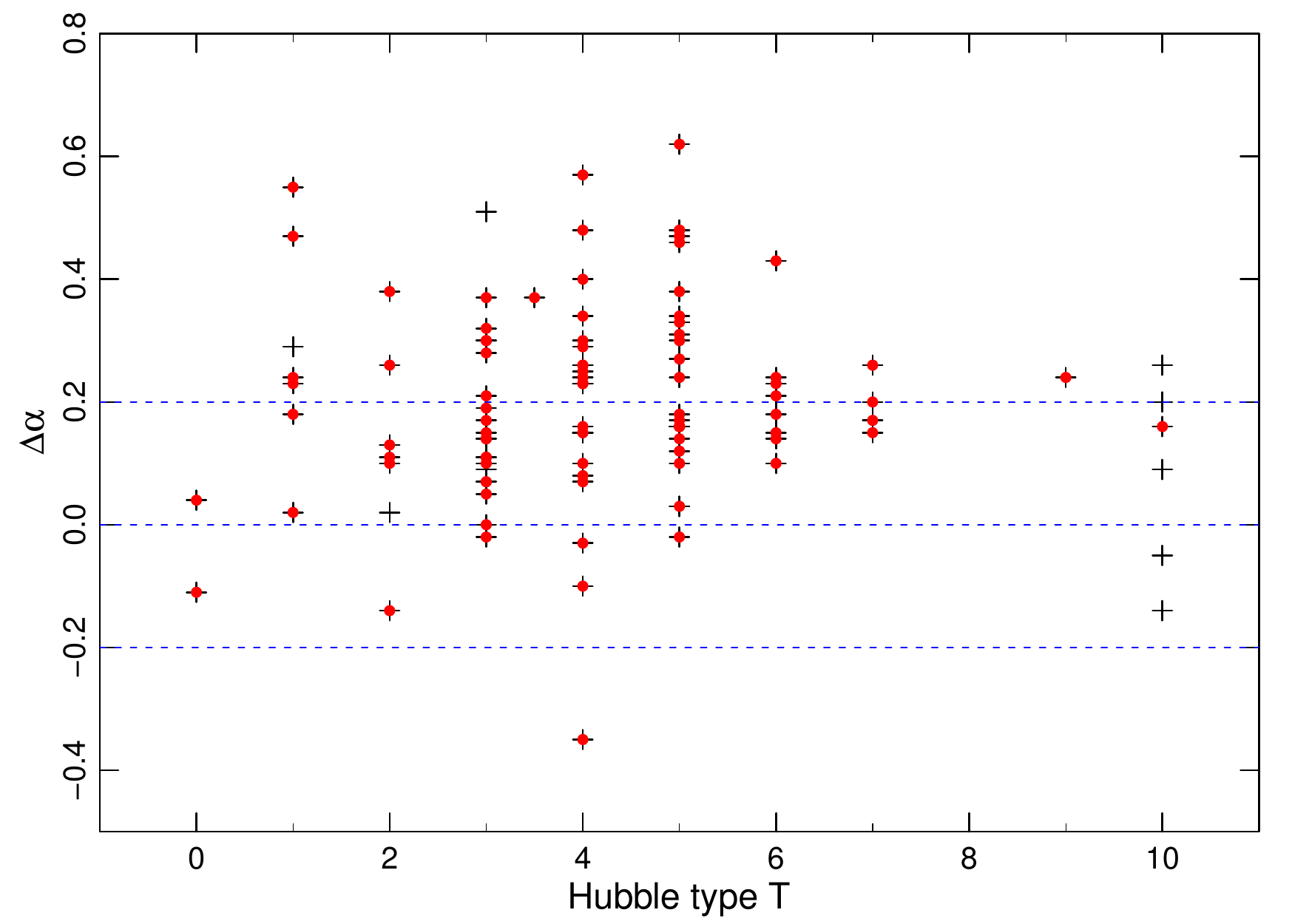}
\caption{Difference between low- and high-frequency spectral indices, $\Delta\alpha$, vs. morphological type $T$ for the sample of  106 galaxies. Galaxies with well-defined disks (93 objects) are indicated by red circles.
}
\label{f:morphology}
\end{figure}

One can also consider a situation where the absorbing thermal gas is situated completely 
outside the synchrotron medium as a foreground layer while still within the galaxy. In that case, the
synchrotron emission would be free-free absorbed, irrespective of the galaxy inclination, and this could potentially explain our results. However, such a configuration is rather unlikely 
as the thermal (\ion{H}{II}) gas has a smaller spatial extent than the nonthermal emission, as in NGC\,6946 \citep{tabatabaei13}, or in NGC\,4254 \citep{chyzy07}. 
Accordingly, we conclude that the flattening observed at low frequencies (Sect. \ref{s:flattening}) is not due to free-free absorption. 
These results contradict the claim of \cite{israel90}, who used a similar diagram to show that galaxy
spectra are flatter for highly inclined objects. As a consequence, we do not see any need for a special low-temperature ionised gas postulated by those authors. We also considered that some galaxies in our sample may not have a well-determined value of viewing angle as their disks are not particularly regular. Such objects could have somehow affected probing the absorption origin of spectral flattening. Therefore, in another approach we excluded mergers (like UGC\,8696), amorphous compact galaxy pairs (like NGC\,5929/30), and dwarf-irregular galaxies (like NGC\,4449) from our sample. These objects are indicated in Table \ref{t:sample} by the flux density flag 2. The remaining 93 objects with well-defined disks are marked red in Fig. \ref{f:alfadiff_incl}. It can be seen again that the distribution of $\Delta \alpha$ does not reveal any systematic dependence on the viewing angle, which is further confirmed by a low value ($-0.04$) of the Kendall correlation coefficient. 

\begin{figure}
\centering
\includegraphics[angle=0,width=9.2cm]{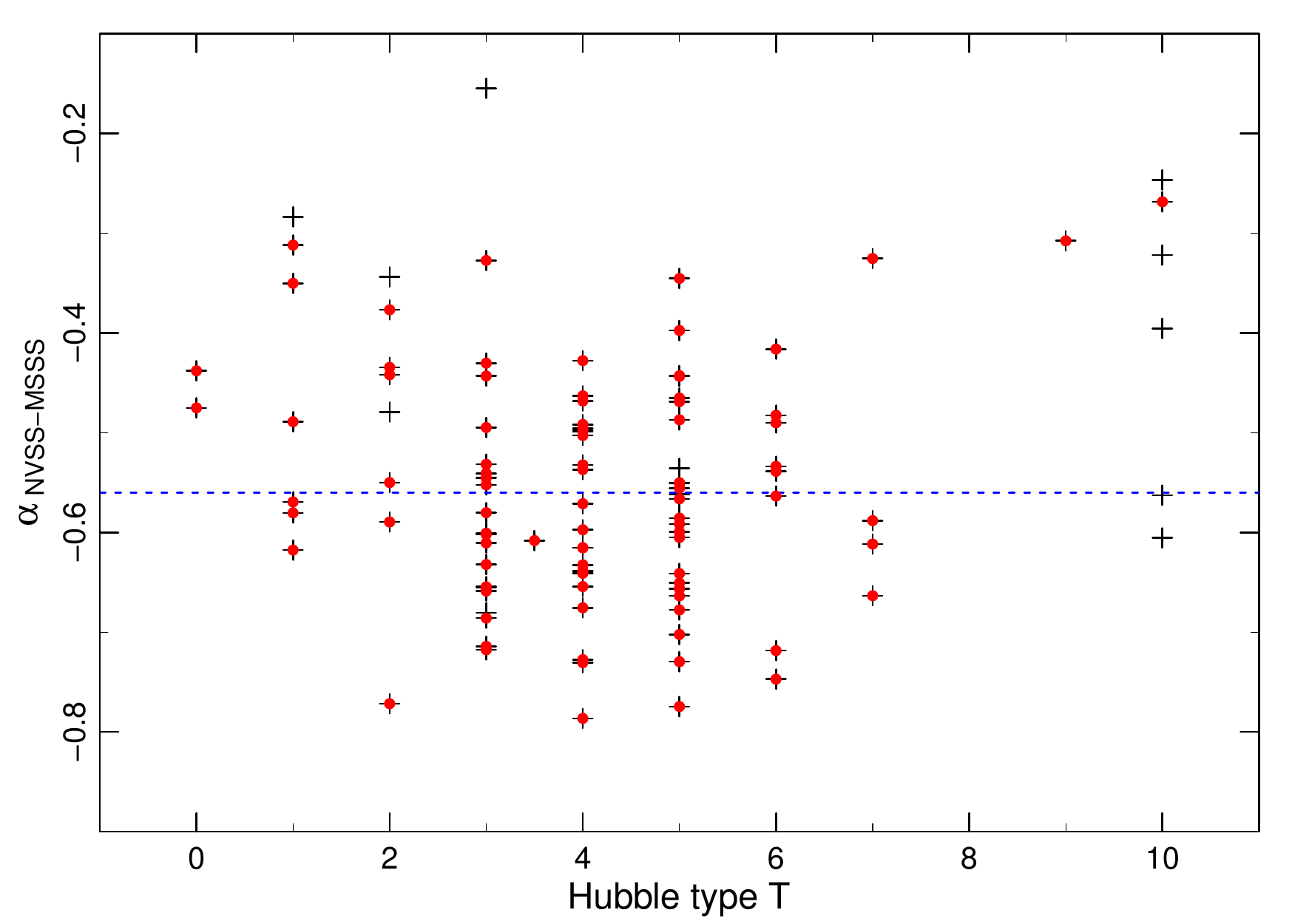}
\caption{Two-point spectral index between 150\,MHz and $1.4$\,GHz, computed from the MSSS and NVSS data, vs. morphological type $T$ for the sample of 106 galaxies. Galaxies with well-defined disks (93 objects) are indicated by red circles.
}
\label{f:globalmorphology}
\end{figure}

Apart from attempting to account for spectral flattening in terms of
orientation, we also investigated its possible relationship with the
morphological type of galaxies. The resulting diagram of the
$\Delta\alpha$ as a function of the Hubble type $T$ is shown in Fig.
\ref{f:morphology}. No dependence of flattening on the galaxy's
morphology is to be seen; although there is a small number of flatter
low-frequency spectra (smaller $\alpha_{\mathrm{low}}$) for late-type
spirals, spectra such as these are also found for early-type objects
(Fig. \ref{f:globalmorphology}). Relatively flat spectra for
dwarf galaxies have already been noticed by \cite{klein18}
and interpreted as low CR confinement in low-mass galaxies.
Considerably lower synchrotron components, and therefore systematically
weaker magnetic fields, have recently also been found in late-type
spiral galaxies, suggesting that a similar process could be at work in those objects as well \citep{chyzy17}. Our results indicate that, apart from these trends that depend on galaxy properties generally associated with different morphological types, the spectral curvature ($\Delta\alpha$) does not correlate with galaxy morphology.

\section{Discussion}
\label{s:discussion}

Our examination of the spectra of 106 galaxies
(Sect.~\ref{s:inclination}) has revealed gentle flattenings at low radio frequencies, which, contrary to the claims of \cite{israel90}, do not seem to be caused by thermal gas absorption. Such a conclusion is well supported by the results of \cite{marvil15}, although they used different selection criteria, including, for instance, elliptical galaxies with radio emission possibly influenced by AGNs. This approach could lead to various effects in their sample, such as a positive spectral index. In order to avoid further complication, we deliberately left out such types of objects from our analysis (Sect. \ref{s:sample}).

If the curved spectra observed in our sample of galaxies do not
originate from absorption, we are facing a contradiction with earlier
studies. Strong thermal absorption seems to be the only reasonable
explanation for the distinct drop of low-frequency radio emission in the central part of M\,82 \citep{{wills97},{varenius15}} and in other starburst galaxies \citep[e.g. NGC\,253][]{kapinska17}. Also, the turnovers found in the spectra of some regions in the centre of the Milky Way \citep{roy06} are well accounted for by absorption effects. So how can we possibly reconcile these results with our analysis of a large sample of galaxies?

\begin{figure*}
\centering
\includegraphics[clip,width=7.0cm,angle=0]{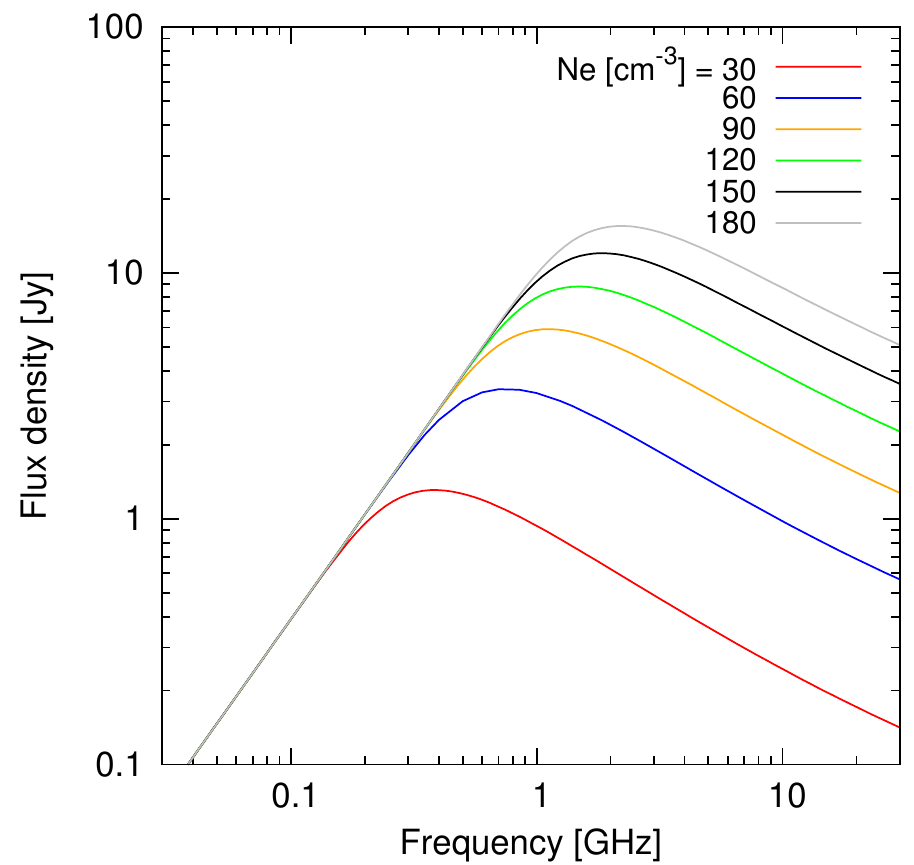}
\hspace{2cm}
\includegraphics[clip,width=7.0cm,angle=0]{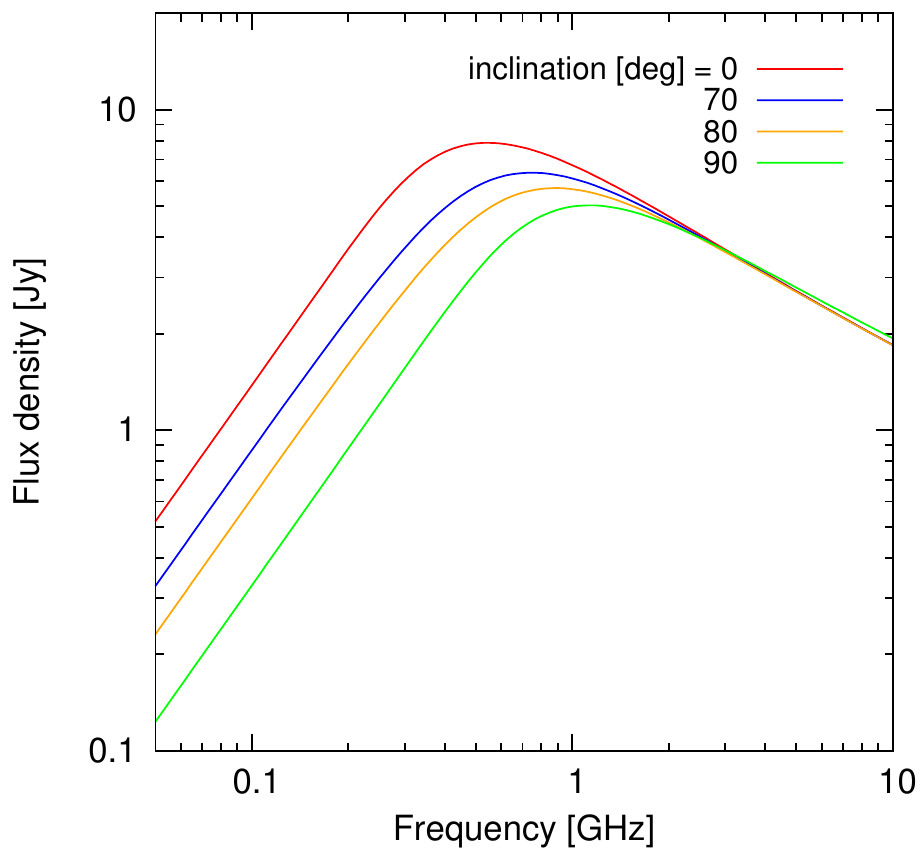}
\caption{Left: Effect of the thermal electron density (determined by the EM) on low-frequency spectra of starburst galaxies seen edge-on. The thermal and nonthermal emitting gases are fully mixed with a constant 10\% thermal fraction at $1.4$~GHz (to be compared with the analytical model of \cite{condon91}. 
Right: Similar model applied for a galaxy seen at a different viewing angle with $N_e=80$\,cm$^{-3}$.
}
\label{f:condon}
\end{figure*}

In order to answer this question, we decided to construct a simple model of galaxy emission allowing us to analyse the effects of absorption with a realistic treatment of projection effects. The modelled radio emission originates from different regions of a galaxy that we assume to be composed of compact \ion{H}{ii} regions, supernovae, and diffuse plasma in both disk and halo. These regions emit and absorb synchrotron and free-free radiation of various amounts, which can be estimated from observational data. We did not take into account the Razin-Tsytovich effect or synchrotron self-absorption, which could affect the galaxy spectra at extremely low frequencies below 10\,MHz \citep{fleishman95}.

\subsection{Numerical model of radio emission}
\label{s:model}

\begin{figure*}
\centering
\includegraphics[width=7.0cm,angle=0]{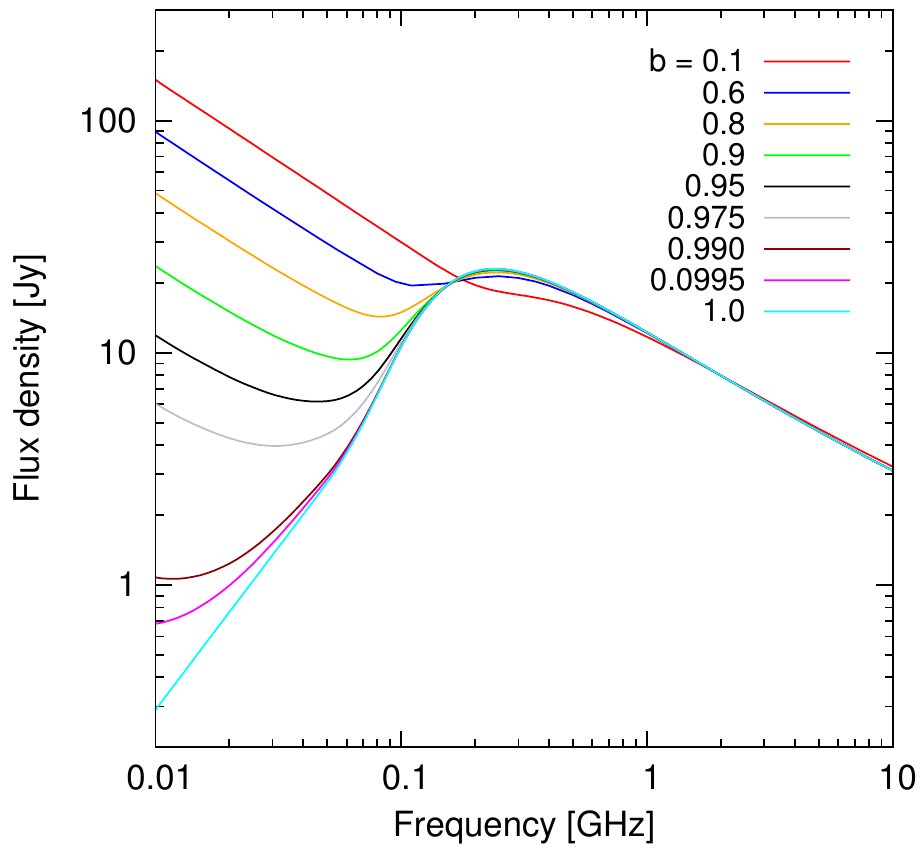}
\hspace{2cm}
\includegraphics[width=7.0cm,angle=0]{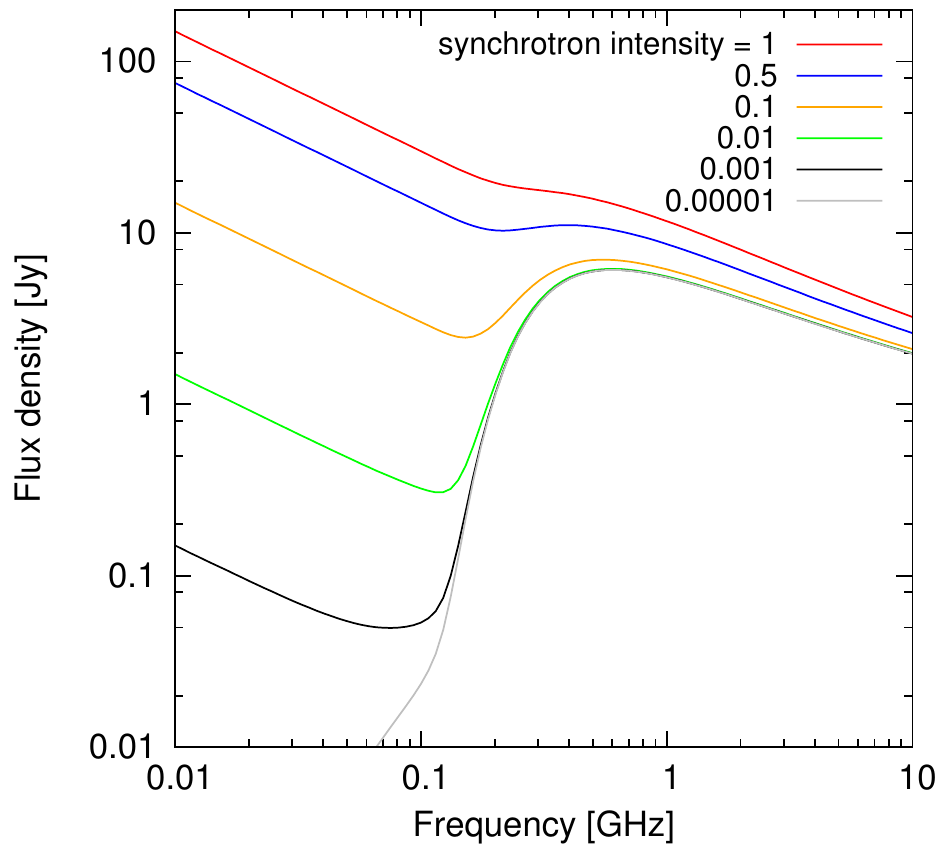}
\caption{Left: Model of a starburst galaxy with a synchrotron halo of varying size, as measured by the ratio $b$ (see text for details). The thermal fraction of the disk emission is assumed to be 10\% at $1.4$~GHz for all models.
Right: Similar model with the same synchrotron halo of $b=0.1$ but different synchrotron intensity.
}
\label{f:condon_halo}
\end{figure*}

In our numerical modelling we considered the 3D shapes of individual components of galaxies (such as the core and thin and thick disks), which were positioned on a 3D grid. We set the parameters describing thermal emission and absorption in each grid element, chose the galaxy inclination angle, and solved the radiative transfer equation along the line of sight at various frequencies. On the path through the source towards the observer, we distinguished two solutions for radiative transfer in the cell with index $n$, depending on the content of this and the preceding cells:
\begin{equation}
 \label{e:s_n}
 S_n  = \left\{
  \begin{array}{l l}
    S_{n-1}\,e^{-\tau_n}+ \left(2kT_e \, \nu^{2} c^{-2} + B_{n} {\tau_n}^{-1} \right)\,  \left(1-e^{-\tau_n}\right) \, \Omega\\
    S_{n-1} + B_{n} \, \Omega\\
  \end{array} \right.
,\end{equation}
where  
\begin{equation}
\label{e:tau_n}
\tau_n=8.235\times 10^{-2} \, \left(\frac{T_e}{K}\right)^{-1.35} \left(\frac{\nu}{GHz}\right)^{-2.1} \left(\frac{EM_n}{\mathrm{pc\,cm}^{-6}}\right)
,\end{equation}
is the optical thickness of thermal gas in the cell with index
  $n$, $T_e$ is the thermal electron temperature, $EM_n=s\,N^2_{en}$ is the
  emission measure, $N_{en}$ is the thermal electron density in the cell,
  $s$ is the the cell size, $B_{n}$ is the synchrotron intensity, and $\Omega$
is the solid angle of the cell. The first solution in Eq.~(\ref{e:s_n})
corresponds to a cell filled with well-mixed synchrotron- and thermal-emitting gas. The radio emission ($S_{n-1}$) from the previous cell along the line of sight is absorbed by thermal gas in the cell with index $n$. The thermal and synchrotron gas components of this cell contribute with their emission minus the thermally absorbed part. The second solution applies to a cell with a solely synchrotron-emitting gas (e.g. in the galaxy halo). By solving the radiative transfer equation for all cells in the grid for a particular viewing angle and at different frequencies, we obtained synthetic maps of radio emission at various frequencies. We then integrated the flux density in the maps to construct the modelled global galaxy spectra for a variety of inclination angles.

First, we considered a simple model for a starburst galaxy represented by a single cylinder with a well-mixed thermal and nonthermal plasma. Our 3D model  well reproduces the results of the analytical 1D modelling by \cite{condon91}. Galaxies with higher thermal gas densities, but with the same thermal fraction (fixed at 10\% at 1.4\,GHz), have a higher turnover frequency in their integrated spectra (Fig.~\ref{f:condon}, left panel). In our modelling, we were also able to simulate what such a starburst would look like at different viewing angles. Less inclined galaxies are stronger radio emitters (less thermally absorbed), and have spectral turnovers at lower frequencies (Fig.~\ref{f:condon}, right panel).

Because the starburst model of \cite{condon91} did not include a synchrotron halo, which is apparently present in such objects \citep[e.g.][]{adebahr13,varenius16}, we constructed another model with synchrotron emission coming also from beyond the thermally emitting volume. In this model, the low-frequency spectra of galaxies strongly depended on the size of the synchrotron halo, measured by the ratio $b$ of the volumes of thermal to non-thermal-emitting regions (Fig.~\ref{f:condon_halo}, left panel). Value $b$=1 corresponds to a region radiating both thermally and nonthermally, whereas $b$=0.1 means that the central part of the galaxy, containing both thermal and synchrotron gas, accounts for only 10\% of the entire synchrotron halo volume. In this model, the thermal fraction at 1.4 GHz was kept constant at 10\% independent of $b$. We found that the resulting spectra depend strongly on $b$ (Fig.~\ref{f:condon_halo}, left panel). Furthermore, we found that even for a fixed halo size at the low value of $b$=0.1, the spectra can still be easily modified by changing either the level of synchrotron intensity in the halo (Fig.~\ref{f:condon_halo}, right panel) or the value of the nonthermal spectral index.

The above examples show that the shape of integrated spectra at low frequencies depends strongly on both the specifics of geometry of the radio emitting regions and the parameters of the thermal and nonthermal emissions. Adding a halo component to a simple starburst region introduces a further potential ambiguity in modelling if only integrated spectra are considered. Therefore, it is not possible to fully interpret integrated spectra without detailed information on the distribution of thermal and nonthermal radio emissions throughout the galaxy. 

Therefore, in the following sections, we carefully investigate those galaxies from our sample (Sect. \ref{s:sample}) for which such details are available. Firstly we model M\,51, a face-on spiral galaxy with relatively low star formation activity, and secondly we model M\,82, a starburst galaxy seen nearly edge-on. With our models, we are able to modify inclination angles and compare the constructed spectra to observed ones. This analysis enables us to draw general conclusions on the role of thermal absorption as well as on the origin of the curved spectra that are observed in nearby galaxies. 

\subsection{Modelling M51-like galaxies} 
\label{s:m51}
 
We modelled \object{M\,51} as an example of a non-starbursting galaxy with a low inclination angle ($i\approx$$20 \degr$). The thermal emission coming from the ionised gas was represented by two vertical components: the thin and thick disk. Just as in the case of the Milky Way, we described the distribution of the thermal electron density ($N_e$) in these components with exponential functions with vertical scale heights of about 100\,pc and 1\,kpc, respectively \citep[cf.][]{cordes04}. The radial dependence of $N_e$ was approximated by inner and outer exponential disks. In the inner disk, we determined the radial profile of $N_e$ from H$\alpha$-derived emission measure maps by \cite{greenawalt98}, while for the outer disk we assumed a radial scale length of 10\,kpc \citep{gutierrez10}. The density of free electrons in the plane of the outer disk is approximated by 0.02\,cm$^{-3}$, like in the models of the Milky Way \citep{ferriere01}. Therefore, the 3D distribution of thermal electrons was modelled as:
\begin{equation}
\label{e:ne_m51}
\begin{aligned}
N_e(r,z)=A^a\,\sqrt{\frac{EM(r)}{100\,\mathrm{pc}}}\,\exp\left(\frac{-|z|}{100\,\mathrm{pc}}\right) + 0.02\,\exp \left( \frac{-r}{10\,\mathrm{kpc}} \right) \times \\
\exp \left( \frac{-|z|}{1\,\mathrm{kpc}}\right),
\end{aligned}
\end{equation}
where $A^a$ is a constant and the superscript $a$ used in this and following equations signifies the modelling of M\,51-like galaxies. The value of $A^a$ was determined so that the integrated radio thermal emission corresponds to a thermal fraction of 28\% at 14.7\, GHz, estimated for M\,51 by \cite{klein84}.
This equation was used to find the $EM_n$ in each model cell and then $\tau_n$ according to Eq.~\ref{e:tau_n}, where we assumed $T_e\approx 10^4$\,K everywhere.

The approximate properties of the nonthermal emission throughout the galaxy were derived by the iteration method outlined below. The multifrequency observations of M\,51 revealed that the radial profiles of the radio intensity and the spectral index between 151\,MHz and 1.4\,GHz vary significantly within the galaxy \citep{mulcahy14}. Furthermore, the local spectra in the central part of M\,51 are relatively flat, but become steeper closer to the galaxy edges, as expected for spectral ageing of CR electrons by synchrotron and inverse Compton radiation. There is also evidence for diffusion of CRs from the star-forming regions into the inter-arm regions and outer parts of the galaxy \citep{mulcahy16}. Accordingly, we modelled the unabsorbed nonthermal intensity $B_n$ as two exponential vertical disks with a radial variation that we described using four different continuous functions:
\begin{equation}
 \label{e:b_m51}
 \begin{aligned}
 B_n(r,z)  = \begin{cases}
  C^a\,\exp(-0.5/R^a_{1}) \, [\exp(-|z|/Z^a_1) + \exp(-|z|/Z^a_2)],  \\
  C^a\,\exp(-r/R^a_1)\, [\exp(-|z|/Z^a_1) + \exp(-|z|/Z^a_2)],     \\
  C^a\,\exp(-r/R^a_2)\, \left[\frac{\exp(-1.7/R^a_1)}{\exp(-1.7/R^a_2)}\right] \, [\exp(-|z|/Z^a_1) \, + \\
  \qquad \qquad \qquad \qquad \qquad \qquad \qquad \exp(-|z|/Z^a_2)],  \\
  C^a\,\exp(-r/R^a_3)\, \left[\frac{\exp(-1.7/R^a_1)}{\exp(-1.7/R^a_2)}\right] \, \left[\frac{\exp(-10/R^a_2)}{\exp(-10/R^a_3)}\right]  \times \\
\qquad \qquad \qquad \qquad \left[\exp(-|z|/Z^a_1) + \exp(-|z|/Z^a_2)\right],
\end{cases}
 \end{aligned}
\end{equation}
where the consecutive equations are for different sections of radial distance from the galactic centre: $r<0.5$\,kpc, $0.5 \le r<1.7$\,kpc, $1.7 \le r<10$\,kpc, and $10 \le r\le 16$\,kpc respectively. The $C^a$ parameter is a scale factor, $R^a_1$, $R^a_2$, $R^a_3$ are scale-lengths of separate exponential profiles describing radial dependence of nonthermal emission in corresponding sections. In the first section ($r<0.5$\,kpc), we kept the emission constant along the radius, because the central part of M\,51 is not resolved in the observational profiles \citep{mulcahy16}; furthermore, an exponential profile would lead to a strong centrally concentrated source, which is not expected, due to CRs diffusion. The parameters $Z^a_1$ and $Z^a_2$ denote exponential scale heights of the vertical thin and thick nonthermal disks, respectively.

\begin{table}
\caption{Best-fit parameter values for the model of M\,51-like galaxy and reduced chi-square values for the models where particular parameters were increased by 20\% and decreased by 20\%.
}
\begin{center}
\begin{tabular}{lccc}
\hline
\hline
Parameter & Best fit & $\chi^2_\nu(+20\%)$ & $\chi^2_\nu(-20\%)$ \\
\hline
$C^a_0$ &  $-4.1$ & $1.58$  & $1.58$ \\
$C^a_1$ & $-0.56$ & $1.77$  & $1.13$ \\
$C^a_2$ &  $-0.05$ & $1.13$  & $1.63$ \\
$C^a_3$ & $-0.01$ & $1.42$  & $1.28$ \\
$R^a_1$ & $1.1$\,kpc & $7.02$  & $2.84$ \\
$R^a_2$ & $5.3$\,kpc & $2.99$  & $1.77$ \\
$R^a_3$ & $2.1$\,kpc & $1.53$  & $1.27$ \\
$Z^a_1$ & $0.3$\,kpc & $3.57$  & $2.03$ \\
$Z^a_{2}$ & $1.8$\,kpc & $3.92$  & $2.28$ \\
$Da$ & $-0.2$ & $1.68$  & $1.13$ \\
$Ea$ & $-0.05$ & $1.50$  & $1.24$ \\
\hline
\end{tabular}
\end{center}
\label{t:modelM51}
\end{table}

\begin{figure*}
\centering
\includegraphics[clip,width=4.25cm,angle=-90, viewport=20 8 499 690]{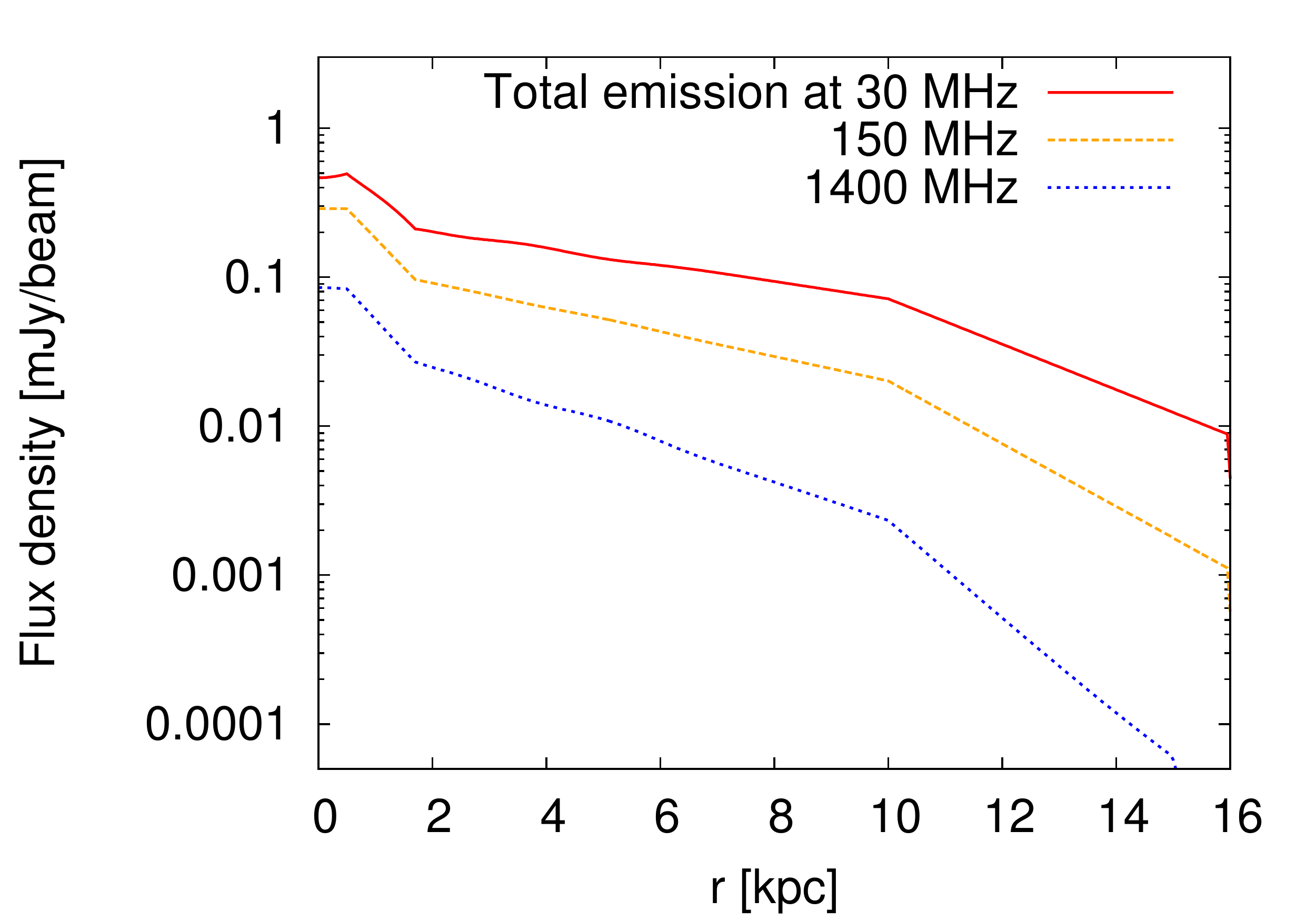}
\includegraphics[clip,width=4.25cm,angle=-90,viewport=20 8 499 684]{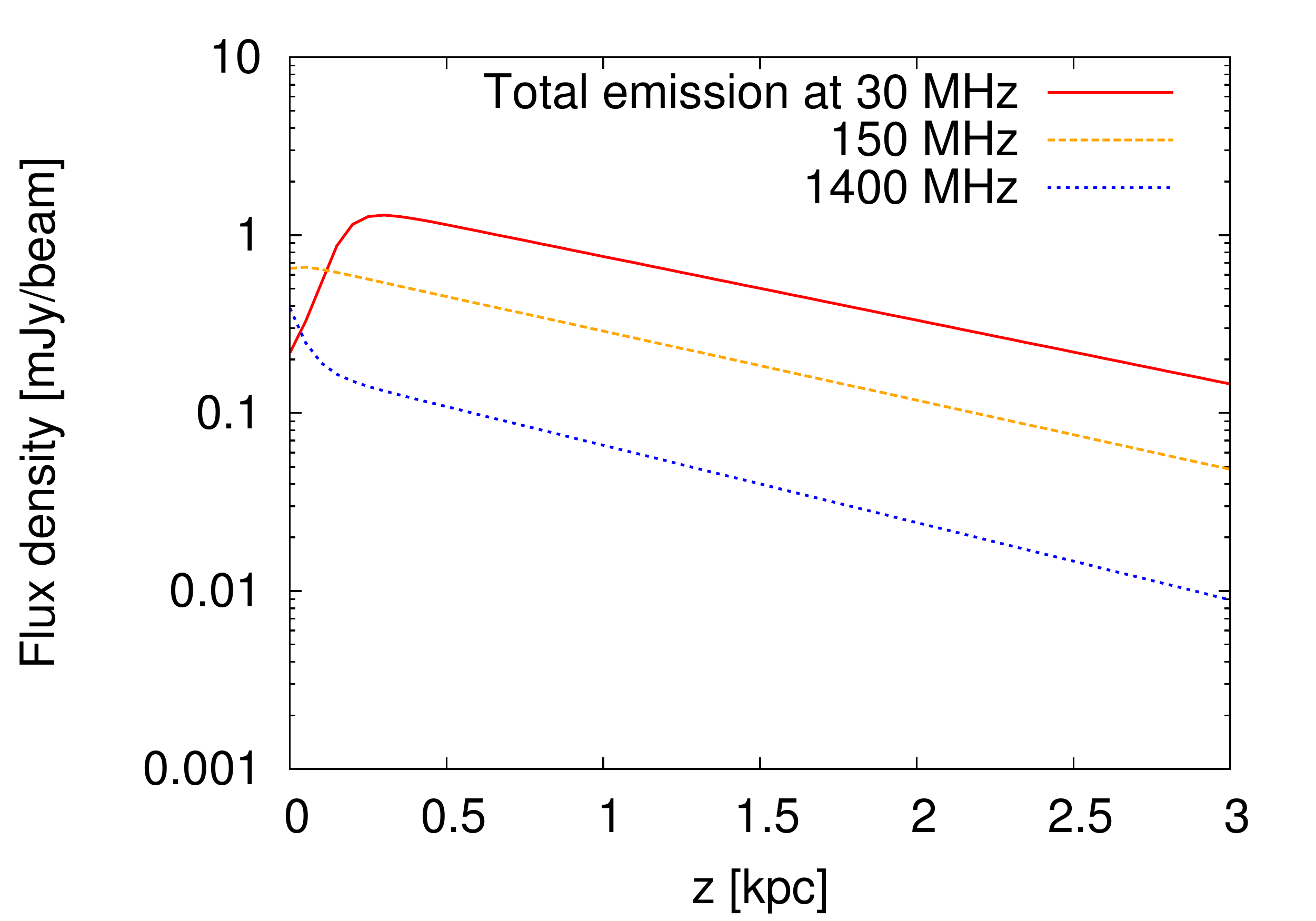}
\includegraphics[clip,width=4.25cm,angle=-90,viewport=20 8 499 684]{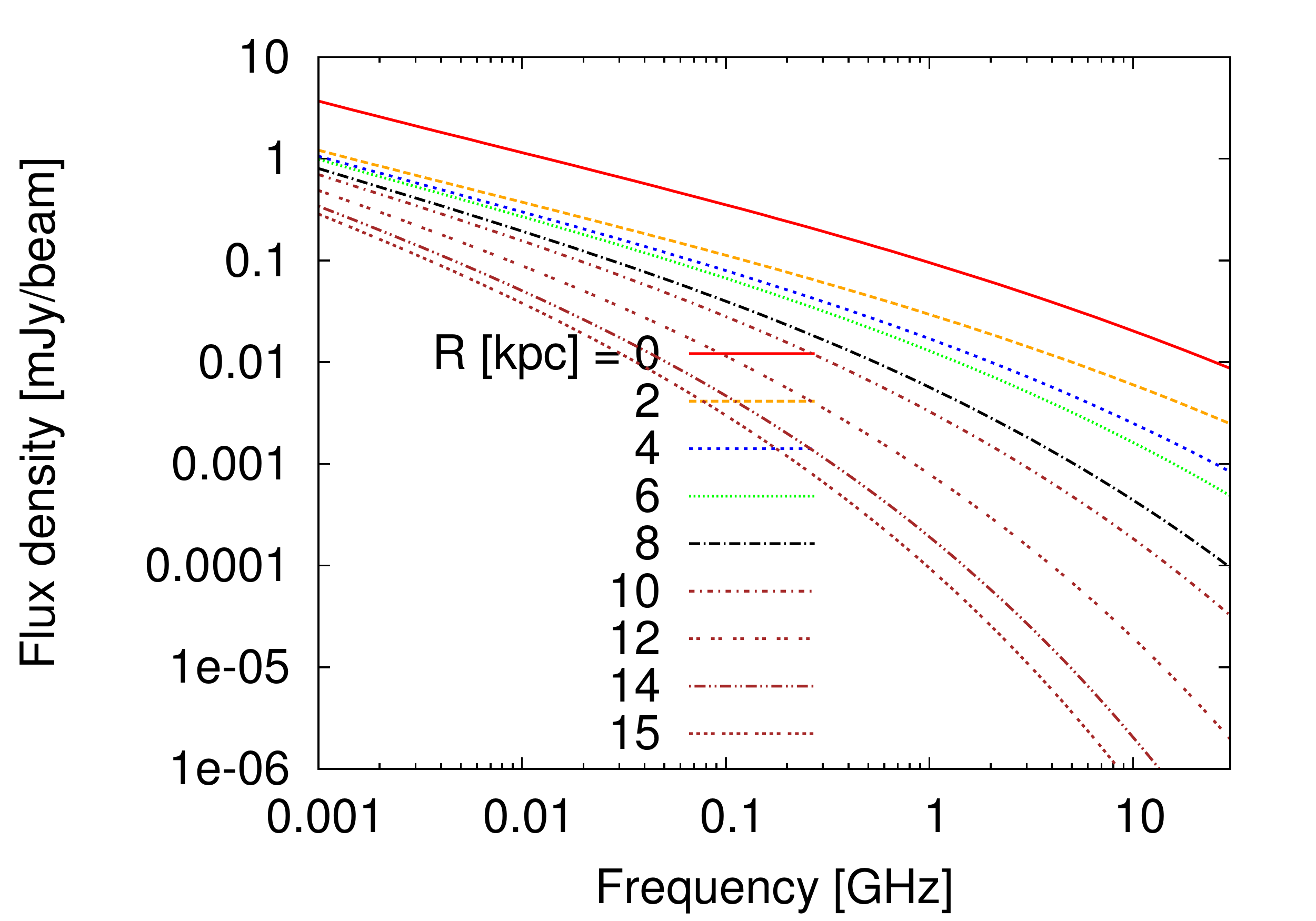}
\caption{Left: Modelled radial profiles of total radio emission of an M51-like galaxy seen face-on at 30, 150, and 1400 MHz. 
Middle: Model vertical profiles of total radio emission of an M51-like galaxy seen edge-on at 30, 150, and 1400 MHz. 
Right: Local synchrotron spectra of the galaxy at different distances along the galaxy major axis from the centre.
}
\label{f:m51_profiles}
\end{figure*}

\begin{figure*}
\centering
\includegraphics[clip,width=4.25cm,angle=-90, viewport=25 9 500 685]{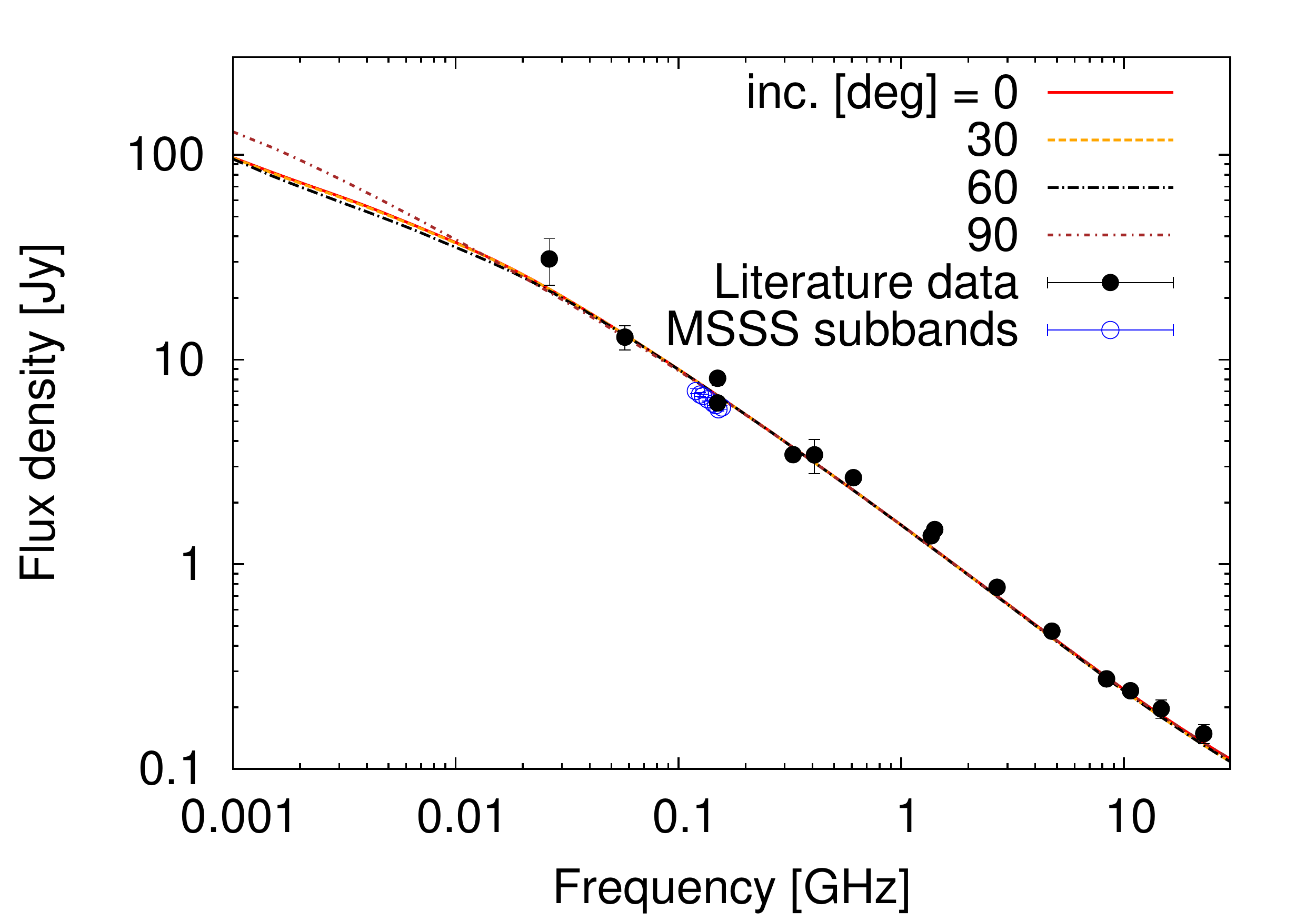}
\includegraphics[clip,width=4.25cm,angle=-90,viewport=25 9 500 685]{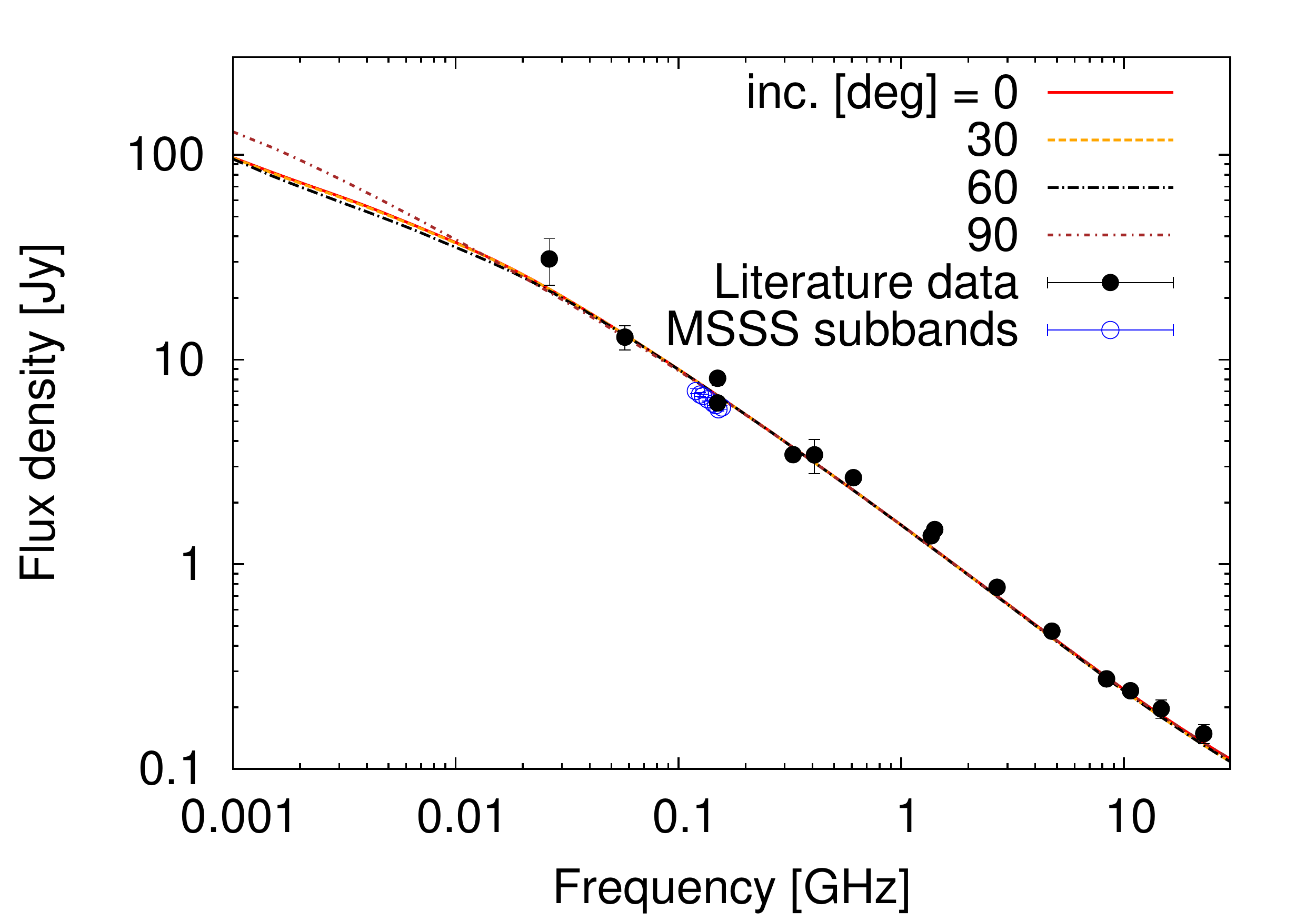}
\includegraphics[width=4.25cm,angle=-90,viewport=25 9 500 685]{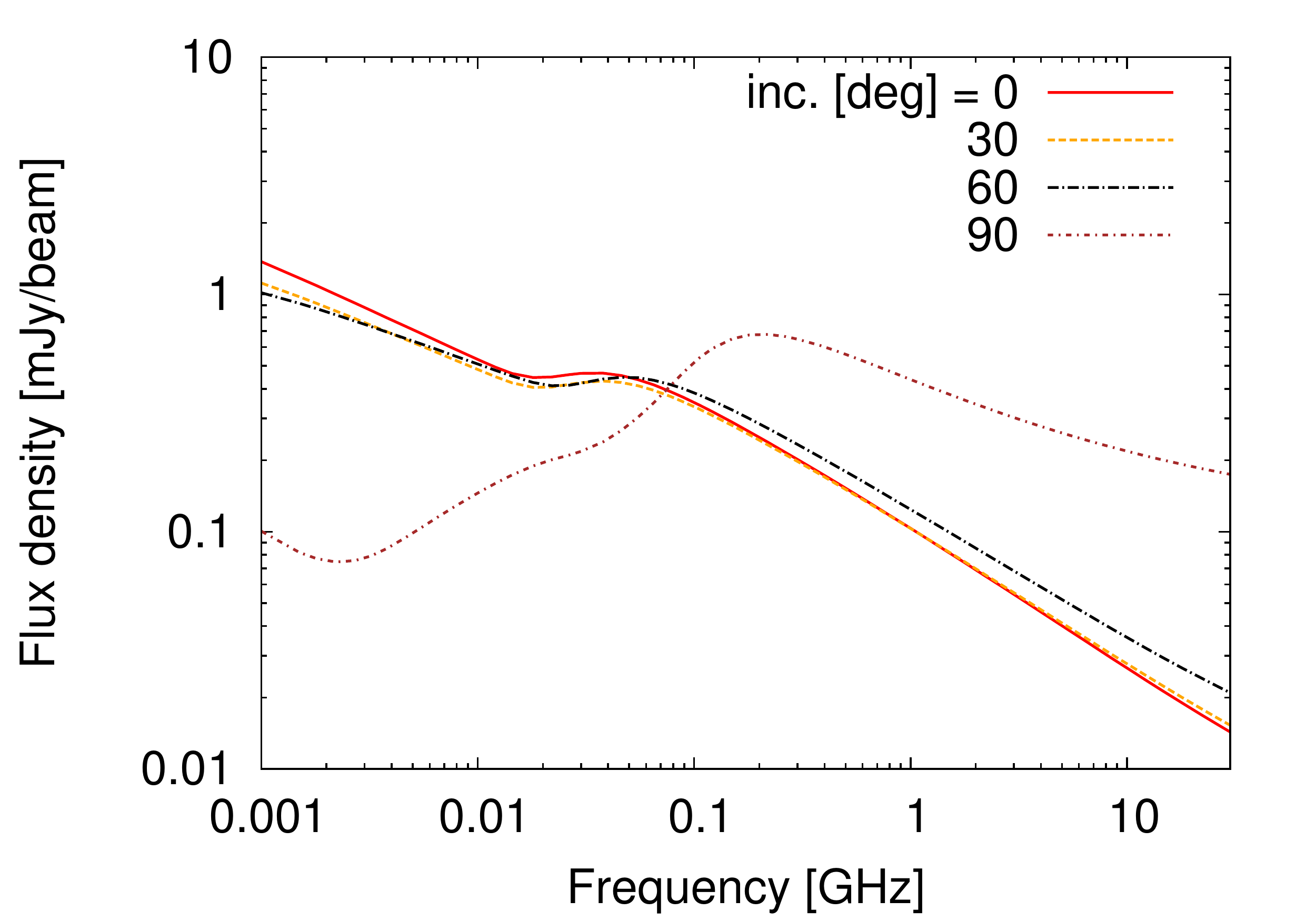}
\caption{Left: Resulting global spectrum of an M\,51-like galaxy from our 3D model as seen face-on (red solid line) with the synchrotron component without absorption (orange dotted line) and thermal free-free emission (black dashed line). Black solid circles denote literature data used during the modelling and the interpolated flux density at 150 MHz from MSSS survey. For the sake of completeness, data from the individual MSSS sub-bands (blue circles) are also shown. Middle: Global spectra for different inclination angles of 0, 30, 60, and 90 degrees, respectively. Right: Local spectrum of an area surrounding the galaxy centre for different inclination angles.
}
\label{f:m51_spectra}
\end{figure*}

In order to reproduce the radial profile of the spectral index of M\,51 \citep{mulcahy14}, the scale heights and lengths of the modelled radio emission profiles have to be frequency-dependent. Such a frequency dependence can be underlined by the various CR propagation processes \citep{{mulcahy14},{krause18}}. 
We assumed the power-law functions: $R^a_2(\nu), R^a_3(\nu)\propto (\nu/\nu_1 )^{Da}$ and $Z^a_1, Z^a_2\propto (\nu/\nu_2)^{Ea}$. The values of $R^a_1$, $R^a_2$, and $R^a_3$ at $\nu_1 = 0.15$ GHz were estimated as: 1.1\,kpc, 5.3\,kpc, and 2.1\,kpc, respectively \citep{mulcahy14}. The values of $Z^a_1$ and $Z^a_2$ at $\nu_2=4.85$ GHz we approximated as 0.3\,kpc and 1.8\,kpc, respectively \citep{krause18}. 
We notice that the scale factor $C^a$ should also depend on frequency, $C^a=f(\nu)$. Initially, we assumed $f$ in the form of a power-law function. However, we found that in order to obtain a better fit of the global spectrum
it was necessary to introduce a third-degree polynomial: $\log C^a(\nu)=C^a_0 + C^a_1\,\log \nu + C^a_2\, {\log}^2 \nu+ C^a_3\, {\log}^3 \nu$. 
 
After assuming some initial values for $Da$, $Ea$, and $C^a$ in Eq. \ref{e:b_m51}, the above values of $R^a_1$, $R^a_2$, $R^a_3$, $Z^a_1$ $Z^a_2$, and values of parameters describing the thermal emission (Eq. \ref{e:ne_m51}), we solved the transfer equation (Eq. \ref{e:s_n}) for various frequencies, and constructed model maps of radio emission. Next, we integrated flux densities using these maps, derived the modelled global spectrum, and compared it with the observed spectrum of M\,51. We tried out a number of models with different values of model parameters. The reduced chi-square value was used as a measure of goodness of the fit. We also qualitatively compared the model results with the radial profile of the radio emission of M\,51 and the spectral index profile from \cite{mulcahy14}. As the final stage, we changed the model parameters by a small value ($\pm 10\%$), identifying the model that gave the best fit. The parameters of the best-fitting model resulting from this procedure   are given in Table \ref{t:modelM51}.

\begin{figure}
\centering
\includegraphics[width=6.5cm,angle=-90]{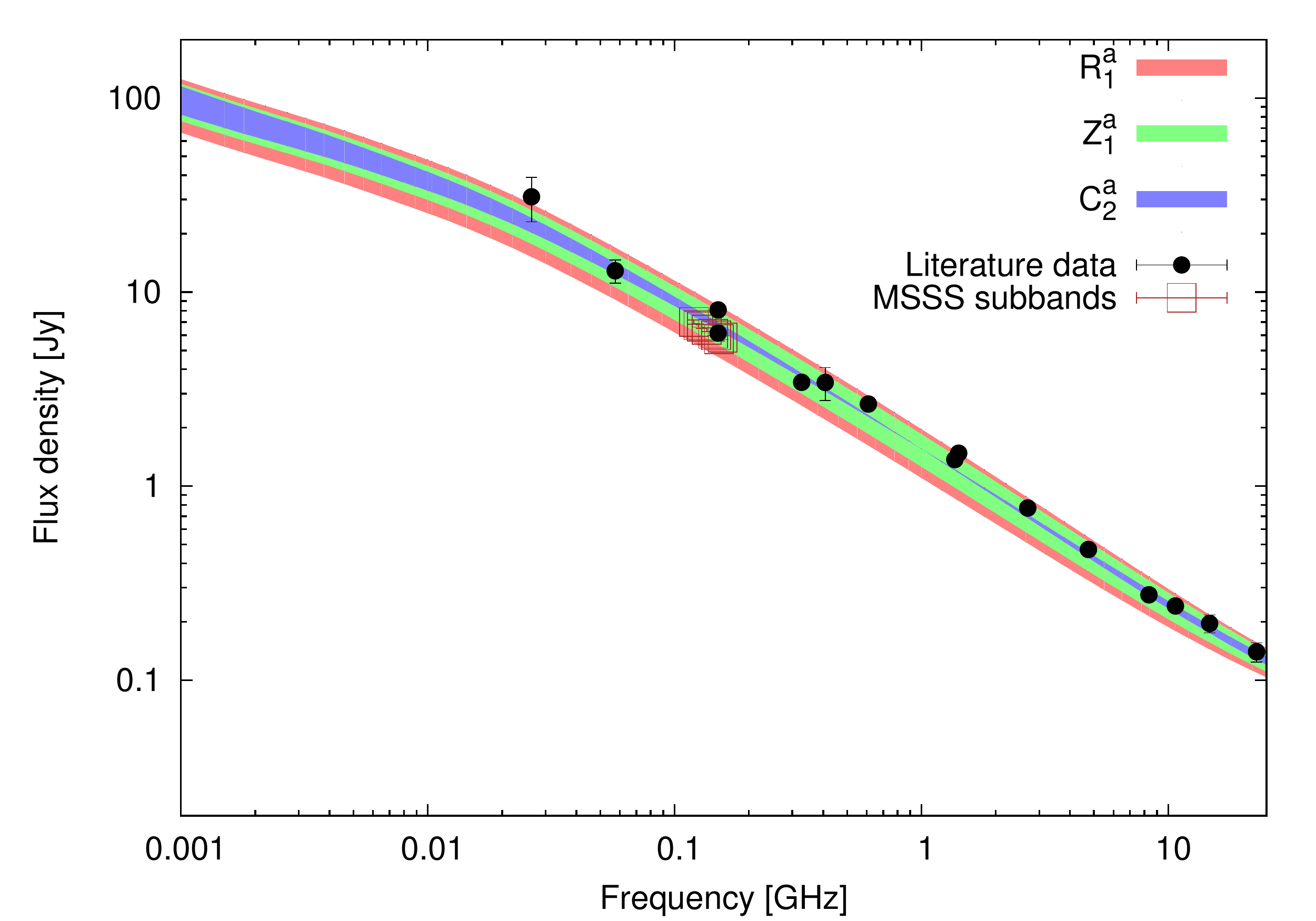}
\caption{Ranges of the modelled global spectrum of M\,51-like galaxy as affected by increasing and decreasing best-fit model parameters $R^a_1$, $Z^a_1$, and $C^a_2$ by $\pm\,20\%$. Black solid circles denote literature data used during the modelling and the interpolated flux density at 150 MHz from the MSSS survey. For the sake of completeness, the data from individual MSSS sub-bands are also shown (brown rectangles).
}
\label{f:m51_model}
\end{figure}

The obtained model (Fig. \ref{f:m51_profiles}) reproduced well the observed radial intensity and spectral index profiles from \cite{mulcahy14}. The break in the radial intensity profile at $r$$\approx$$2$\,kpc defines the transition from the inner to the outer star-forming disk. Another break at $r$$\approx$$10$\,kpc corresponds to the virtual disappearance of CR sources at larger galactocentric radii where, as suggested by \citet{mulcahy14}, the outer edge of the star-forming disk is located. The resulting global spectrum for the modelled galaxy also matches well the literature data (Fig.~\ref{f:m51_spectra}, left panel). 

We calculated the extent to which the model parameters affect the fitted spectrum and the goodness of the model fit by changing the obtained best-fit values by +20\% and -20\%. The resulting chi-square values are presented in Table \ref{t:modelM51}. The influence is also illustrated in Fig. \ref{f:m51_model} for several selected parameters. The ranges of the fitted model spectra are narrow, comparable to the data spread. The results were similar for all the other parameters.

\begin{figure*}
\includegraphics[clip,angle=0,width=1.0\textwidth]{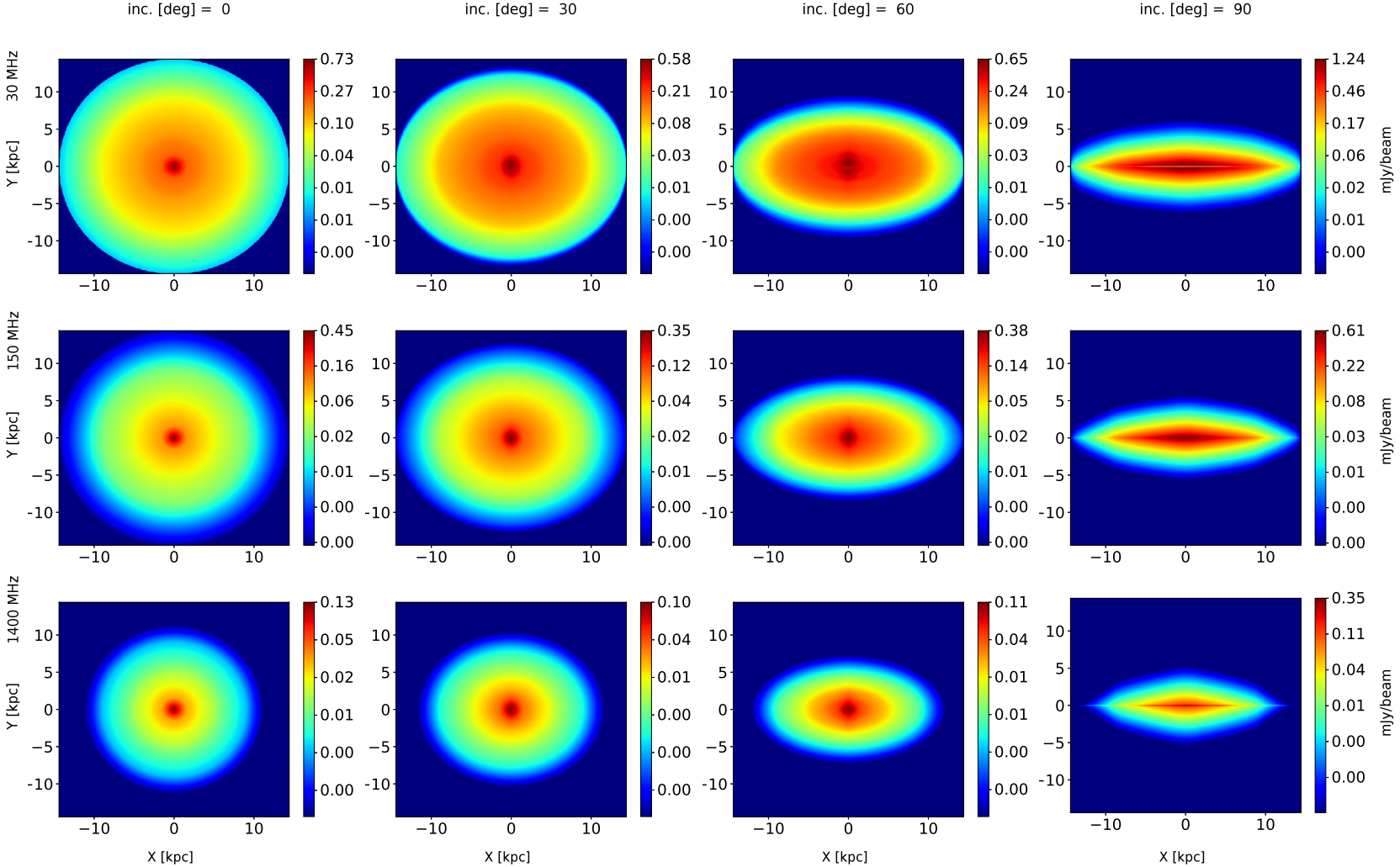}
\caption{Synthetic radio emission maps of a nonstarburst M\,51-like galaxy; from top to bottom, rows show results at 30 MHz, 150 MHz, and 1400 MHz; from left to right, columns show results for inclination angles of 0, 30, 60, and 90 degrees. The colour scale is in mJy/beam.
}
\label{f:m51_maps}
\end{figure*}

Using the obtained model parameters to best fit the M\,51 observations, we then solved the transfer equation for a number of viewing angles and frequencies. The results are shown in the form of synthetic maps of the radio continuum emission in Fig.~\ref{f:m51_maps}. Our modelling revealed no significant absorption effects at 150\,MHz, that is, the frequency of the MSSS survey, independently of viewing angle. This result corresponds well to the observed global spectrum of the edge-on galaxy NGC\,891 recently obtained by \cite{mulcahy18}, in which the effect of thermal absorption is also barely visible. Strongly reduced emission due to absorption was only found in our maps at 30 MHz for highly inclined galaxies ($i\ge 60\degr$).

The effects of thermal absorption on the global spectrum are very limited; they can only be seen as a weak flattening below 20\,MHz (Fig.~\ref{f:m51_spectra}, middle panel). This is due to the relatively small scale-height of thermal electron distribution in M\,51 (100\,pc of the thin disk and 1\,kpc in the thick disk, see Eq. \ref{e:ne_m51}) as compared to the synchrotron scale-height (0.3\,kpc of the thin disk and 1.8\,kpc of the thick disk, see Table \ref{t:modelM51}). Moreover, thermal emission has scale-heights that are half the size of the mentioned scales of thermal electrons due to the direct proportionality of thermal emission to $N_e^2$. A more quantitative analysis of absorption is presented in Table \ref{t:tauM51}, which contains the calculated optical depths at different radial distances from the centre of the modelled face-on galaxy. We note that only in the centre of the galaxy is the optical depth significant ($\tau=0.032$) giving a reduction of synchrotron emission at 150\,MHz by only $(1-\exp(-0.3))\times 100=3\%$. Our model predicts that at a much lower frequency of 30\,MHz much stronger absorption, by up to 30\% (for $\tau=0.36$), is to be observed, but only in the central region up to about 1\,kpc radius (Table \ref{t:modelM51}). 

\begin{table}
\caption{Optical depth $\tau$ along the line of sight in the modelled M\,51-like galaxy at different radial distances $r$ from the galactic centre at two different frequencies.}
\begin{center}
\begin{tabular}{lccc}
\hline
\hline
$r$[kpc] & $EM(r)$ [cm$^{-6}$pc] & $\tau$(150\,MHz) & $\tau$(30\,MHz) \\
\hline
0 & 1800 & 0.032 & 0.93 \\
1 & 700 & 0.012 & 0.36 \\
10 & 14 & 0.0002 & 0.007 \\
\hline
\end{tabular}
\end{center}
\label{t:tauM51}
\end{table}

Therefore, we expect no spectral turnovers due to absorption in the {\em global} spectra of galaxies of this type at any observation frequency available from the Earth. Our modelling properly accounts for the results of our statistical analysis of the observed spectra of nearby galaxies (Sect. \ref{s:inclination}): the statistically weak effect of thermal absorption on the global galaxy spectra was suggested by the radio-FIR relation and confirmed by the spectral index versus inclination angle diagram (Fig. \ref{f:alfadiff_incl}). Our analysis therefore indicates that for typical late-type spiral galaxies with only moderate stellar activity, such as M\,51, any observed curvature visible in the global spectrum at frequencies above 20\,MHz is the result of curved local synchrotron spectra rather than of internal thermal absorption. Such curved nonthermal spectra can be explained by the combined effects of CR electron energy losses and CR transport.

As a case in point, observations of the Milky Way spectrum at very low frequencies \citep[e.g.][]{brown73, ellis82} show a turnover at about 3\,MHz, interpreted as the effect of thermal absorption. This is different to our modelling of M\,51-like galaxies, where no turnover is found down to 1\,MHz. However, it should be noted that the measurements of the Milky Way spectrum are by necessity made from the inside, while external galaxies are seen from the outside. It is likely that due to our location within the plane of the Galaxy's disk, we can observe efficient absorption of radio emission by the thermally ionised gas in the thin disk. If the Milky Way does not differ too much from M\,51, our modelling would therefore indicate that there would be no turnover in the Galaxy's spectrum as seen from outside.

We also noticed that there could be another effect associated with the Milky Way. It can be expected that its thermal gas absorbs the incoming emission of external galaxies, which could be seen as spectral turnovers at about 3\,MHz for objects outside the Galactic plane. Very-low-frequency observations of galaxies distributed over the celestial sphere and studies of their low-frequency spectra would also contribute substantially to our knowledge of the Milky Way's ISM and facilitate modelling of its detailed structure.

From our modelling, we also obtained local spectra of the central parts of M\,51-like galaxies viewed at a different inclination angle (Fig.~\ref{f:m51_spectra}, right panel). The spectra of these compact regions are different from the modelled integrated galaxy spectra (Fig.~\ref{f:m51_spectra}, middle panel), because the shapes of local spectra highly depend on the galaxy tilt. In the case of edge-on galaxies, they show distinct turnovers at frequencies of about $100-200$\,MHz due to strong absorption in the thin disk. The observed local spectra in the star-forming regions in the disk of NGC\,891 seem to be indeed clearly affected by thermal absorption, showing low nonthermal spectral index, below the injection value \citep{mulcahy18}.

For the less inclined galaxies ($i<60\degr$), turnovers become less pronounced, and at very low frequencies (especially below $10-20$\,MHz) a rise of spectra is observed due to unabsorbed synchrotron emission in the thick galactic disk. Our modelling therefore indicates that observations of compact regions in galaxies at high resolution and very low frequencies would enable us to constrain models of their constituents as well as properties of the ISM. Unfortunately, such observations would require radio instruments located in space or on the Moon.

\subsection{Modelling of M82-like galaxies and relation to sample galaxies} 
\label{s:m82}

We also modelled \object{M\,82}, a starburst galaxy visible at a high inclination angle ($i\approx 80\degr$). It shows an extremely complex ISM, due to strong gravitational interaction with M\,81. Numerous young massive star clusters and supernovae in the central starburst region drive an H$\alpha$- and X-ray-bright superwind \citep{westmoquette09}. In the circumnuclear starburst region (with a radius of about 250\,pc) the emission measures in the compact sources are typically about $10^5\,$cm$^{-6}$\,pc \citep{varenius15, wills97}. Pervasive diffuse H$\alpha$ emission at low levels is found throughout the halo of the galaxy up to about 1\,kpc height \citep{shopbell98}. Therefore, in our model of M\,82, we approximated the 3D distribution of thermal electrons in the form of thin and thick exponential disks and estimated their characteristic scales by fitting a double-exponential function to vertical profiles of H$\alpha$ emission obtained from the H$\alpha$ map of M\,82 from the SINGS survey \citep{kennicutt03}, which gave the following formula:
\begin{equation}
\begin{aligned}
\label{e:m82_ne}
N_e(r,z)=A^b\, \exp\left(\frac{-r}{0.4\,\mathrm{kpc}}\right)\,\exp\left(\frac{-|z|}{0.04\,\mathrm{kpc}}\right)+ \\
 \qquad \qquad \qquad \qquad B^b\,\exp \left( \frac{-r}{0.4\,\mathrm{kpc}} \right) \exp \left( \frac{-|z|}{0.22\,\mathrm{kpc}}\right),
\end{aligned}
\end{equation}
where $A^b$ and $B^b$ are scale factors and the superscript $b$ denotes, in this and the following equations, the modelling of M\,82-like galaxies. The values of these factors were determined so that the integrated radio thermal emission should correspond to a thermal fraction of 3\% at 1.4\,GHz, as estimated for M\,82 by \cite{basu12}. The peak of the emission measure in the central part of the disk, resulting from the modelled distribution of $N_e$ given by Eq.~\ref{e:m82_ne}, is $2.3\times 10^6\,$cm$^{-6}$\,pc.

\begin{table}[t]
\caption{Best-fit parameter values for the model of M\,82-like galaxy and reduced chi-square values for the models where particular parameters were increased by 20\% and decreased by 20\%.
}
\begin{center}
\begin{tabular}{lccc}
\hline
\hline
Parameter       & Best fit & $\chi^2_\nu(+20\%)$ & $\chi^2_\nu(-20\%)$ \\
\hline
$C^b_0$ & $-4.2$ & $2.46$  & $2.18$ \\
$C^b_1$ & $-0.49$ & $1.74$  & $1.27$ \\
$C^b_2$ & $-0.02$ & $0.23$  & $0.25$ \\
$C^b_3$ &  $-0.01$& $0.24$  & $0.23$ \\
$F_b$ & $0.003$ & $0.21$  & $0.22$ \\
$R^b_1$ & $0.4$\,kpc & $6.83$  & $4.28$ \\
$R^b_2$ & $0.8$\,kpc & $0.33$  & $0.23$ \\
$Z^b_1$ & $0.13$\,kpc & $2.26$  & $1.34$ \\
$Z^b_2$ & $0.6$\,kpc & $0.30$  & $0.22$ \\
$Db$ & $-0.2$ & $0.27$  & $0.23$ \\
$Eb$ & $-0.05$ & $0.35$  & $0.23$ \\
\hline
\end{tabular}
\end{center}
\label{t:modelM82}
\end{table}

\begin{figure*}
\centering
\includegraphics[clip,width=4.25cm,angle=-90,viewport=25 8 500 690]{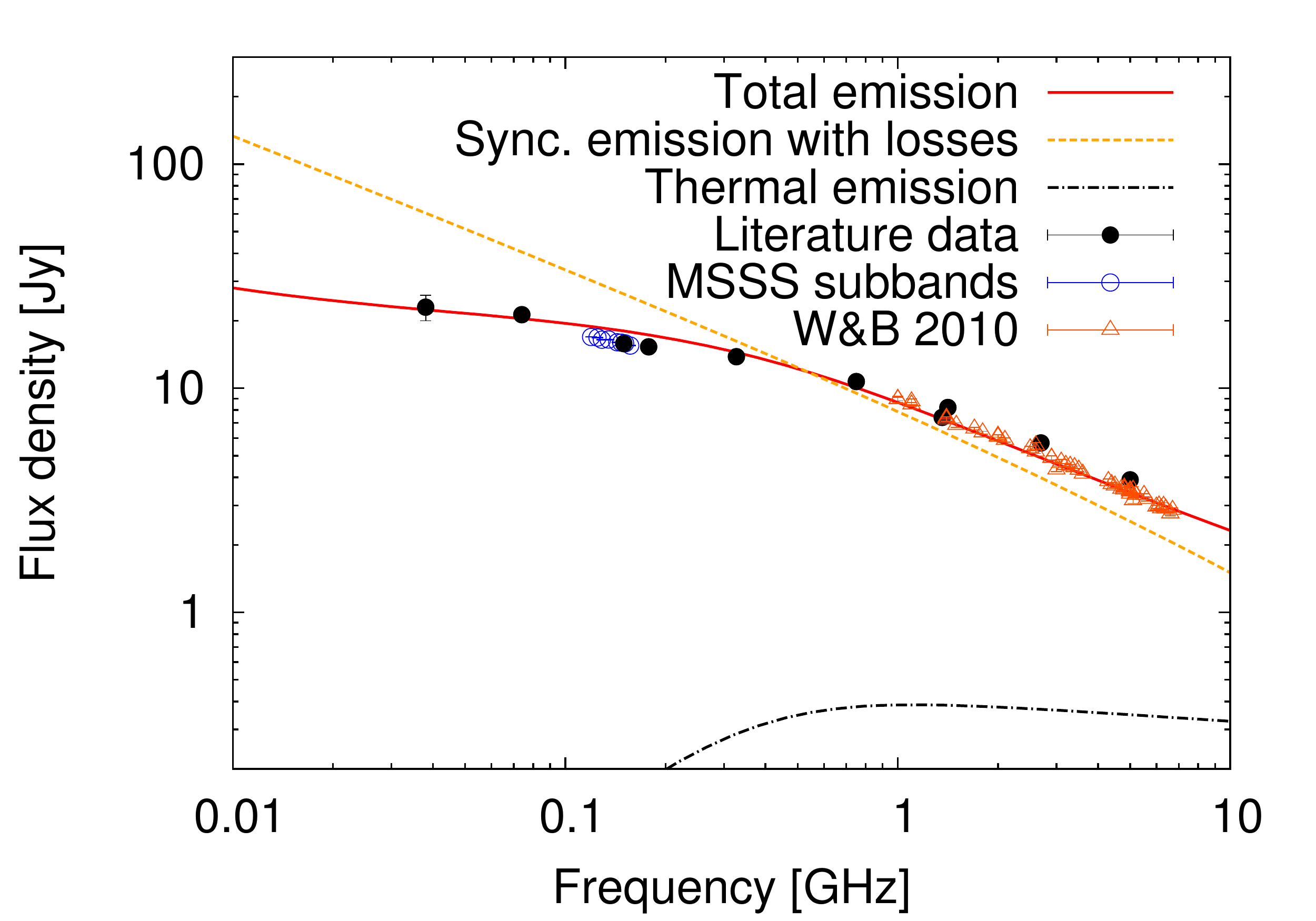}
\includegraphics[clip,width=4.25cm,angle=-90, viewport=25 8 500 690]{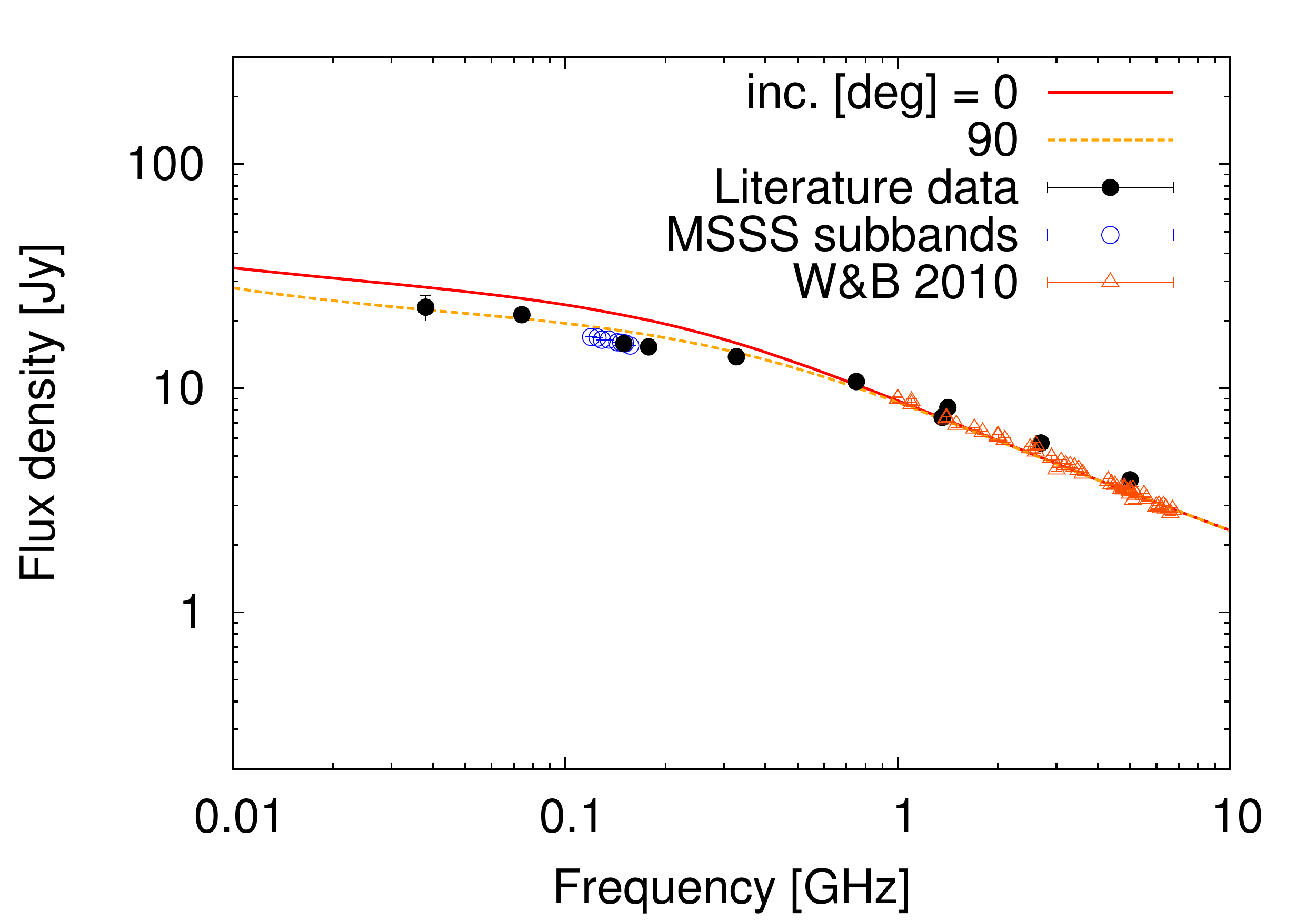}
\includegraphics[width=4.25cm,angle=-90,viewport=25 8 500 680]{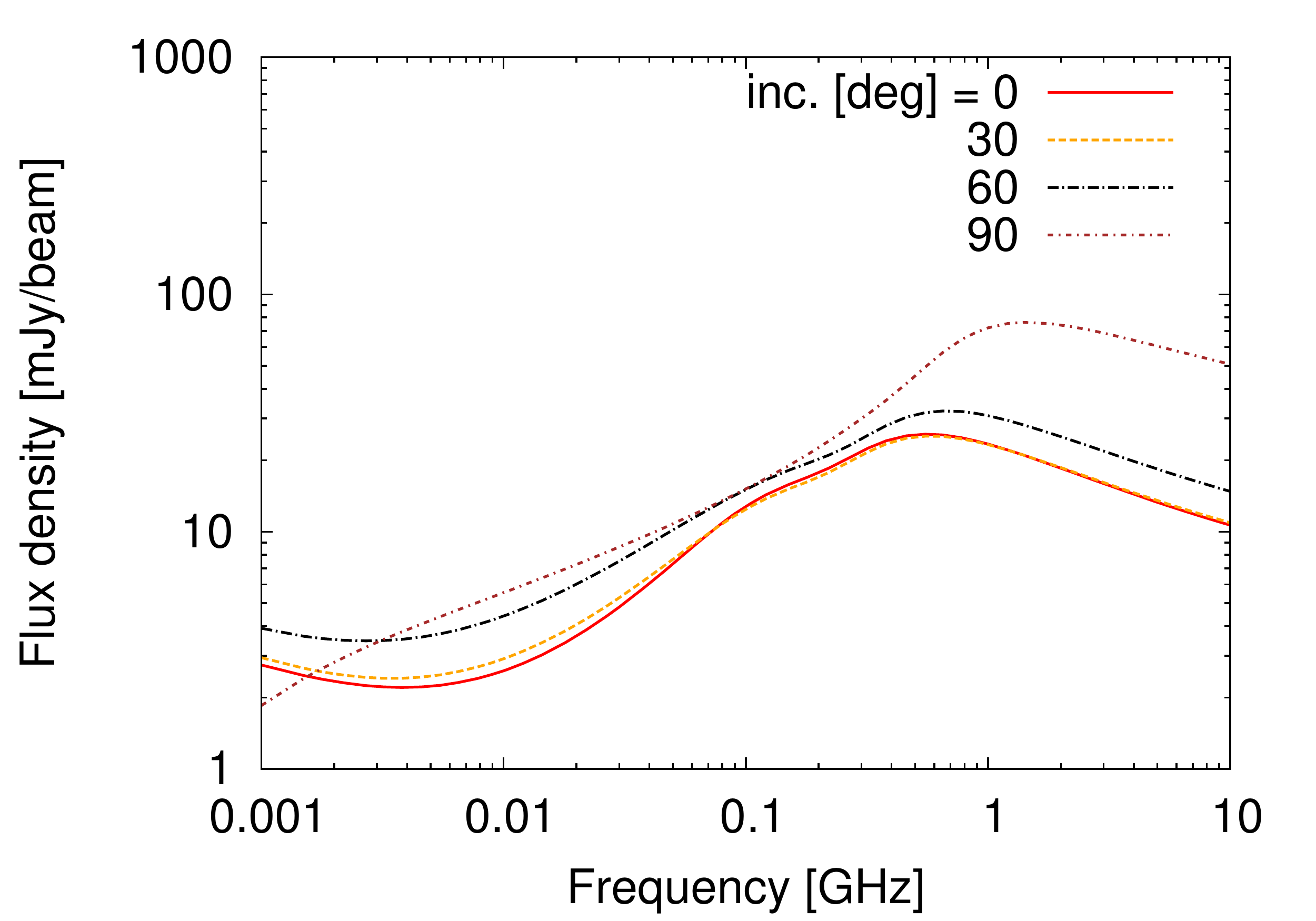}
\caption{Global spectra from our 3D model of an M\,82-like galaxy. Left: Global spectrum of a highly inclined galaxy (red solid line) with the synchrotron component without absorption (orange dotted line) and thermal free-free emission with absorption (black dashed line). Black solid circles denote literature data used during modelling, including the interpolated flux density at 150 MHz from the MSSS survey. For the sake of completeness, data from individual MSSS sub-bands  (blue circles) and flux densities determined during the commissioning phase of the 42-element Allen Telescope
Array \citep[][]{williams10}(orange triangles) are also shown.
Middle: Comparison of global spectra for edge-on (orange dotted line)  and face-on galaxies (red solid line).  Right: Local spectra from the galaxy centre for different inclination angles.
}
\label{f:m82_spectra}
\end{figure*}

\begin{figure}
\centering
\includegraphics[width=6.4cm,angle=-90]{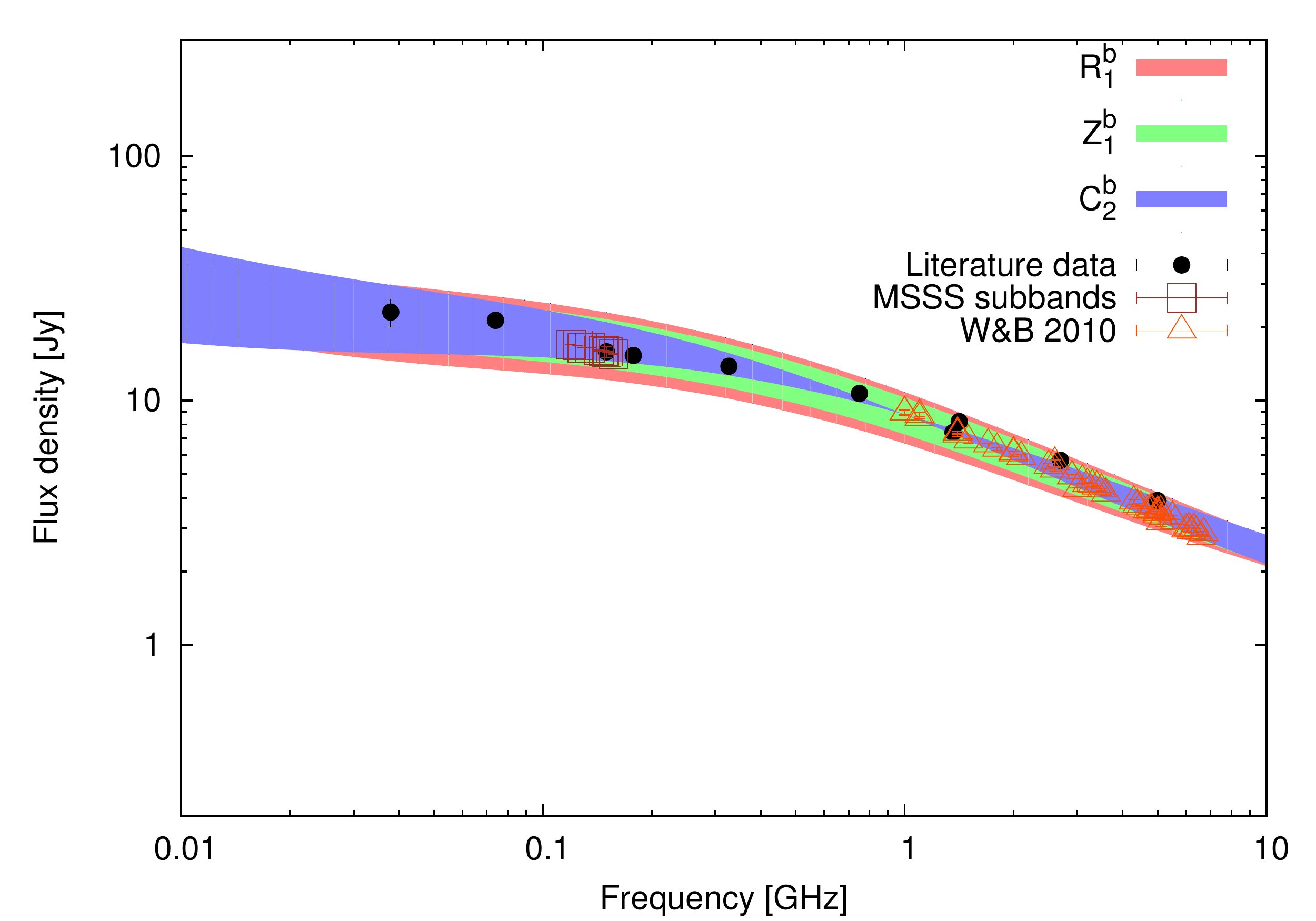}
\caption{Ranges of the modelled global spectrum of M\,82-like galaxy as affected by increasing and decreasing best-fit model parameters $R^b_1$, $Z^b_1$, and $C^b_2$ by $\pm\,20\%$. Black solid circles denote literature data used during modelling including the interpolated flux density at 150 MHz from the MSSS survey. For the sake of completeness data from the individual MSSS sub-bands (brown rectangles) and flux densities  determined during commissioning phase of the 42-element Allen Telescope Array \citep[][]{williams10} (orange triangles) are also shown.
}
\label{f:m82_model}
\end{figure}

In the nonthermal part of the model, we introduced a thick disk with
the scale height similar to the scale length. Such a synchrotron halo
extending up to a height of about $3\farcm 5$ ($\approx$3.56\,kpc) is
seen in the observations of M\,82 at 330\,MHz and 1.4\,GHz \citep{adebahr13}. In our model, we did not include details such
as the asymmetries between the northern and southern parts of the
galaxy, which could be the result of faster CR electron energy losses or
slower wind speeds in the southern outflow. Neither did we take into
account the emission from individual supernovae and star clusters. Therefore, the nonthermal radio emission is represented in the model in the following way:
\begin{equation}
\begin{aligned}
 \label{e:b_m82}
B_n(r,z) =C^b\,\left[\exp\left(\frac{-r}{R^b_1}\right)\, \exp\left(\frac{-|z|}{Z^b_1}\right) + F^b\,\exp\left(\frac{-r}{R^b_2}\right)\,%\times \\
\exp\left(\frac{-|z|}{Z^b_2}\right)\right]
\end{aligned}
,\end{equation}
where $C^b$ and $F^b$ are scale factors, $R^b_1$ and $R^b_2$ denote exponential scale-lengths of inner and outer synchrotron disks, and $Z^b_1$ and $Z^b_2$ are scale-heights of exponential thin and thick disks, respectively. Similar constituents of M\,82 were involved in the modelling of \citet{yoast13}, but in a simplified cylindrical geometry.

We employed a similar fitting procedure as for M\,51, comparing modelled and observed radio emission profiles along the major and minor axes, using the data of \citet{adebahr13} as reference. We assumed power-law functions:  $R^b_1(\nu), R^b_2(\nu)\propto (\nu/\nu_1 )^{Db}$ and $Z^b_1, Z^b_2\propto (\nu/\nu_1)^{Eb}$, where $\nu_1$ refers to 0.33 GHz. The values obtained for the scale lengths $R_1$ and $R_2$ were 0.4 and 0.8\,kpc, respectively, and for the scale heights $Z_1$ and $Z_2$ we found values of 0.13 and 0.6\,kpc, respectively (see Table \ref{t:modelM82}). 

As for M\,51 (Sect. \ref{s:m51}), the integrated spectra obtained from the range of M\,82 models were compared with the observed spectrum by measuring the goodness of fit with the chi-square value. The best-fit spectrum is shown in Fig. \ref{f:m82_spectra} (left panel) and the corresponding values of model parameters are given in Table \ref{t:modelM82}. We also present how the goodness of fit was affected when model parameters were changed by $\pm20\%$. In Fig. \ref{f:m82_model}, we illustrate this influence for three selected parameters. The range of the obtained modelled spectra is narrow (close to the data spread), and its shape confirms spectral flattening at low frequencies. 

\begin{figure*}
\includegraphics[clip,angle=0,width=1.0\textwidth]{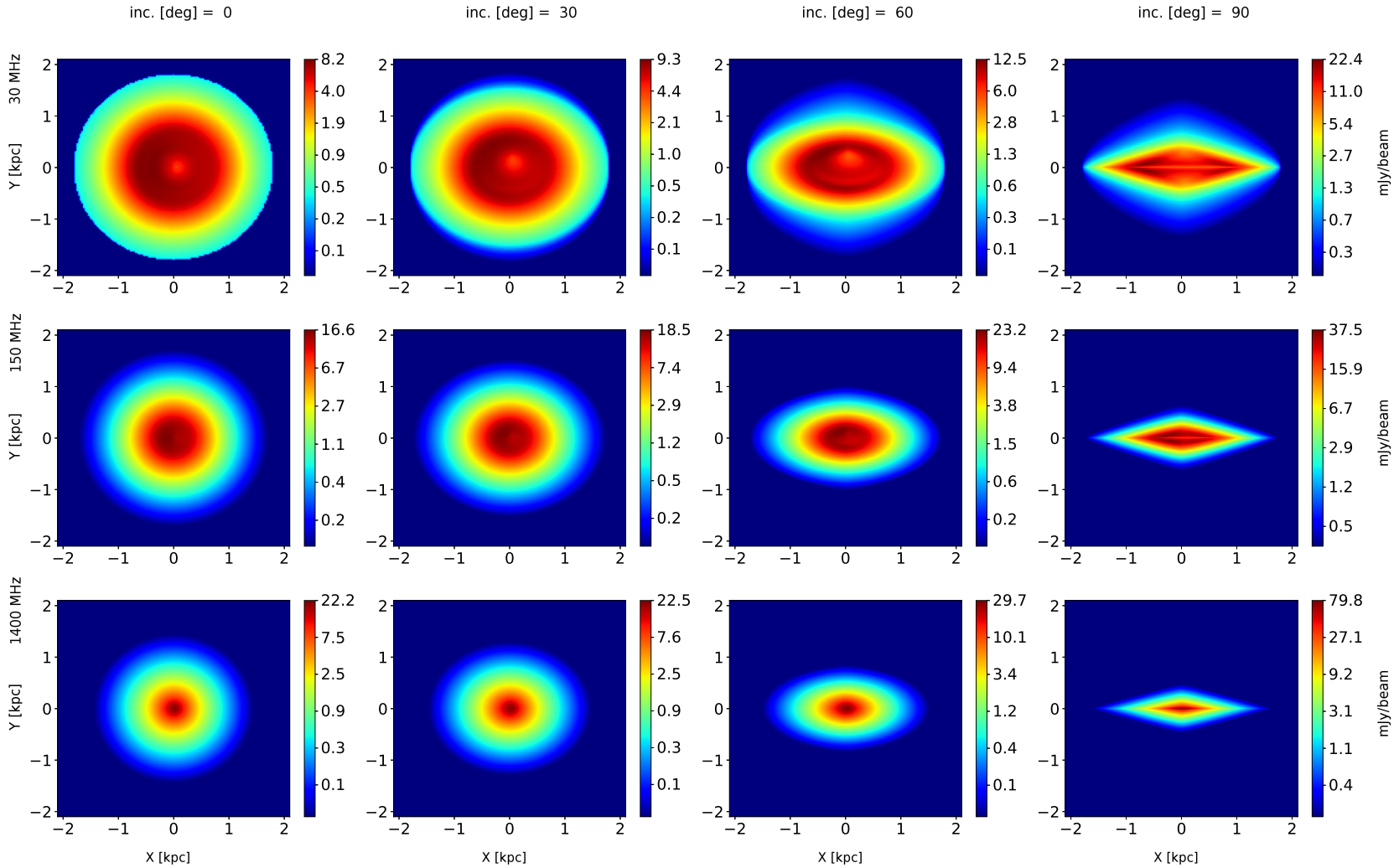}
\caption{Synthetic radio emission maps of a starburst M\,82-like galaxy; from top to bottom, rows show results at 30 MHz, 150 MHz, and 1400 MHz; from left to right, columns show results for inclination angles of 0, 30, 60, and 90 degrees. The colour scale is in mJy/beam.
}
\label{f:m82_maps}
\end{figure*}

Some modelled maps that we obtained by solving the radiative transfer equation (\ref{e:s_n}) for various frequencies and viewing angles are presented in Fig.~\ref{f:m82_maps}. The effect of free-free absorption can already be seen in the maps at 150\,MHz as a significant drop of the emission in the thin star-forming disk. This region corresponds to a modelled emission measure of several $10^5\,$cm$^{-6}$\,pc, which well reproduces the results of \citet{wills97}. The LOFAR high-resolution observations of M\,82 by \citet{varenius15} reveal a similar decrease of emission in the galactic disk. While in our model maps the drop at 150\,MHz is relatively localised, the maps at 30\,MHz predict a much larger depression all around the galactic centre (Fig.~\ref{f:m82_maps}, top-right panel). The existence of such a depression could be verified by future observations of starburst galaxies with the LOFAR low-band antenna (LBA).

\begin{figure}
\centering
\includegraphics[width=6.4cm,angle=-90]{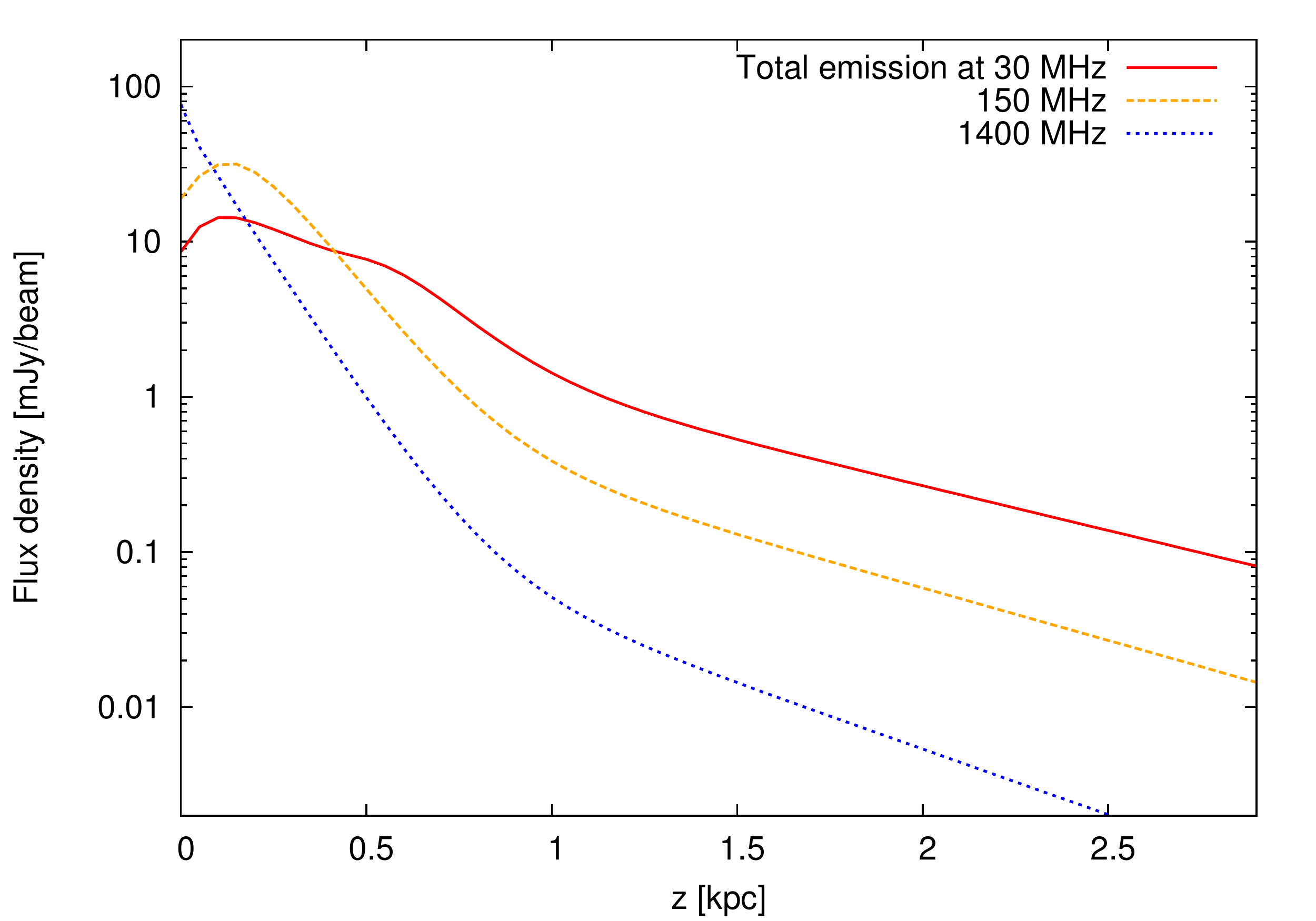}
\caption{Modelled vertical profiles of the total radio emission of an M\,82-like galaxy seen edge-on at 30, 150, and 1400\,MHz.
}
\label{f:m82_proiles}
\end{figure}

Unlike typical nearby galaxies like M\,51, starbursting objects like M\,82 show strong absorption effects that appear in the integrated spectra for all viewing angles (Fig.~\ref{f:m82_spectra}, left and middle panels). 
Towards lower frequencies, the modelled integrated spectrum of M\,82 first flattens but then slightly rises again below frequencies of approximately 10\,MHz due to its synchrotron halo. Depending on the viewing angle of the modelled sources, a spectral flattening due to absorption is observed between 200 and 400\,MHz. As can be inferred from Fig. \ref{f:m82_proiles}, the global spectrum is mainly shaped by the emission from the inner part of the galaxy, at least to a height of about 500\,pc. At heights in excess of about 700\,pc, the profiles show no indication of spectral flattening.

Modelling M\,82 as a single starburst region involving mixed thermal and nonthermal gas components, as in the geometrically simplified starburst model presented in Fig.~\ref{f:condon}, cannot account for the observational data at frequencies below 200\,MHz (Fig. \ref{f:m82_spectra}, left panel and Fig. \ref{f:m82_model}). One needs to add a synchrotron envelope with layers in front of the disk that are not thermally absorbed to increase the integrated flux densities at the lowest frequencies \citep[see][]{yoast13}. Such an effect was achieved even in our geometrically simplified model after introducing a halo, as presented in Fig. \ref{f:condon_halo}.

The local spectra of the modelled compact starburst core show strong turnovers that move from $\approx$1\,GHz for edge-on galaxies to $\approx$500\,MHz for face-on objects. For M\,51-like galaxies (Sect.~\ref{s:m51}), due to the lower emission measure in the galactic centre, the turnover appears at lower frequencies (of about 200\,MHz) and for the edge-on objects only (Fig.~\ref{f:m51_spectra}, right panel). 

The SFR of M\,82 is relatively high as compared to most of the
galaxies in our sample, while not being particularly extreme
among starburst galaxies. Accordingly, we can predict even stronger
absorption effects in the integrated spectra of violently star-forming
galaxies as well as of more distant galaxies. The local spectra of
compact star-forming regions within these objects should reveal
turnovers at even higher frequencies (around or even exceeding
1\,GHz). Starburst galaxies are in fact rare in the local Universe,
which can explain why we found just very weak spectral curvatures in
our sample of galaxies (Sect.~\ref{s:spectra}). However, we note that \citet{calistro-rivera17} searched for
evolution of spectral curvature in star-forming galaxies out to high
redshift (which in their flux-limited sample corresponds to
luminosity) and did not find any evidence for it, albeit with much
sparser sampling of the radio spectra than we present here. Clear evidence of thermal absorption in spectra of starburst galaxies was recently reported by \citet{galvin18}.

Our modelling of both M\,51 and M\,82 required the introduction of locally curved nonthermal spectra, which could be the effect of spectral ageing of the CR electrons due to for example synchrotron losses growing stronger at high CR energies. Such losses become most pronounced at high radio frequencies and are likely to be best observed in galactic outskirts due to CR transport \citep{mulcahy14}. A difference between the spectral indices at low and high frequencies could also be influenced by other CR electron energy losses: nonthermal bremsstrahlung, and ionisation, which is most effective at low energies \citep{murphy09}. For typical field strengths of $10\,\mu$G and atomic hydrogen densities of 1\,cm$^{-3}$, the transition energy above which synchrotron cooling begins to dominate over bremsstrahlung and ionisation cooling is at about 1\,GeV, which corresponds to a spectral break at about 200\,MHz. For stronger fields (e.g. $15\,\mu$G) or less dense ISM, synchrotron cooling can dominate over CR electron losses down to frequencies of about 100\,MHz.

On the other hand, \citet{klein18} claim that the integrated spectra of most spiral and dwarf galaxies can be fitted with a break or an exponential decline in their high-frequency part ($>1$\,GHz). Such features could be due either to synchrotron and inverse-Compton losses accompanied by advective transport in a
galactic wind or due to a cutoff in recent times of the CR production
rate. These results are different from those of \citet{niklas97} and \citet{tabatabaei17}, who, in
spite of not considering breaks in the synchrotron spectrum in their
models, were still able to fit most galaxy spectra quite well. Our
modelling shows that in general the validity of the approach used in
all the aforementioned studies, which are based entirely on the integrated spectra of galaxies, is very limited. As local synchrotron spectra are different in different galaxy regions (as can be seen e.g. in M\,51; Fig. \ref{f:m51_profiles}, right panel), the assumption of a constant break frequency or thermal fraction in the entire confinement volume is not correct. Different break/flattening frequencies due to locally different CR losses or thermal absorption are smeared out after adding the emission in constructing the global spectrum \citep[see][]{lisenfeld00, basu15}. Spatially resolved radio observations of galaxies and modelling of spatial composition of the ISM components are advisable in further studies of radiation processes and CR transport as well as for a proper interpretation of the global spectra of galaxies. For most of our sample galaxies, such high-resolution observations at low frequencies are not yet available; however, they will soon become available from the ongoing LOFAR Two-metre Sky Survey \citep[LoTSS;][]{shimwell17}. 

\section{Summary and conclusions} 
\label{s:conclusions}

We compiled a sample of 129 nearby star-forming galaxies from the LOFAR MSSS survey and measured their integrated flux densities at 150\,MHz (Table \ref{t:sample}). Combining them with flux densities at various frequencies from the literature, we constructed global radio spectra for 106 objects. A statistical analysis of the data led us to the following conclusions. 

\begin{itemize}

\item The low-frequency radio-FIR relation resembles the high-frequency one, which confirms (i) the good quality of the MSSS data, (ii) the proper removal of AGN-dominated galaxies from the sample, and (iii) the tight relation between dust emission and the almost pure synchrotron radiation at low radio frequencies. It may also indicate that the effect of thermal absorption in the sample is relatively weak at 150\,MHz.
\item In general, the global spectra of galaxies in our sample are statistically flatter at lower frequencies, or equivalently, steeper at higher frequencies (Fig. \ref{f:alfa_hist}). The sample median value of the low-frequency spectral index $\alpha_\mathrm{low}$, measured between $\approx$50\,MHz and 1.5\,GHz, is $-0.57\pm0.01$, which is significantly larger than the corresponding value of the high-frequency spectral index $\alpha_\mathrm{high}=-0.77\pm0.03$, calculated between 1.3\, and 5\,GHz. 
For each galaxy we estimated the spectral curvature as the difference $\Delta\alpha=\alpha_\mathrm{low}-\alpha_\mathrm{high}$ and found that its median value is $0.18\pm0.02$, indicating a small but statistically significant spectral flattening at low frequencies.
\item There is no tendency for highly inclined galaxies to have more flattened low-frequency spectra (Fig. \ref{f:alfadiff_incl}). Hence the observed flattening is not due to thermal absorption, which contradicts the suggestion by \citet{israel90}. We show that the flattening does not depend on galaxy morphology either. 
\end{itemize}

Interpretation of these results and discussion of processes to shape the low-frequency spectra were performed with a numerical model of radio emission involving absorption and projection effects. Our modelling of M\,51-like galaxies, which represent low-SFR objects in the local Universe, suggests that local depressions of emission due to thermal absorption can only be seen in spatially resolved radio maps at frequencies below $\approx$ 30\,MHz. We show that thermal absorption can influence the global spectrum at frequencies only below 20\,MHz. A weak flattening observed in some galaxy spectra above this frequency (cf. Fig. \ref{f:spectra}) can be accounted for by the curvature in the synchrotron spectrum due to for example CR electron energy losses and diffusion of CRs from their regions of acceleration. We also predict a possible influence of the Milky Way foreground gas on external galaxy spectra at frequencies below 10\,MHz.

Our modelling of M\,51-like galaxies indicates that local radio spectra of central parts of galaxies may differ much from their global spectra, and may even show turnovers due to thermal absorption in edge-on objects. Such turnovers could appear at frequencies of about 100-200\,MHz (Fig. \ref{f:m51_spectra}) and high-resolution observations are necessary to reveal them. Our models also show that even galaxies with simple, power-law-like global spectra can locally show significantly curved spectra, especially far away from the galactic centres. Therefore, integrated spectra alone cannot be properly interpreted without supplementary data on the properties of the local ISM within the galaxies.

Moreover, we modelled galaxies with higher SFRs based on observations of M\,82. 
Compared to M\,51, the modelled M\,82-like galaxies show much stronger absorption effects irrespective of the galaxy tilt. At 150\,MHz, strong absorption was found mainly along the thin disk, while at 30\,MHz our model predicts a more extensive depression of emission around the galaxy centre (Fig. \ref{f:m82_maps}). Towards low frequencies, the integrated spectra of modelled starbursts first flatten and then slightly increase again at about 10\,MHz due to presence of the synchrotron halo. This could explain why simple modelling based on just one mixed region of thermal and nonthermal gas is not able to reproduce the observed spectra of M\,82.

In the modelled local spectra of the central starbursts strong turnovers were found below about 1\,GHz for galaxies viewed edge-on and below about 500\,MHz for galaxies viewed face-on (Fig. \ref{f:m82_spectra}). Therefore one can expect much stronger free-free absorption effects and significant changes in the radio spectra of LIRGs and more distant galaxies with high SFRs. Since starburst galaxies are not frequent in the local Universe we observed mostly weak spectral curvatures in our sample.

In this work we primarily discussed the impact of free-free absorption
on the low-frequency spectra of galaxies. The forthcoming, more
sensitive LoTSS HBA survey \citep{shimwell17}, and
particularly the planned LBA survey at $\approx 50$\,MHz, as well as
Murchison Widefield Array data covering a wide frequency range \citep[e.g.][]{hurley17} are highly desirable to distinguish
between a number of processes shaping the spectra in galaxies at
frequencies below 100\,MHz (Sect. \ref{s:intro}) as well as to verify
our conjectures. Broad-band polarization observations spanning low and
high frequencies would also help to constrain the thermal content of
galaxies.

\begin{acknowledgements}
WJ acknowledges support by the Polish National Science Centre grant
No. 2013/09/N/ST9/02511. MJH acknowledges support from the UK Science
and Technology Facilities Council [ST/M001008/1]. GG acknowledges the CSIRO OCE Postdoctoral Fellowship. AOC and AMS gratefully acknowledge support from the European Research Council (ERC) under grant ERC-2012-StG-307215 LODESTONE. RJvW acknowledges support from the ERC Advanced Investigator programme NewClusters 321271. RPB acknowledges funding from the ERC under the European Union's Horizon 2020 research and innovation programme (ERC Starting Grant agreement 715051 ``Spiders''). We acknowledge the
help of the Academic Computer Centre CYFRONET AGH in Krakow, Poland,
where our numerical analyses were partly performed with the use of the
computing cluster Prometheus. We acknowledge the use of the HyperLeda
(http://leda.univ-lyon1.fr), NED (http://nedwww.ipac.caltech.edu), and
SINBAD (http://simbad.u-strasbg.fr) databases.
We thank Fatemeh Tabatabaei and an anonymous referee for helpful comments on the manuscript. This paper is based (in part) on data obtained with the International LOFAR Telescope (ILT). LOFAR \citep{haarlem13} is the Low Frequency Array designed and constructed by ASTRON. It has observing, data processing, and data storage facilities in several countries, that are owned by various parties (each with their own funding sources), and that are collectively operated by the ILT foundation under a joint scientific policy. The ILT resources have benefited from the following recent major funding sources: CNRS-INSU, Observatoire de Paris and Universit\'e d'Orl\'eans, France; BMBF, MIWF-NRW, MPG, Germany; Science Foundation Ireland (SFI), Department of Business, Enterprise and Innovation (DBEI), Ireland; NWO, The Netherlands; The Science and Technology Facilities Council, UK, Ministry of Science and Higher Education, Poland. 

\end{acknowledgements}

\bibliographystyle{aa} %style aa.bst
\bibliography{bibliography} %file .bib

\longtab{
\begin{longtable}{lccccccc}
\caption{\label{t:sample} Main properties of a sample of 129 nearby galaxies. For 106 objects global continuum spectra and spectral indexes ($\alpha_\mathrm{low}$, $\alpha_\mathrm{high}$) were obtained.}\\
\hline
Name & T-Type & Inclination & Distance [Mpc] & $S_\mathrm{MSSS}$ [Jy] & $\alpha_\mathrm{low}$ & $\alpha_\mathrm{high}$ & Flags \\
\hline
\endfirsthead
\caption{Continued}\\
\hline
Name & T-Type & Inclination & Distance [Mpc]  & $S_\mathrm{MSSS}$ [Jy]& $\alpha_\mathrm{low}$ & $\alpha_\mathrm{high}$ & Flags \\
\hline
\endhead
\hline
\endfoot
IC 10 & 10 & 31.1 & 0.66 & $0.72 \pm 0.11$ & $-0.38 \pm 0.08$ & $-0.33 \pm 0.06$ &  1  2  2  1 \\
IC 342 & 6 & 18.5 & 3.3 & $11.20 \pm 0.56$ & $-0.64 \pm 0.06$ & $-0.85 \pm 0.12$ & 1  1  1  1 \\
IC 883/UGC8387 & 10 & 61.4 & 91.9 & $0.18 \pm 0.01$ & $-0.43 \pm 0.07$ & $-0.63 \pm 0.06$ & 2  2  1  1 \\
NGC 278 & 3 & 12.8 & 8.6 & $0.30 \pm 0.05$ & $-0.45 \pm 0.04$ & $-0.64 \pm 0.05$ & 1  1  2  1 \\
NGC 520 & 1 & 75.7 & 28.8 & $0.33 \pm 0.09$ & $-0.34 \pm 0.08$ & $-0.63 \pm 0.10$ & 3  2  2  1 \\
NGC 693 & 0 & 66.8 & 20.9 & $0.14 \pm 0.03$ & $-0.47 \pm 0.05$ & $-0.36 \pm 0.18$ & 1  1  2  2 \\
NGC 660 & 1 & 78.7 & 11.3 & $0.85 \pm 0.05$ & $-0.39 \pm 0.03$ & $-0.63 \pm 0.05$ & 1  1  1  1 \\
NGC 772 & 3 & 59.6 & 32.8 & $0.33 \pm 0.03$ & $-0.59 \pm 0.13$ & $-1.10 \pm 0.49$ & 3  2  2  1 \\
NGC 828 & 1 & 44.8 & 71.7 & $0.38 \pm 0.05$ & $-0.62 \pm 0.08$ & $-0.64 \pm 0.01$ & 1  1  1  1 \\
NGC 876/77 & 4 & 41.1 & 52.4 & $0.50 \pm 0.06$ & $-0.70 \pm 0.01$ & $-0.84 \pm 0.02$ & 2  3  2  2 \\
NGC 891 & 3 & 90 & 7 & $3.03 \pm 0.16$ & $-0.65 \pm 0.03$ & $-0.82 \pm 0.04$ & 1  1  1  1 \\
NGC 935/IC1801 & 6 & 52.9 & 55.2 & $0.26 \pm 0.03$ & $-0.72 \pm 0.02$ & $-1.16 \pm 0.28$ & 3  3  1  1 \\
NGC 972 & 2 & 65.8 & 20.6 & $0.94 \pm 0.13$ & $-0.46 \pm 0.04$ & $-0.84 \pm 0.11$ & 1  1  1  1 \\
NGC 1055 & 3 & 62.5 & 13.4 & $0.83 \pm 0.06$ & $-0.66 \pm 0.02$ & $-0.77 \pm 0.10$ & 1  1  1  1 \\
NGC 1530 & 3 & 58.3 & 32.8 & $0.23 \pm 0.05$ & $-0.64 \pm 0.03$ & $-0.74 \pm 0.04$ & 1  1  2  1 \\
NGC 1569 & 10 & 66.5 & 2.9 & $0.75 \pm 0.04$ & $-0.34 \pm 0.04$ & $-0.50 \pm 0.04$ & 1  1  1  1 \\
NGC 1961 & 5 & 47 & 52.5 & $1.00 \pm 0.06$ & $-0.74 \pm 0.01$ & $-0.90 \pm 0.04$ & 2  1  1  1 \\
NGC 2276 & 5 & 39.8 & 32.1 & $1.22 \pm 0.08$ & $-0.75 \pm 0.08$ & $-1.09 \pm 0.07$ & 1  1  2  2 \\
NGC 2339 & 4 & 52.6 & 30 & $0.43 \pm 0.04$ & $-0.59 \pm 0.02$ & $-0.93 \pm 0.03$ & 1  1  2  1 \\
NGC 2403 & 6 & 61.3 & 3.6 & $0.97 \pm 0.05$ & $-0.47 \pm 0.03$ & $-0.57 \pm 0.05$ & 1  1  2  1 \\
NGC 2623 & 3 & 83 & 73.4 & $0.14 \pm 0.03$ & $-0.21 \pm 0.04$ & $-0.30 \pm 0.08$ & 2  2  1  1 \\
NGC 2633 & 3 & 53.8 & 28.9 & $0.27 \pm 0.03$ & $-0.54 \pm 0.05$ & $-0.59 \pm 0.15$ & 1  1  1  1 \\
NGC 2639 & 1 & 44.7 & 43 & $0.46 \pm 0.05$ & $-0.65 \pm 0.04$ & $-0.83 \pm 0.07$ & 1  1  1  1 \\
NGC 2683 & 3 & 82.8 & 5.5 & $0.24 \pm 0.03$ &  $-0.51 \pm 0.06$ & $-0.79 \pm 0.16$ &  1  1  2  1 \\
NGC 2748 & 4 & 68.1 & 19.9 & $0.14 \pm 0.01$ & N/A & N/A & 1  1  3  2 \\
NGC 2782 & 1 & 45.2 & 34 & $0.37 \pm 0.04$ & $-0.53 \pm 0.06$ & $-0.76 \pm 0.01$ & 1  1  1  1 \\
NGC 2841 & 3 & 65.2 & 8.4 & $0.42 \pm 0.03$ & $-0.66 \pm 0.01$ & $-0.81 \pm 0.23$ & 1  1  1  1 \\
NGC 2903 & 4 & 67.1 & 7.2 & $1.38 \pm 0.07$ & $-0.51 \pm 0.02$ & $-0.91 \pm 0.01$ & 1  1  1  1 \\
NGC 2964 & 4 & 42.9 & 18 & $0.28 \pm 0.03$ & $-0.49 \pm 0.05$ & $-0.64 \pm 0.13$ & 1  1  1  1 \\
NGC 2976 & 5 & 60.5 & 3.6 & $0.19 \pm 0.02$ & $-0.48 \pm 0.11$ & $-0.62 \pm 0.22$ & 1  1  1  2 \\
NGC 2985 & 2 & 37.9 & 17 & $0.35 \pm 0.04$ & $-0.81 \pm 0.13$ & $-0.67 \pm 0.48$ & 1  1  2  1 \\
NGC 3034(M82) & 7 & 76.9 & 3.6 & $15.84 \pm 0.79$ & $-0.33 \pm 0.02$ & $-0.53 \pm 0.06$ & 1  1  1  1 \\
NGC 3044 & 5 & 90 & 17.2 & $0.36 \pm 0.13$ & $-0.56 \pm 0.01$ & $-0.73 \pm 0.03$ & 1  1  2  2 \\
NGC 3079 & 7 & 90 & 14.9 & $3.74 \pm 0.19$ & $-0.69 \pm 0.02$ & $-0.86 \pm 0.06$ & 1  1  1  1 \\
NGC 3147 & 4 & 31.2 & 37 & $0.42 \pm 0.05$ & $-0.67 \pm 0.13$ & $-0.57 \pm 0.04$ & 1  1  1  1 \\
NGC 3184 & 6 & 14.4 & 12.1 & $0.30 \pm 0.03$ & $-0.56 \pm 0.10$ & $-0.79 \pm 0.17$  & 1  1  2  1 \\
NGC 3221 & 6 & 65.7 & 54.5 & $0.38 \pm 0.06$ & $-0.59 \pm 0.05$ & $-0.73 \pm 0.01$ & 2  1  2  1 \\
NGC 3226/27 & 1 & 68.3 & 10 & $0.22 \pm 0.05$ & $-0.52 \pm 0.24$ & $-0.57 \pm 0.01$ &  2  3  2  1 \\
NGC 3294 & 5 & 60.7 & 20.9 & $0.19 \pm 0.02$ & $-0.46 \pm 0.09$ & $-0.58 \pm 0.17$ & 1  1  2  1 \\
NGC 3310 & 4 & 16.1 & 13.2 & $1.19 \pm 0.06$ & $-0.54 \pm 0.02$ & $-0.74 \pm 0.02$ & 3  3  1  1 \\
NGC 3359 & 5 & 47.2 & 13.5 & $0.13 \pm 0.03$ & $-0.42 \pm 0.01$ & $-1.04 \pm 0.21$ & 1  1  2  1 \\
NGC 3395/96 & 6 & 57.8 & 21.7 & $0.30 \pm 0.03$ & $-0.54 \pm 0.08$ & $-0.65 \pm 0.11$ & 3  3  2  1 \\
NGC 3424 & 3 & 79.2 & 18.9 & $0.23 \pm 0.03$ & $-0.53 \pm 0.05$ & $-0.64 \pm 0.01$ & 1  1  2  1 \\
NGC 3432 & 9 & 90 & 8.1 & $0.16 \pm 0.08$ & $-0.30 \pm 0.01$ & $-0.54 \pm 0.12$ & 1  1  2  2 \\
NGC 3437 & 5 & 2.8 & 17 & $0.23 \pm 0.04$ & $-0.52 \pm 0.06$ & $-0.62 \pm 0.02$ & 1  1  2  1 \\
NGC 3448 & 4 & 79.2 & 18.1 & $0.16 \pm 0.03$ & $-0.47 \pm 0.03$ & $-0.62 \pm 0.03$ & 3  2  1  1 \\
NGC 3486 & 5 & 46 & 9.1 & $0.20 \pm 0.03$ & $-0.46 \pm 0.13$ & $-0.79 \pm 0.44$ &  1  2  3  2 \\
NGC 3504 & 2 & 26.1 & 20.1 & $0.80 \pm 0.05$ & $-0.53 \pm 0.05$ & $-0.73 \pm 0.03$ & 1  2  2  1 \\
NGC 3556(M108) & 6 & 67.5 & 9.3 & $0.91 \pm 0.05$ & $-0.56 \pm 0.05$ & $-0.99 \pm 0.05$ & 1  1  2  1 \\
NGC 3583 & 3 & 58.7 & 28.4 & $0.25 \pm 0.02$ & $-0.64 \pm 0.03$ & $-0.67 \pm 0.03$ & 1  1  2  2 \\
NGC 3593 & 0 & 86.7 & 9.2 & $0.25 \pm 0.04$ & $-0.44 \pm 0.04$ & $-0.48 \pm 0.18$ & 1  1  2  1 \\
NGC 3627(M66) & 3 & 67.5 & 9.7 & $1.38 \pm 0.08$ & $-0.53 \pm 0.04$ & $-0.85 \pm 0.18$ & 1  1  1  1 \\
NGC 3628 & 3 & 79.3 & 11.2 & $1.80 \pm 0.10$ & $-0.51 \pm 0.04$ & $-0.72 \pm 0.04$ & 1  1  1  1 \\
NGC 3631 & 5 & 34.7 & 15.4 & $0.42 \pm 0.04$ & $-0.69 \pm 0.02$ & $-0.67 \pm 0.06$ & 2  1  1  1 \\
NGC 3646 & 4 & 63.9 & 57.76 & $0.36 \pm 0.04$ & $-0.77 \pm 0.04$ & $-0.85 \pm 0.05$ & 1  1  1  1 \\
NGC 3655 & 5 & 47.1 & 19.8 & $0.29 \pm 0.04$ & $-0.61 \pm 0.04$ & $-0.85 \pm 0.08$ & 1  1  1  1 \\
NGC 3735 & 5 & 83.5 & 35.9 & $0.34 \pm 0.04$ & $-0.66 \pm 0.05$ & $-0.82 \pm 0.21$ & 1  1  2  1 \\
NGC 3799/3800 & 3 & 75.7 & 44 & $0.21 \pm 0.04$ & $-0.64 \pm 0.01$ & $-0.62 \pm 0.01$ & 1  3  1  1 \\
NGC 3810 & 5 & 48.2 & 13.2 & $0.31 \pm 0.04$ & $-0.47 \pm 0.15$ & $-0.93 \pm 0.26$ & 1  1  2  1 \\
NGC 3893 & 5 & 59.8 & 12 & $0.40 \pm 0.04$ & $-0.50 \pm 0.03$ & $-0.97 \pm 0.05$ & 1  1  1  1 \\
NGC 3938 & 5 & 14.1 & 16.9 & $0.31 \pm 0.03$ & $-0.66 \pm 0.05$ & $-0.99 \pm 0.14$ & 1  1  1  1 \\
NGC 3949 & 4 & 56.5 & 12 & $0.37 \pm 0.03$ & $-0.50 \pm 0.04$ & $-0.95 \pm 0.03$ & 1  1  2  2 \\
NGC 3982 & 3 & 29.9 & 15.8 & $0.28 \pm 0.03$ & $-0.71 \pm 0.05$ & $-0.78 \pm 0.22$ & 1  1  1  1 \\
NGC 3987 & 3 & 90 & 61.2 & $0.20 \pm 0.04$ & N/A & N/A & 1  1  3  1 \\
NGC 4041 & 4 & 22 & 16.3 & $0.37 \pm 0.04$ & $-0.58 \pm 0.01$ & $-0.81 \pm 0.01$ & 1  1  1  1 \\
NGC 4051 & 4 & 30.2 & 9.7 & $0.35 \pm 0.03$ & $-0.60 \pm 0.04$ & $-0.83 \pm 0.03$ & 1  1  2  1 \\
NGC 4088 & 4 & 71.2 & 12 & $0.89 \pm 0.06$ & $-0.63 \pm 0.04$ & $-0.86 \pm 0.16$ & 3  1  1  1 \\
NGC 4096 & 5 & 80.5 & 12 & $0.15 \pm 0.02$ & $-0.50 \pm 0.08$ & $-0.80 \pm 0.15$ & 1  1  2  1 \\
NGC 4100 & 4 & 79.6 & 12 & $0.15 \pm 0.04$ & $-0.58 \pm 0.16$ & $-0.57 \pm 0.01$ & 1  1  2  1 \\
NGC 4102 & 3 & 58.2 & 12 & $0.72 \pm 0.05$ & $-0.54 \pm 0.03$ & $-0.91 \pm 0.05$ & 1  1  1  1 \\
NGC 4157 & 3 & 90 & 12 & $0.82 \pm 0.05$ & $-0.55 \pm 0.07$ & $-0.53 \pm 0.07$ & 1  1  1  1 \\
NGC 4194 & 10 & 52.6 & 33.7 & $0.21 \pm 0.03$ & $-0.37 \pm 0.07$ & $-0.63 \pm 0.14$ & 2  2  2  1 \\
NGC 4217 & 3 & 81 & 17.1 & $0.40 \pm 0.03$ & $-0.59 \pm 0.04$ & $-0.73 \pm 0.10$ & 1  1  1  1 \\
NGC 4254(M99) & 5 & 20 & 14.4 & $1.86 \pm 0.14$  & $-0.66 \pm 0.05$ & $-0.93 \pm 0.05 $ & 1 1 1 1\\
NGC 4321(M100) & 4 & 23.4 & 16 & $0.97 \pm 0.10$ & $-0.52 \pm 0.04$ & $-1.09 \pm 0.05$ & 1  1  2  1 \\
NGC 4414 & 5 & 56.6 & 9.6 & $0.91 \pm 0.06$ & $-0.58 \pm 0.03$ & $-0.89 \pm 0.07$ & 1  1  1  1 \\
NGC 4449 & 10 & 63.5 & 3.8 & $0.93 \pm 0.05$ & $-0.60 \pm 0.05$ & $-0.46 \pm 0.06$ & 1  2  1  1 \\
NGC 4490 & 7 & 79 & 7.5 & $2.88 \pm 0.15$ & $-0.58 \pm 0.03$ & $-0.84 \pm 0.02$ & 3  1  1  1 \\
NGC 4532 & 10 & 90 & 17.5 & $0.48 \pm 0.28$ & $-0.61 \pm 0.01$ & $-0.70 \pm 0.05$ & 1  2  2  2 \\
NGC 4559 & 6 & 64.8 & 10.9 & $0.22 \pm 0.02$ & $-0.55 \pm 0.09$ & $-0.79 \pm 0.24$ &  1  1  3  1 \\
NGC 4565 & 3 & 90 & 16.4 & $0.61 \pm 0.04$ & $-0.67 \pm 0.07$ & $-0.81 \pm 0.05$ & 1  1  2  1 \\
NGC 4631 & 7 & 90 & 8.3 & $4.71 \pm 0.26$ & $-0.55 \pm 0.02$ & $-0.70 \pm 0.01$ & 1  1  1  2 \\
NGC 4736(M94) & 2 & 31.7 & 4.1 & $0.71 \pm 0.05$ & $-0.39 \pm 0.04$ & $-0.65 \pm 0.07$ & 1  1  1  2 \\
NGC 4793 & 5 & 61.1 & 32.7 & $0.50 \pm 0.04$ & $-0.63 \pm 0.05$ & $-0.79 \pm 0.08$ & 1  1  2  1 \\
NGC 4826 & 2 & 63.9 & 5.5 & $0.24 \pm 0.03$ & $-0.38 \pm 0.04$ & $-0.48 \pm 0.06$ & 1  1  2  1 \\
NGC 5005 & 4 & 77.1 & 13.6 & $0.76 \pm 0.05$ & $-0.64 \pm 0.04$ & $-0.93 \pm 0.04$ & 1  1  1  1 \\
NGC 5033 & 5 & 64.6 & 11.7 & $0.75 \pm 0.05$ & $-0.54 \pm 0.05$ & $-0.72 \pm 0.48$ & 1  1  1  1 \\
NGC 5055(M63) & 4 & 54.9 & 6.6 & $1.60 \pm 0.08$ & $-0.62 \pm 0.04$ & $-0.88 \pm 0.05$ & 1  1  1  1 \\
NGC 5194(M51) & 4 & 32.6 & 6.3 & $5.89 \pm 0.30$ & $-0.64 \pm 0.04$ & $-0.94 \pm 0.04$ & 3  1  1  2 \\
NGC 5248 & 4 & 56.4 & 15.4 & $0.69 \pm 0.05$ & $-0.66 \pm 0.05$ & $-0.31 \pm 0.07$ & 1  1  2  2 \\
NGC 5256 & 10 & N/A & 109.9 & $0.42 \pm 0.04$ & $-0.59 \pm 0.07$ & $-0.88 \pm 0.05$ & 2  3  2  1 \\
NGC 5371 & 4 & 54 & 34 & $0.21 \pm 0.04$ & $-0.64 \pm 0.02$ & $-0.61 \pm 0.03$ & 1  1  1  1 \\
NGC 5457(M101) & 6 & 16 & 6.7 & $2.64 \pm 0.16$ & $-0.57 \pm 0.01$ & $-0.72 \pm 0.02$ & 1  1  1  2 \\
NGC 5676 & 4 & 66.2 & 28.1 & $0.34 \pm 0.04$ & $-0.65 \pm 0.07$ & $-0.91 \pm 0.04$ & 1  1  2  1 \\
NGC 5678 & 3 & 61.6 & 25.7 & $0.48 \pm 0.04$ & $-0.66 \pm 0.05$ & $-0.96 \pm 0.01$ & 1  1  2  1 \\
NGC 5775 & 5 & 83.2 & 22.3 & $0.61 \pm 0.12$ & $-0.36 \pm 0.03$ & $-0.84 \pm 0.05$ & 1  1  1  1 \\
NGC 5900 & 3 & 74.7 & 34 & $0.26 \pm 0.04$ & N/A & N/A & 1  1  3  1 \\
NGC 5929/30 & 2 & 24.4 & 34 & $0.45 \pm 0.04$ & $-0.66 \pm 0.02$ & $-0.73 \pm 0.03$ & 3  3  2  1 \\
NGC 5936 & 3 & 19.3 & 53.1 & $0.53 \pm 0.05$ & $-0.65 \pm 0.02$ & $-0.72 \pm 0.09$ & 2  1  1  1 \\
NGC 5953/54 & 1 & 44 & 26.4 & $0.27 \pm 0.05$ & $-0.49 \pm 0.07$ & $-0.62 \pm 0.14$ & 3  3  2  1 \\
NGC 5962 & 5 & 50.8 & 26.2 & $0.29 \pm 0.04$ & $-0.57 \pm 0.02$ & $-0.73 \pm 0.02$ & 1  1  1  1 \\
NGC 6052 & 5 & 39.6 & 60.5 & $0.37 \pm 0.07$ & $-0.57 \pm 0.01$ & $-0.60 \pm 0.03$ & 2  2  1  1 \\
NGC 6181 & 5 & 68.2 & 31.6 & $0.29 \pm 0.04$ & $-0.43 \pm 0.07$ & $-0.46 \pm 0.02$ & 1  1  2  1 \\
NGC 6217 & 4 & 44.8 & 18.3 & $0.22 \pm 0.03$ & N/A & N/A & 1  1  3  1 \\
NGC 6285/6 & 3 & 23.7 & 74.7 & $0.48 \pm 0.03$ & $-0.63 \pm 0.04$ & $-0.91 \pm 0.05$ &  3  3  2  1 \\
NGC 6574 & 4 & 41.5 & 30.1 & $0.34 \pm 0.04$ & $-0.58 \pm 0.02$ & $-0.82 \pm 0.06$ & 1  1  1  1 \\
NGC 6643 & 5 & 62.7 & 19.9 & $0.37 \pm 0.04$ & $-0.60 \pm 0.02$ & $-1.08 \pm 0.07$ & 1  1  1  1 \\
NGC 6701 & 1 & 17.3 & 53.1 & $0.33 \pm 0.04$ & $-0.60 \pm 0.05$ & $-1.15 \pm 0.01$ & 2  1  2  1 \\
NGC 6764 & 3.5 & 63.9 & 31.7 & $0.43 \pm 0.03$ & $-0.62 \pm 0.04$ & $-0.99 \pm 0.14$ & 1  1  1  1 \\
NGC 6951 & 4 & 50.8 & 18.7 & $0.21 \pm 0.03$ & $-0.51 \pm 0.07$ & $-0.61 \pm 0.06$ & 1  1  1  1 \\
NGC 7331 & 3 & 70 & 10.9 & $1.43 \pm 0.09$ & $-0.66 \pm 0.02$ & $-0.96 \pm 0.20$ & 1  1  1  1 \\
NGC 7448 & 4 & 70.1 & 28.8 & $0.42 \pm 0.09$ & $-0.55 \pm 0.06$ & $-0.71 \pm 0.11$ &  1  1  2  2 \\
NGC 7479 & 5 & 43 & 31.6 & $0.32 \pm 0.11$ & N/A & N/A & 1  1  3  1 \\
NGC 7541 & 4 & 74.8 & 35.7 & $0.84 \pm 0.14$ & $-0.79 \pm 0.02$ & $-0.94 \pm 0.10$ & 1  1  2  1 \\
NGC 7625 & 1 & 15 & 21.6 & $0.12 \pm 0.05$ & $-0.31 \pm 0.01$ & $-0.78 \pm 0.01$ & 1  1  2  2 \\
NGC 7674 & 4 & 26.7 & 116 & $0.73 \pm 0.07$ & $-0.55 \pm 0.01$ & $-0.80 \pm 0.03$ & 2  1  1  1 \\
NGC 7678 & 5 & 43.7 & 46.4 & $0.17 \pm 0.04$ & N/A & N/A & 1  1  3  1 \\
NGC 7770/1 & 1 & 66.7 & 56.7 & $0.44 \pm 0.10$ & $-0.48 \pm 0.04$ & $-0.78 \pm 0.11$ & 2  3  2  1 \\
UGC 1845 & 2 & 82.3 & 62.4 & $0.17 \pm 0.03$ & $-0.47 \pm 0.06$ & $-0.60 \pm 0.01$ & 2  1  1  1 \\
UGC 2855 & 5 & 68.2 & 16 & $1.58 \pm 0.08$ & $-0.60 \pm 0.03$ & $-0.90 \pm 0.08$ & 1  1  1  1 \\
UGC 3293 & 6 & 61.7 & 63.1 & $0.20 \pm 0.04$ & $-0.58 \pm 0.13$ & $-0.76 \pm 0.01$ & 1  1  2  2 \\
UGC 3350/1 & 2 & 54.6 & 59.3 & $0.51 \pm 0.04$ & $-0.62 \pm 0.03$ & $-0.73 \pm 0.09$ & 1  1  1  1 \\ 
UGC 3405/10 & 3 & 74.2 & 50.5 & $0.27 \pm 0.02$ & N/A & N/A & 2  3  3  2 \\
UGC 5376 & 8 & 90 & 27.7 & $0.16 \pm 0.04$ & N/A & N/A & 1  1  3  2 \\
UGC 5613 & 8 & 78.9 & 128.6 & $0.21 \pm 0.02$ & N/A & N/A & 1  1  3  1 \\
UGC 8696 & 2 & 67 & 149.1 & $0.31 \pm 0.03$ & $-0.37 \pm 0.08$ & $-0.39 \pm 0.11$ & 2  2  1  1 \\
UGC 9794 & 8 & 90 & 86.5 & $0.30 \pm 0.03$ & N/A & N/A & 1  1  3  2 \\
UGC 11041 & 2 & 56.3 & 64 & $0.15 \pm 0.05$ & N/A & N/A & 2  2  3  1 \\
UGC 12914/15 & 6 & 54.2 & 58.3 & $0.65 \pm 0.06$ & $-0.71 \pm 0.02$ & $-1.05 \pm 0.05$   & 3  3  1  1 \\
\end{longtable}
}

\end{document}